\documentclass[a4paper,fleqn]{cas-sc}
\usepackage[dvipsnames]{xcolor}
\usepackage{lineno}
\usepackage{hyperref}
\usepackage{subcaption}
\usepackage{makecell}
\usepackage{adjustbox}
\usepackage{diagbox}
\usepackage{gensymb}
\usepackage{amssymb}
\hypersetup{
citebordercolor=white,
linkbordercolor=white,
urlbordercolor=blue}
\usepackage{amsmath}
\usepackage{mathtools}
\usepackage{amsfonts}
\usepackage{graphicx}
\usepackage[authoryear]{natbib}
\usepackage[amsstyle]{cases}
\newcommand{\al}{$\rm ^{26}Al$ }
\newcommand{\fe}{$\rm ^{60}Fe$ }
\newcommand{\Xs}{$\rm X_{S,0}$ }
\newcommand{\feratio}{$\rm ^{60}Fe/^{56}Fe$ }
\newcommand{\rcmf}{$\phi_{\mm{C}}$ }
\newcommand{\etar}{$\eta_0$ }
\newcommand{\Rem}{$Re_\mm{m}$ }
\newcommand{\fcmb}{$F_{\rm CMB}$ }
\newcommand{\mfrac}{$m_{\rm frac}$ }
\newcommand{\mm}[1]{\mathrm{#1}} 
\newcommand{\alp}{$\rm ^{26}Al$}
\newcommand{\fep}{$\rm ^{60}Fe$}
\newcommand{\Xsp}{$\rm X_{S,0}$}
\newcommand{\feratiop}{$\rm ^{60}Fe/^{56}Fe$}
\newcommand{\rcmfp}{$\phi_{\mm{C}}$}
\newcommand{\etarp}{$\eta_0$}
\newcommand{\Remp}{$Re_\mm{m}$}
\newcommand{\fcmbp}{$F_{\rm CMB}$}
\newcommand{\mfracp}{$m_{\rm frac}$}

\begin{document}
\let\WriteBookmarks\relax
\def\floatpagepagefraction{1}
\def\textpagefraction{.001}

\shorttitle{A refined, versatile model for unlocking planetesimal magnetic field histories}

\shortauthors{H. Sanderson et~al.}

\title [mode = title]{Unlocking planetesimal magnetic field histories: a refined, versatile model for thermal evolution and dynamo generation}                      
%
\author[1]{Hannah R. Sanderson}[orcid = 0000-0001-5842-6985]

\cormark[1]


\ead{hannah.sanderson@earth.ox.ac.uk}

\credit{Conceptualization, Methodology, Software, Writing - Original Draft}

\affiliation[1]{organization={Department of Earth Sciences},
    addressline={University of Oxford, South Parks Road}, 
    city={Oxford},
    postcode={OX1 3AN}, 
    country={UK}}

\author[1]{James F.J. Bryson}[orcid = 0000-0002-5675-8545]
\credit{Conceptualization, Writing - Review \& Editing, Supervision}
\author[1]{Claire I.O. Nichols}[orcid = 0000-0003-2947-5694]
\credit{Conceptualization, Writing - Review \& Editing, Supervision}

\author[2]{Christopher J. Davies}[orcid = 0000-0002-1074-3815]
\credit{Writing - Review \& Editing}

\affiliation[2]{organization={School of Earth and Environment},
    addressline={University of Leeds}, 
    city={Leeds},
    postcode={LS2 9JT}, 
    country={UK}}

\cortext[cor1]{Corresponding author}

\begin{abstract}
The thermal and magnetic histories of planetesimals provide unique insights into the formation and evolution of Earth's building blocks. These histories can be gleaned from meteorites by using numerical models to translate measured properties into planetesimal behaviour. 
In this paper, we present a new 1D planetesimal thermal evolution and dynamo generation model. This magnetic field generation model is the first of a differentiated, mantled planetesimal that includes both mantle convection and sub-eutectic core solidification. We have improved fundamental aspects of mantle heat transport by including a more detailed viscosity model and stagnant lid convection parametrisations consistent with internal heating. We have also added radiogenic heating from \fe in the metallic Fe-FeS core. Additionally, we implement a combined thermal and compositional buoyancy flux, as well as the latest magnetic field scaling laws to predict magnetic field strengths during the planetesimal's thermal evolution until core solidification is complete. 
We illustrate the consequences of our model changes with an example run for a 500\,km radius planetesimal. These effects include more rapid erosion of core thermal stratification and longer duration of mantle convection compared to previous studies. The additional buoyancy from core solidification has a marginal effect on dynamo strength, but for some initial core sulfur contents it can prevent cessation of the dynamo when mantle convection ends.
Our model can be used to investigate the effects of individual parameters on dynamo generation and constrain properties of specific meteorite parent bodies. Combined, these updates mean this model can predict the most reliable and complete magnetic field history for a planetesimal to date, so is a valuable tool for deciphering planetesimal behaviour from meteorite properties.
\end{abstract}


\begin{highlights}
\item Our model can predict a complete magnetic field generation history for a planetesimal.
\item We can model dynamo generation by inward sub-eutectic core solidification.
\item We have refined the description of planetesimal mantle convection and viscosity.

\end{highlights}


\maketitle
\section{Introduction}
Magnetic field generation in rocky bodies depends on conditions in their interiors, such as the extent of differentiation, and the vigour of core and mantle convection. Planetesimals were small, rocky bodies, which accreted in the first few million years (Ma) after Solar System formation and formed the building blocks of the terrestrial planets. Some planetesimals survive today as asteroids, and meteorites are the fragments of them that reach the Earth. Magnetisations recorded in meteorites can provide key insights into the evolution of planetesimals, as well as processes in the protoplanetary disk. Planetesimal thermal evolution and magnetic field generation models are crucial for interpreting these remanences in terms of processes in these bodies and in the solar nebula that generated these fields.  

In order to generate a magnetic field by dynamo action, a rocky body must have a partially molten, metallic core, that is in vigorous motion. Core flow can be driven by thermal or compositional convection, or a combination of the two, or by mechanical forcing. However, the latter is thought not to be relevant for planetesimals \citep{dodds_thermal_2021}. Thermal convection is driven by temperature-induced density differences and requires that the heat flux across the core-mantle boundary (CMB) is superadiabatic. Compositional convection is driven by chemically-induced density differences resulting from core solidification. For example, at the present day the Earth's magnetic field is driven by partitioning of buoyant, incompatible light elements from the solidifying inner core into the liquid outer core, which rise and drive convection in the liquid outer core \citep{braginsky_structure_1963,gubbins_energetics_1977,nimmo_energetics_2007}. A record of past internally generated magnetic fields on a planetesimal tells us that this body met the requirements for thermal \citep[e.g.][]{wang_lifetime_2017} and/or compositional \citep[e.g.][]{maurel_long-lived_2021} dynamo generation at some point in its history.

Chondritic and achondritic meteorites from the inner and outer Solar System have had their paleomagnetic remanences measured \citep[e.g][]{carporzen_magnetic_2011,cournede_early_2015,wang_lifetime_2017,weiss_nonmagnetic_2017}. Some of these measurements relate to bulk rocks \citep[e.g][]{carporzen_magnetic_2011}, while others are measurements of individual phases, such as dusty olivines in chondrules \citep[e.g][]{tarduno_evidence_2012,fu_solar_2014} or cloudy zones in metal-rich meteorites \citep[e.g][]{nichols_time-resolved_2021,maurel_long-lived_2021}. These measurements combined with numerical modelling provide evidence that differentiated \citep[e.g.][]{fu_ancient_2012} and partially differentiated bodies \citep[e.g.][]{elkins-tanton_chondrites_2011,bryson_paleomagnetic_2019} generated magnetic fields. If a paleomagnetic remanence can be dated, numerical models can be used to predict a range of parent body sizes and properties that generate a magnetic field at the identified time \citep[e.g][]{bryson_paleomagnetic_2019}. If a remanence cannot be dated, numerical models can be used to estimate the age at which the magnetisation was acquired \citep[e.g.][]{tarduno_evidence_2012,bryson_long-lived_2015,nichols_time-resolved_2021}. 

Some meteorites have been interpreted to record magnetisations from the nebula field, which threaded the entire protoplanetary disk until 4--5\,Ma after CAI formation \citep{wang_lifetime_2017}. The nebula field affected stellar accretion from the protoplanetary disk \citep{wardle_magnetic_2007,weiss_history_2021} and may have influenced disk structure \citep{hu_nonideal_2019}. Paleomagentic records of the nebula field can help us understand disk lifetime \citep{wang_lifetime_2017,borlina_lifetime_2022}, structure and strength of the disk field \citep{cournede_early_2015,borlina_paleomagnetic_2021,bryson_unified_2023}, movement of solids within the disk \citep{bryson_constraints_2020,bryson_evidence_2020}, and stellar accretion rate \citep{weiss_history_2021}. Randomly oriented paleomagnetic directions between individual chondrules \citep{fu_solar_2014,fu_weak_2020,borlina_paleomagnetic_2021} can provide evidence these chondrule remanences are pre-accretionary, and therefore must have acquired a nebula field record prior to planetesimal accretion. In contrast, ancient, bulk paleomagnetic remanences require planetesimal dynamo generation models to help determine whether these remanences were imparted by dynamo or nebula fields \citep[e.g.][]{cournede_early_2015,maurel_4565-my-old_2024}.

There have been several previous models of dynamo generation in planetesimals \citep{elkins-tanton_chondrites_2011,SterenborgCrowley2013,bryson_constraints_2019,dodds_thermal_2021}. Each model iteration included a new degree of complexity, such as parametrised mantle convection \citep{SterenborgCrowley2013}, multi-stage accretion \citep{bryson_constraints_2019}, gradual accretion \citep{dodds_thermal_2021}, or core thermal stratification \citep{dodds_thermal_2021}. These models have predicted the range of planetesimal sizes that can generate a dynamo \citep[e.g. $>340$\,km][]{bryson_constraints_2019} and the timing and duration of the dynamo \citep[4--30\,Ma after CAI formation and 5--25\,Ma depending on size,][]{dodds_thermal_2021}. Additionally, these models have been used to constrain the properties of meteorite parent bodies, including the CV chondrites, H chondrites, and angrites \citep{bryson_constraints_2019,dodds_thermal_2021}. Other numerical models have been developed that focus purely on dynamo generation by core solidification and compositional convection. These models either have a fixed entropy dissipation in the core \citep{nimmo_energetics_2009} or focus on the IVA iron meteorites so have no mantle \citep{scheinberg_core_2016,neufeld_top-down_2019}. Here, we present our refined planetesimal thermal evolution and dynamo generation model, which can be used to understand the controls on planetesimal dynamo generation and recover meteorite parent body properties.

While each previous model added new behaviour, some fundamental aspects of mantle heat transport have remained unchanged between models and are reconsidered here. Magnetic field generation requires rapid core cooling (superadiabatic CMB heat flux) or core solidification, for which heat must be moved from the core to the mantle. Therefore, the timing of magnetic field generation is intrinsically linked to mantle cooling. Unlike previous models, our chosen mantle convection scaling laws are suitable for a body with internal heating and surface and basal heat fluxes and have been benchmarked against 2D numerical simulations \citep{deschamps_scaling_2021,thiriet_scaling_2019}. Mantle convection depends strongly on viscosity, so we have refined the viscosity law from previous models. Our chosen mantle convection scaling laws also respond consistently to changes to viscosity parameters, which enables a full investigation of the effects of viscosity on dynamo generation. Moreover, our criterion for the cessation of mantle convection ensures smooth and more physically realistic behaviour in the CMB heat flux and the mantle temperature profiles when the mantle becomes fully conductive. In the core, we have also included heating from radiogenic \fep, which may increase the core temperature and the heat flux into the mantle at early times.

Alongside mantle convection and variable viscosity, we also consider sub-eutectic core solidification. Modelling dynamo generation from planetesimal core solidification is challenging, because there are multiple possible mechanisms of core solidification including: growth of stable or unstable iron dendrites; iron snow \citep{scheinberg_core_2016}; or viscous delamination of solid iron from the CMB \citep{neufeld_top-down_2019}. Due to the uncertainty in solidification mechanism, some models only studied early thermal dynamos until the end of mantle convection \citep{SterenborgCrowley2013,dodds_thermal_2021} or used the time taken for eutectic core solidification as a proxy for when a compositional dynamo could be generated even though eutectic solidification cannot drive a compositional dynamo \citep{bryson_constraints_2019}. Also, \citet{bryson_constraints_2019} used different magnetic field scaling laws for thermal compared to compositional dynamos and were not able to accommodate both mechanisms driving core flow simultaneously. Models that have included sub-eutectic core solidification either focus on unmantled planetesimals \citep{scheinberg_core_2016,neufeld_top-down_2019} or neglect mantle convection in the early thermal history of the body \citep{nichols_time-resolved_2021}. Here, we have developed a model for magnetic field generation by sub-eutectic core solidification, which also considers the relative contribution of both thermal and compositional buoyancy to magnetic field generation during core solidification. 

In this paper, we describe our refined, 1D, spherically symmetric thermal evolution model in Section \ref{model}. We show the results of an example run and discuss the effects of our model changes in Section \ref{exrun}. We justify our assumptions about accretion and differentiation, the mantle melting, and core solidification in Sections \ref{acc-diff}, \ref{mantle} and \ref{csm}, respectively. Future applications of the model and areas for further refinement are addressed in Section \ref{outlook}, and we conclude in Section \ref{conc}.

\section{Thermal Evolution Model}\label{model}
\subsection{Overview}
\begin{figure}
    \centering
    \includegraphics[width=1\textwidth]{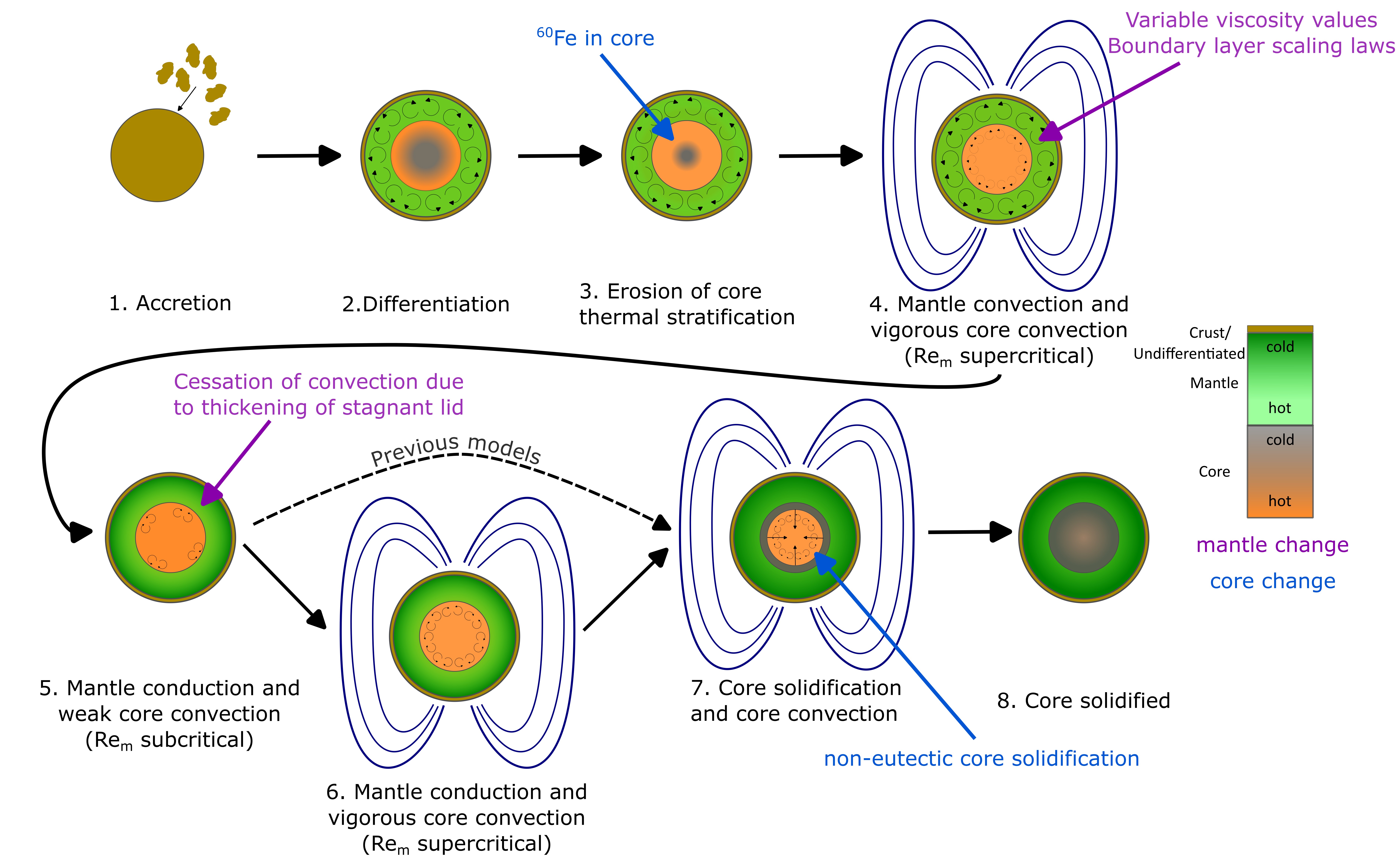}
    \caption{Schematic depicting the stages in our planetesimal thermal evolution. Blue (core) and purple (mantle) labels highlight significant refinements adopted in our model compared to \citet{SterenborgCrowley2013,bryson_constraints_2019} and \citet{dodds_thermal_2021}. Previous models only considered magnetic field generation after the end of mantle convection via core solidification (stage 7). However, our model allows for the possibility of two epochs of magnetic field generation prior to core solidification. Magnetic field generation can occur in stages 4, 6, and 7 if the magnetic Reynolds number exceeds a critical value. For low initial core sulfur contents, the core can start solidifying before the cessation of mantle convection and there may not be a pause in dynamo generation (stage 5).}
    \label{fig:lifecycle}
\end{figure}

The key stages in planetesimal thermal evolution are shown in Figure \ref{fig:lifecycle}. After accretion, the planetesimal will heat up due to the decay of radiogenic \al \citep[$t_{1/2}=0.717$\,Ma,][]{neumann_differentiation_2012}. If accreted sufficiently early \citep[within 1.5--2.5\,Ma,][]{neumann_differentiation_2012,monnereau_differentiation_2023}, the planetesimal reaches a high enough temperature that it can melt and differentiate, forming a core and mantle. During differentiation \al partitions into the mantle, while radiogenic \fe \citep[$t_{1/2}=2.62$\,Ma,][]{ruedas_radioactive_2017} partitions into the core. These two isotopes then heat the mantle and the core, respectively, and the mantle becomes unstable to convection. The adiabatic gradient is small in planetesimals due to their small size, so we neglect the core adiabat in our core convection criterion and when determining the direction of heat flow across the CMB. If the mantle becomes hotter than the core (e.g. due to strong internal heating and low surface heat flux) the core is heated by the mantle from above and becomes thermally stratified at shallow depths \citep{dodds_thermal_2021}. Once the mantle is cooler than the core, the core also begins to cool. Since the CMB heat flux, \fcmbp, is positive, core convection begins. If the magnetic Reynolds number, \Remp, is greater than a critical value this convection can lead to the onset of the dynamo. As the mantle cools, it becomes more viscous and mantle convection ceases and heat in the mantle is transported by conduction. Initially, the cessation of mantle convection decreases \fcmb and can lead to a subcritical \Rem and the cessation of the dynamo. Later, during mantle conduction, if \fcmb increases sufficiently, \Rem can become supercritical again and the dynamo can restart. Once the core cools below its liquidus, core solidification begins. This provides an additional buoyancy source for the core dynamo due to the ejection of light elements (e.g. carbon, sulfur) and can lead to a second period (or extend the first/second period) of dynamo generation. Dynamo generation is no longer possible once the core is fully solidified. 

We describe our 1D, spherically symmetric model in detail in the following sections. We begin with our mantle viscosity model, then explain each stage in the thermal evolution and finish with magnetic field generation and the numerical implementation. Model assumptions are discussed in Sections \ref{acc-diff}, \ref{mantle} and \ref{csm}.

\subsection{Mantle viscosity}\label{model-eta} 
Mantle viscosity varies over many orders of magnitude during a planetesimal's thermal evolution, due to variations in mantle temperature and melt fraction, and has a large impact on mantle cooling. It controls mantle boundary layer thicknesses during convection and the transition from mantle convection to conduction, which affects CMB heat flux and dynamo generation. In this model, we define a four-piece viscosity law to capture changing rheological behaviour with temperature, and ensure the viscosity law and convection parametrisations are self-consistent and can be fully adjusted to reflect uncertainties in viscosity parameters.

\begin{figure}
    \centering
    \includegraphics[width=1\textwidth]{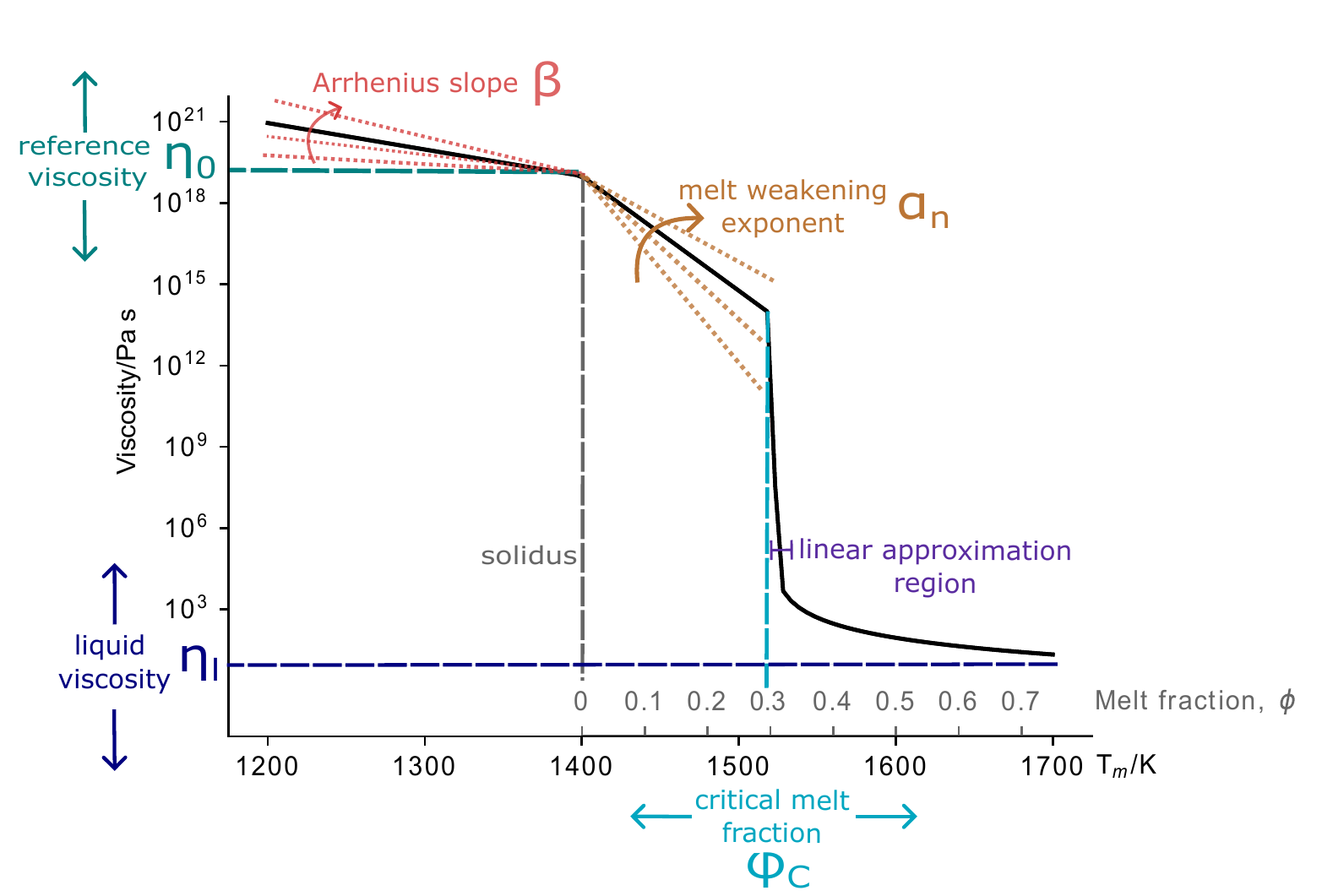}
    \caption{Variation of mantle viscosity with temperature (solid black line) for \etar$=10^{19}$\,Pa\,s, \rcmf=0.3, $\beta=0.0225\,K^{-1}$ $\alpha_\mm{n}=30$, and $\eta_l=10$\,Pa\,s. The effect of varying each parameter is shown by the coloured arrows and lines. The grey, upper ticks on the horizontal axis indicates the equivalent melt fraction for temperatures above the solidus for mantle solidus and liquidus temperatures of 1400\,K and 1800\,K, respectively. }
    \label{fig:viscosity}
\end{figure}

The temperature dependence of mantle viscosity is defined piece-wise with five control variables (Figure \ref{fig:viscosity}): critical melt fraction, \rcmfp, Arrhenius slope, $\beta$, melt weakening exponent, $\alpha_\mm{n}$, reference viscosity at the solidus, \etarp, and liquid viscosity, $\eta_l$:
\begin{subnumcases}{\label{eq:eta} \eta =}
\eta_0 \text{exp}\left(-\beta(T-T_{\rm{m,s}})\right)  & T \leq T_{\rm{m,s}} \label{eq:eta1}\\   
\eta_0 \text{exp}\left(-(\beta+\frac{\alpha_\mm{n}}{T_{\rm{m,l}}-T_{\rm{m,s}}})(T-T_{\rm{m,s}})\right)  &  T_{\rm{m,s}} < T \leq T_{\phi_c} \label{eq:eta2}\\
\eta_{\mm{C}} 10^{\frac{\Delta\eta}{w}(T-T_{\phi_c})} &  T_{\phi_c} < T < T_{\phi_c}+w \label{eq:eta3}\\
 \eta_\mm{l}\left(\frac{\phi-\phi_C}{1-\phi_C}\right)^{-2.5(1-\phi_C)} & T_{\phi_c} +w \leq T. \label{eq:eta4}
\end{subnumcases}
Here $T_{\rm{m,s}}$ and $T_{\rm{m,l}}$ are the mantle solidus and liquidus temperatures, respectively, $T_{\phi_c}$ is the temperature at \rcmf and $\eta_{\mm{C}}=\eta(T_{\phi_c})$ is the viscosity at \rcmfp, which is calculated using Equation \ref{eq:eta2}. The melt fraction, $\phi$ is calculated assuming a linear liquidus approximation: $\phi=\frac{T-T_{\rm{m,s}}}{T_{\rm{m,l}}-T_{\rm{m,s}}}$.  We do not specify a mantle composition and assume the same silicate solidus, $T_{\rm{m,s}}=1400$\,K, and liquidus, $T_{\rm{m,l}}=1800$\,K, values as \citet{bryson_constraints_2019} and \citet{dodds_thermal_2021} (see Section \ref{acc-diff-solid}). The mantle solidus and liquidus are pressure-independent constants, because the pressure variation in planetesimal mantles is small ($<$50\,MPa). 

Below the solidus, the viscosity has an Arrhenius temperature dependence (Equation \ref{eq:eta1}). This dependence is modelled using the Frank-Kamenetskii approximation at a reference temperature, $T_{\mm{ref}}$, and reference viscosity, \etarp, $\frac{E}{RT}\approx \frac{E(T-T_{\mm{ref}})}{R_\mm{g}T_{\mm{ref}}^2}=\beta(T-T_{\rm{m,s}})$, where $E$ is the activation energy, $R_\mm{g}=8.31 \rm J K^{-1} mol^{-1}$ is the gas constant, and the mantle solidus is taken as the reference temperature \citep{frank-kamenetskii_diffusion_1969}. Above the solidus, but below the critical melt fraction, this kinetic Arrhenius dependence has an additional melt weakening dependence $\alpha_\mm{n}\phi=\alpha_\mm{n}\frac{T-T_{\rm{m,s}}}{T_{\rm{m,l}}-T_{\rm{m,s}}}$ (Equation \ref{eq:eta2}). There is a sharp drop in viscosity at \rcmfp, because melt surrounds any remaining solid phases and the material disaggregates. Above \rcmfp, viscosity is described by the Krieger-Dougherty relation \citep{faroughi_generalized_2015,sturtz_birth_2022} (Equation \ref{eq:eta4}), which tends to a constant liquid viscosity as the melt fraction approaches unity.

Equation \ref{eq:eta3} is an approximation using $\frac{\mm{dlog}_{10}\eta}{\mm{d}T}=\frac{\Delta \eta}{w}$ to ensure numerical stability at the rapid drop in viscosity at \rcmfp. $w$ is the width of the region covered by this approximation, and is set to 5\,K to minimise the approximated region, but ensure numerical stability. $\Delta\eta=\mm{log}_{10}(\eta(T_{\phi_c}+w))-\mm{log}_{10}(\eta(T_{\phi_c}))$ is the logarithmic difference in the viscosity at \rcmfp, determined by Equation \ref{eq:eta2}, and the viscosity at $T_{\phi_c}+w$, determined by Equation \ref{eq:eta4}.

The Arrhenius temperature dependence (Equation \ref{eq:eta1}) determines the viscous temperature scale, $\frac{\Delta T}{f_{\mm{rh}}}$, which determines the fraction of the temperature difference across the convecting region that will control convection \citep{michaut_formation_2022,davaille_transient_1993} 
\begin{equation}
    f_{\mm{rh}} = -\frac{\mm{d}\eta(T_\mm{m})/dT}{\eta(T_\mm{m})}\Delta T.
\label{eq:frh}\end{equation}
In this definition, $T_\mm{m}$ is the temperature of the convecting region (the mantle) and $\Delta T$ is the temperature difference between the interior and the boundary (for the mantle $\Delta T = T_\mm{m} - T_s$).

Compared to the three piece model of \citet{bryson_constraints_2019}, this model has an order of magnitude narrower region approximated for numerical stability, $w$, and applies the Krieger-Dougherty relation to viscosity beyond \rcmf rather than assuming a constant value. Unlike the $\tanh$ approximation employed by \citet{dodds_thermal_2021}, it captures the change in slope at the solidus, has a steeper gradient in viscosity at \rcmfp, and can be adjusted using physically meaningful parameters. 

\subsection{Accretion} \label{model-ac} 
Our model begins with instantaneous accretion of an undifferentiated mixture of silicates and Fe-FeS at 200\,K at a specified time, $t_{acc}$, after CAI formation. The planetesimal surface temperature is fixed at an equilibrium temperature of 200\,K throughout the simulation, because the range of planetesimal surface heat fluxes is small enough that its effect on the equilibrium surface temperature is negligible \citep[$<10$\,K;][]{dodds_thermal_2021}. At the plantesimal centre, $\frac{\partial T}{\partial r}\big|_{r=0}=0$ throughout the simulation.

\subsection{Before differentiation}\label{model-bd}
After accretion, the planetesimal heats up due to decay of radiogenic \al and \fep. Heat is initially transported through the body by conduction
\begin{equation}
    \rho_{\mm{ch}} c_{\rm p,ch, eff} \frac{\partial T}{\partial t} = \frac{1}{r^2}\frac{\partial}{\partial r}\left(k_{\mm{ch}} r^2 \frac{\partial T}{\partial r}\right)+\rho_{\mm{ch}} H,
\label{eq:diff-conduction}\end{equation} where $\rho$ is the density, $c_p$ is the specific heat capacity, $T$ is the temperature, $r$ is the radius, $t$ is the time, and $k$ is the thermal conductivity. The subscript ch denotes the values for the undifferentiated, chondritic material. The undifferentiated planetesimal is assumed to have the same thermal properties as the silicate mantle \citep[e.g.][]{elkins-tanton_chondrites_2011,bryson_constraints_2019} except for the density which uses the bulk density of the planetesimal. Compaction and sintering of the planetesimal prior to differentiation are neglected and a constant thermal diffusivity is used throughout, because these processes only affect the planetesimal as it initially heats up from 200\,K to 700\,K \citep{yomogida_multiple_1984} and have a minimal effect on the overall thermal evolution. $H$ is the radiogenic heating power per unit mass
\begin{equation} H = H_{\rm Al,0}f_{^{26}Al}0.01X_{Al}e^{-\frac{ln(2)t}{t_{1/2,Al}}}+H_{\rm Fe,0}f_{^{60}Fe}0.01X_{\mm{Fe}}e^{-\frac{ln(2)t}{t_{1/2,Fe}}},  \label{eq:heat}\end{equation}
where $H_{i,0}$ is the heating power per unit mass of isotope $i$ at the time of CAI formation, $f_i$ is the radiogenic isotope abundance as a ratio to its most common stable isotope, $X_i$ is the elemental abundance in wt \% in the accreting material, and $t_{1/2,i}$ is the radiogenic isotope half life.  The values chosen for each parameter are summarised in Table \ref{tab:parameters}. Melting of metal and silicate as the planetesimal heats up is accounted for using a modified specific heat capacity \citep[adapted from][for a mixture of metal and silicate]{merk_numerical_2002,dodds_thermal_2021}.
\begin{equation}
    c_{\rm p,ch, eff} = \begin{cases}
    c_{\mm{p,ch}} & T<T_{\mm{c,s}} \; \& \; T<T_{\rm{m,s}} \\
    c_{\mm{p,ch}}\left(1+\frac{0.01X_{\mm{Fe}}L_c}{c_{\mm{p,ch}}(T_{\mm{c,l}}-T_{\mm{c,s}})}\right) & T_{\mm{c,s}} < T < T_{\mm{c,l}} \; \& \; T<T_{\rm{m,s}} \\
    c_{\mm{p,ch}}\left(1+\frac{0.01X_{\mm{Fe}}L_c}{c_{\mm{p,ch}}(T_{\mm{c,l}}-T_{\mm{c,s}})}+\frac{0.01X_{\mm{Si}}L_\mm{m}}{c_{\mm{p,ch}}(T_{\rm{m,l}}-T_{\rm{m,s}})}\right) & T_{\mm{c,s}} < T < T_{\mm{c,l}} \; \& \; T_{\rm{m,s}}<T<T_{\rm{m,l}} \\
    c_{\mm{p,ch}}\left(1+\frac{ 0.01X_{\mm{Si}}L_\mm{m}}{c_{\mm{p,ch}}(T_{\rm{m,l}}-T_{\rm{m,s}})}\right) & T_{\mm{c,l}} > T  \; \& \; T_{\rm{m,s}}<T<T_{\rm{m,l}} \\
    \end{cases}
\label{eq:cp-diff}\end{equation}
Here $X_{\mm{Fe}}$ and $X_{\mm{Si}}$ are the Fe-FeS and silicate fraction in wt \% in the accreted material, $L_{\mm{C}}$ and $L_{\mm{m}}$ are the latent heats of fusion of Fe-FeS and silicate, $T_{\mm{c,l}}$ and $T_{\rm{m,l}}$ are the Fe-FeS and silicate liquidi, and $T_{\mm{c,s}}$ and $T_{\rm{m,s}}$ are the Fe-FeS and silicate solidi. The Fe-FeS and silicate melt fractions are both calculated using the linear liquidus approximation. The sulfur content of the Fe-FeS system is set as an input parameter. For planetesimals with eutectic Fe-FeS compositions, the temperature is held fixed once the planetesimal temperature reaches the Fe-FeS solidus, until all the Fe-FeS is melted. In this scenario, the time evolution of the Fe-FeS melt fraction, $\phi_{\mm{Fe}}$, is calculated by replacing the term on the left hand side of Equation \ref{eq:diff-conduction} by $\rho_{\mm{ch}}X_{\mm{Fe}}L_{\mm{Fe}}\frac{\partial \phi_{\mm{Fe}}}{\partial t}$  where $\rho_{\mm{ch}}X_{\mm{Fe}}$ is the density of Fe-FeS in the undifferentiated material.  

Solid-state stagnant lid convection (Section \ref{model-mantle}) could begin prior to differentiation, as the body heats up. In this mode of convection, the stagnant lid is a conductive, immobile boundary layer at the surface of the planetesimal with a thickness $\delta_0$ (Equation \ref{eq:d0}) and the interior beneath the lid convects (Figure \ref{fig:sketch-conv}). The interior is assumed to be isothermal, because pressure variation in the planetesimal is small enough for the adiabatic temperature gradient to be neglected. The interior temperature, $T_i$, evolves according to 
\begin{equation}
     \rho_{\mm{ch}} c_{\mm{p,ch}} V_i \frac{\partial T_i}{\partial t} = - F_{\rm lid}A_{\rm lid} + \rho_{\mm{ch}} V_i H,
\label{eq:diff-convection}\end{equation} 
where $F_{\rm lid} = -k_{\mm{ch}}\frac{\mm{d}T}{\mm{d}r}\big|_{r=R-\delta_0}$, $V_i$ is the volume of the interior, and $A_{\rm lid}$ is the surface area of the base of the stagnant lid \citep{solomatov_scaling_1995}. The stagnant lid continues to transport heat following Equation \ref{eq:diff-conduction}, assuming equal abundances of radiogenic elements in the interior and the lid, and decreases in thickness as the planetesimal heats up. If eutectic Fe-FeS melting occurs while the planetesimal is convecting, the left hand side of Equation \ref{eq:diff-convection} is replaced by $\rho_{\mm{ch}}X_{\mm{Fe}}L_{\mm{Fe}}\frac{\partial \phi_{\mm{Fe}}}{\partial t}$. We assume solid-state stagnant lid convection can occur if $\delta_0$  is less than 99\% of the planetesimal radius, $R$.

Once the silicate melt fraction, $\phi$, reaches the critical melt fraction, \rcmfp, there is a steep drop in the silicate viscosity and differentiation proceeds via rain-out, where the more dense, molten Fe-FeS settles through the less dense, low-viscosity silicate via Stokes settling. This process is rapid ($10^4$ years for mm size droplets, see S5) and can be approximated as instantaneous. Differentiation via percolation prior to $\phi=$ \rcmf is neglected due to the uncertainties in the timescales for this process (see Section \ref{acc-diff} and Section S5). Due to the small internal pressure gradient, the body is almost isothermal and, for simplicity, the entire planetesimal is assumed to differentiate at the same time. Our model assumes all Fe-FeS is fully molten before differentiation and that the core is initially liquid. In order to satisfy this assumption, the liquidus temperature for the initial core sulfur content of the Fe-FeS must be less than or equal to the temperature of \rcmf. This limits the range of initial core sulfur contents (see Section \ref{acc-diff-xs}). 

At the point of differentiation, the body is assumed to instantly form an Fe-FeS core, with a radius half that of the body, $r_c=\frac{R}{2}$, with an overlying silicate mantle (Figure \ref{fig:sketch-conv}). The core and the mantle are isothermal at the temperature of the critical melt fraction, $T_{\phi_C}$. All iron is assumed to partition into the core and all aluminium is assumed to stay in the mantle. This increases the concentration of \al in the mantle and \fe in the core compared to initial conditions. The initial core sulfur content, \Xsp, is set by an input parameter and is the same as the sulfur content in Fe-FeS in the undifferentiated planetesimal.

\subsection{After differentiation}\label{model-ad}
\subsubsection{Crust}
The presence of a crust is simplified to the fixed surface temperature boundary condition. Only the surface node does not become hot enough to differentiate. Therefore, the thickness of any porous regolith is below the resolution of this model and has been neglected (Section \ref{acc-diff-sint}). 

\subsubsection{Mantle}\label{model-mantle}
\begin{figure}
    \centering
    \includegraphics[width=0.3\textwidth]{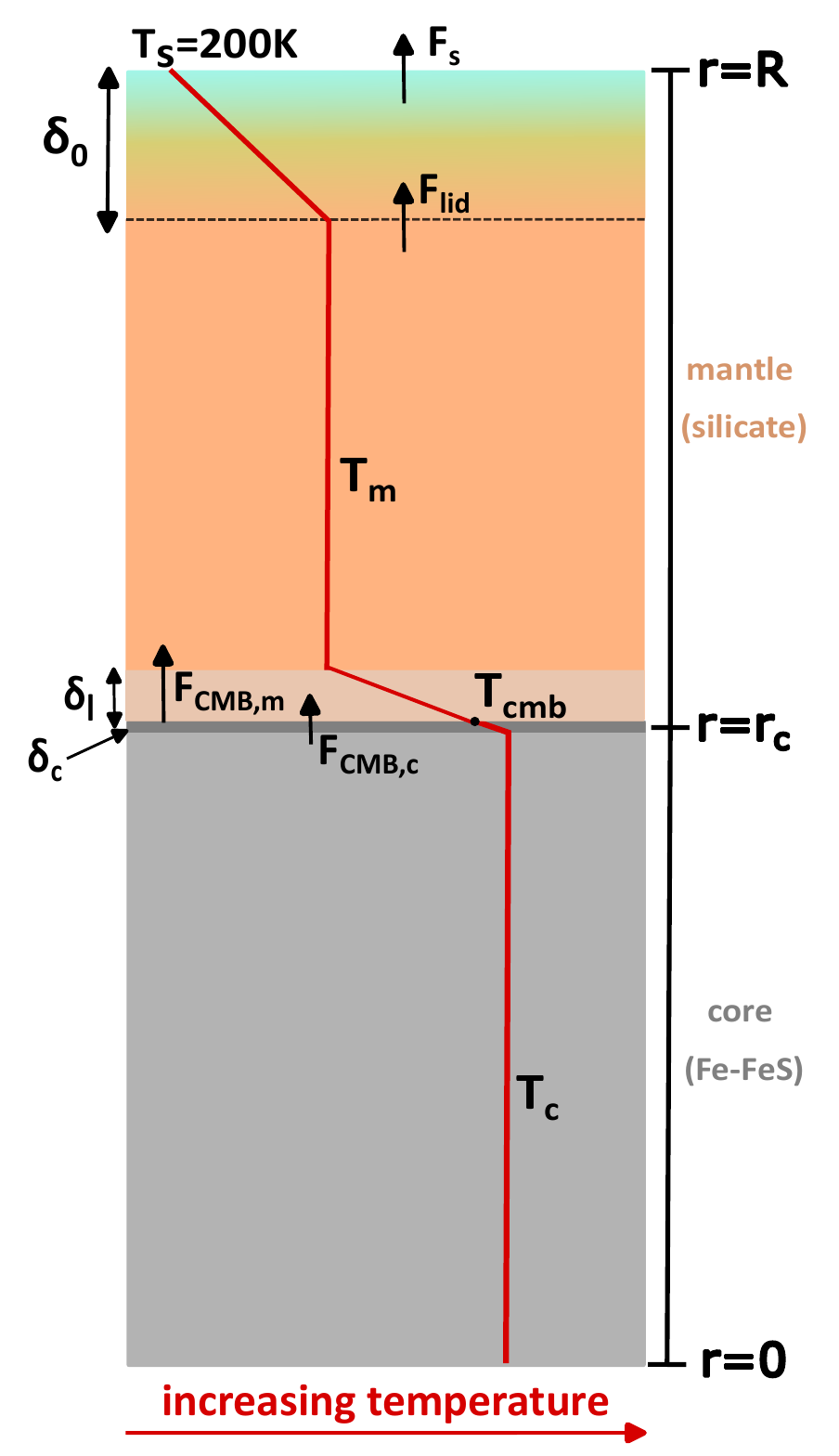}
    \caption{Schematic depicting our model setup for a convecting silicate mantle and a convecting Fe-FeS core. The core and mantle are two coupled reservoirs. The crust is approximated by the fixed surface temperature, $T_s=200$\,K. The red line is the temperature profile in the planetesimal and temperature increases to the right. In this diagram, the mantle is convecting in the stagnant lid regime: the interior is isothermal and there are conductive boundary layers at the CMB, $\delta_l$, and the surface, $\delta_0$. The black dashed line denotes the base of the stagnant lid. The core has a thermal boundary layer at the CMB, $\delta_c$. When the mantle and core are conductive, there is a temperature gradient throughout each layer instead. The balance of CMB heat fluxes ($F_{\mm{CMB,m}}=F_{\mm{CMB,c}}$) is used to calculate the CMB temperature $T_{\mm{CMB}}$. The meaning of the symbols is given in Sections \ref{model-mantle} and \ref{model-core}.}
    \label{fig:sketch-conv}
\end{figure}
The mantle transports heat either via conduction or stagnant lid convection. In stagnant lid convection, the mantle is assumed to have an isothermal, convecting interior with a conductive boundary layer (`stagnant lid') below the surface and another above the CMB (Figure \ref{fig:sketch-conv}). In our adopted parametrisation, the stagnant lid thickness includes the thin mobile boundary layer at the base of the lid (see section S1.2.2). We assume an empirically determined expression for stagnant lid thickness, $\delta_0$, for a system with internal heating, heat flux at the top and bottom boundaries, temperature-dependent viscosity, and free-slip boundary conditions \citep{deschamps_scaling_2021} 
\begin{equation}
    \delta_0 = a_{\rm lid}(R-r_c)f_{\mm{rh}}^{1.21}Ra^{-0.27},
\label{eq:d0}\end{equation} where $a_{lid}$ is a empirically determined constant ($0.633$ for $\rm Ur<1$ and $0.667$ for $\rm Ur>1$). Here Ur is the Urey ratio, which is the ratio of the power of internal heating, $Q_H$, to the power of cooling from the surface, $Q_S$, \begin{equation}
    {\rm Ur} = \frac{Q_H}{Q_S}=\frac{\rho_\mm{m} V_\mm{m} H_{\rm Al}}{F_sA_s}, 
\end{equation} 
where $V_\mm{m}$ is the volume of the mantle, $F_s$ and $A_s$ are the surface heat flux and surface area, respectively. In Equation \ref{eq:d0}, the Rayleigh number, $Ra$, is calculated using the interior mantle temperature
   \begin{equation}
    Ra = \frac{\rho_\mm{m} \alpha_\mm{m} g(R) \Delta T (R-r_c)^3}{\kappa_\mm{m} \eta}.
\label{eq:Ra}\end{equation} Here $\alpha_\mm{m}$, $\rho_\mm{m}$, and $\kappa_\mm{m}$ are the thermal expansivity, density, and diffusivity of the mantle respectively, $\Delta T = T_\mm{m}-T_s$ is the difference between the interior convective temperature, $T_\mm{m}$, and the surface temperature $T_s$, $g(R)$ is the gravitational field strength at radius $r=R$, and the viscosity is calculated for the convecting interior temperature $\eta=\eta(T_\mm{m})$.  In this model, we adopt the middle value of silicate diffusivity, $9\times10^{-7} \rm m^2 s^{-1}$, used in previous work \citep[6, 9 and 12$\times10^{-7} \rm m^2 s^{-1}$,][]{bryson_constraints_2019}, which corresponds to an ordinary chondrite composition. Equation \ref{eq:d0} was derived neglecting melt weakening, therefore $f_{\mm{rh}}=-\frac{\mm{d}\eta(T_\mm{m})/\mm{d}T}{-\eta(T_\mm{m})}\Delta T = \beta(T_\mm{m}-T_s)$ based on the first piece of the viscosity law (Equation \ref{eq:eta1}).

The thickness of the CMB boundary layer at the base of the mantle, $\delta_l$, is difficult to parametrise because it could be affected by downwelling material from the upper boundary layer interfering with its thermal structure \citep{thiriet_scaling_2019}. $\delta_l$ has been approximated in previous planetesimal models in two ways. First, $\delta_l$ has been assumed to be equal thickness to the mobile layer in the stagnant lid at the surface \citep{bryson_constraints_2019}. Second, $\delta_l$ has been calculated with the same parametrisation for lid thickness as the top boundary layer, but using $\Delta T = T_{CMB}-T_m$ rather than $\Delta T = T_m-T_s$ \citep{dodds_thermal_2021}. We have chosen to implement the CMB boundary layer scaling following Equation 13 in \citet{thiriet_scaling_2019}, which has been benchmarked against 2D and 3D mantle convection simulations \begin{equation}
    \delta_l = (R-\delta_0-r_c)\left(\frac{Ra_{bl,crit}}{Ra_{bl}}\right)^{\frac{1}{3}}.
\end{equation} Here $Ra_{bl}$ is the Rayleigh number in the boundary layer and $Ra_{bl,crit}$ is the boundary layer critical Rayleigh number. Using equations 14 and 16 in \citet{thiriet_scaling_2019} to substitute for $Ra_{bl}$ and $Ra_{bl,crit}$ gives \begin{equation}
  \delta_l = 0.65(R-r_\mm{c})^{0.21}\left(\frac{\eta(T_{bl})}{(T_{\mm{CMB}}-T_\mm{m})g_c}\right)^{\frac{1}{3}}\left(\frac{\eta(T_\mm{m})}{(T_\mm{m}-T_s)g(R)}\right)^{-0.07}\left(\frac{\kappa_\mm{m}}{\alpha_\mm{m}\rho_\mm{m}}\right)^{0.26},
\end{equation} where $T_{\mm{CMB}}$ is the CMB temperature, and $g_c$ is the gravitational field strength at the CMB. The temperature profile through the lower boundary layer is conductive and the temperature in the midpoint of the layer, $T_{bl}=\frac{T_\mm{m}+T_{\mm{CMB}}}{2}$, is used to calculate $Ra_{bl}$.
These boundary layer thicknesses control the heat flux from the CMB into the mantle during convection, $F^{\mm{conv}}_{\mm{CMB,m}}$ 
\begin{equation}
    F^{\mm{conv}}_{\mm{CMB,m}} = \frac{-k_\mm{m}(T_\mm{m}-T_{\mm{CMB}})}{\delta_l}
\label{eq:Fcmb-mconv}\end{equation}
and to the surface, $F_s$
\begin{equation}
    F_s = \frac{-k_\mm{m}(T_s-T_\mm{m})}{\delta_0}.
\end{equation}
Altogether, the thermal evolution of the convecting portion of the mantle is given by \begin{equation}
    \rho_\mm{m} c_{\rm p, eff}V_\mm{m}\frac{\partial T_\mm{m}}{\partial t} = -F_{\rm lid}A_{\rm lid} + F^{\mm{conv}}_{\mm{CMB,m}}A_{\mm{CMB}} + \rho_\mm{m}V_\mm{m}H_{\rm Al}.
\end{equation} Here, $V_\mm{m}$ is the volume of the convecting region; the silicate modified specific heat capacity, $c_{\rm p,eff}$, is calculated at $T_\mm{m}$; $A_{\rm lid}$ and $A_{\mm{CMB}}$ are the surface areas at the base of the stagnant lid and CMB, respectively; and $F_{\rm lid}=-k\frac{\mm{d}T}{\mm{d}r}\big|_{r=R-\delta_0}$
The silicate modified specific heat capacity, $c_{\rm p, eff}$, accounts for melting and solidification of silicate.
\begin{equation}
    c_{\rm p,eff} = \begin{cases}
    c_{p,m} & T_\mm{m}<T_{\rm{m,s}}
    \\ c_{p,m}\left(1+\frac{L}{c_{p,m}(T_{\rm{m,l}}-T_{\rm{m,s}})}\right) & T_{\rm{m,s}} < T_\mm{m} < T_{\rm{m,l}}.
    \end{cases}
\end{equation}
Due to efficient heat loss by convection in the model, the mantle is never hotter than its liquidus temperature.

Heat is transported by conduction in the stagnant lid and the CMB boundary layer. As the mantle cools, both these layers thicken until the combined boundary layer thickness equal the thickness of the mantle, i.e., $\delta_0+\delta_l = (R-r_c)$. From this point, mantle convection ceases and the entire mantle cools conductively, such that the thermal evolution is dictated by
\begin{equation}
    \rho_\mm{m}c_{\mm{p,eff}}\frac{\partial T}{\partial t}=\frac{1}{r^2}\frac{\partial}{\partial r}\left(k_\mm{m}r^2\frac{\partial T}{\partial r}\right)+\rho_\mm{m} H_{\rm Al}.
\label{eq:mantle-cond}\end{equation}
In the conductive regime, the heat flux across the CMB into the mantle is given by \begin{equation}
    F^{\mm{cond}}_{\mm{CMB,m}}=-k_\mm{m}\frac{\mm{d}T}{\mm{d}r}\bigg|_{r=r_c^+}=-k_\mm{m}\frac{T^\mm{m}_{\mm{CMB}+1}-T_{\mm{CMB}}}{\Delta r},
\label{eq:Fcmb-mcond}\end{equation} where $T^\mm{m}_{\mm{CMB}+1}$ and $r_c^+$ are the temperature and radius one node above the CMB, respectively, and $\Delta r$ is the length of one grid cell in the simulation. $T_{\mm{CMB}}$ is determined by balancing the heat fluxes across the CMB ($F_{\mm{CMB,m}}=F_{\mm{CMB,c}}$).

\subsubsection{Core}\label{model-core}
The thermal evolution of the core is coupled to the mantle by \fcmbp. Due to the low pressure in planetesimal cores, the adiabatic temperature gradient is negligible ($T(r=r_c)=0.995T(r=0)$ for a 250km core). Therefore, we assume the core is isothermal when convecting and that the criteria for core convection is \fcmb$>0$.
At the beginning of the thermal evolution, the core and mantle are isothermal. If the abundance of \fe in the core is low, the mantle experiences stronger radiogenic heating (from \alp) than the core. This increases the mantle temperature relative to the core and heat flows from the mantle to the core ($F_{\mm{CMB}}<0$). The top of the core becomes thermally stratified, which inhibits core convection. While the core is thermally stratified, heat is transferred conductively as described by 
\begin{equation}
    \rho_cc_{p,c}\frac{\partial T}{\partial t}=\frac{1}{r^2}\frac{\partial}{\partial r}\left(k_cr^2\frac{\partial T}{\partial r}\right)+\rho_\mm{m} H_{\rm Fe}, 
\label{eq:core-cond}\end{equation} where $c_{p,c}$ is core specific heat capacity, $\rho_c$ is core density, $k_c$ is core thermal conductivity, and\\ $H_{\mm{Fe}}=H_{\rm Fe,0}f_{^{60}\mm{Fe}}0.01X_{\mm{Fe}}e^{-\frac{ln(2)t}{t_{1/2,Fe}}}$ is heating from \fep. An effective specific heat capacity is not used because the core begins molten and core solidification causes core convection so is treated separately (see Section \ref{model-core-solid}). The boundary conditions are a fixed CMB temperature using the temperature determined at the previous timestep (which is a reasonable approximation given the small timestep) and no temperature gradient at the centre of the body, i.e., $\frac{\mm{d}T}{\mm{d}r}\big|_{r=0}=0$. The conductive heat flux from the core to the CMB is given by
\begin{equation}
    F^{\mm{cond}}_{\mm{CMB,c}}= -k_\mm{c}\frac{\mm{d}T}{\mm{d}r}\bigg|_{r=r_\mm{c}^-}= -k_\mm{c}\frac{T_{\mm{CMB}}-T^\mm{c}_{\mm{CMB}-1}}{\Delta r},
\label{eq:Fcmb-ccond}\end{equation} 
where $T^\mm{c}_{\mm{CMB}-1}$ and $r_\mm{c}^-$ are the temperature and radius one node below the CMB, respectively.
\begin{figure}
    \centering
    \includegraphics[width=1\textwidth]{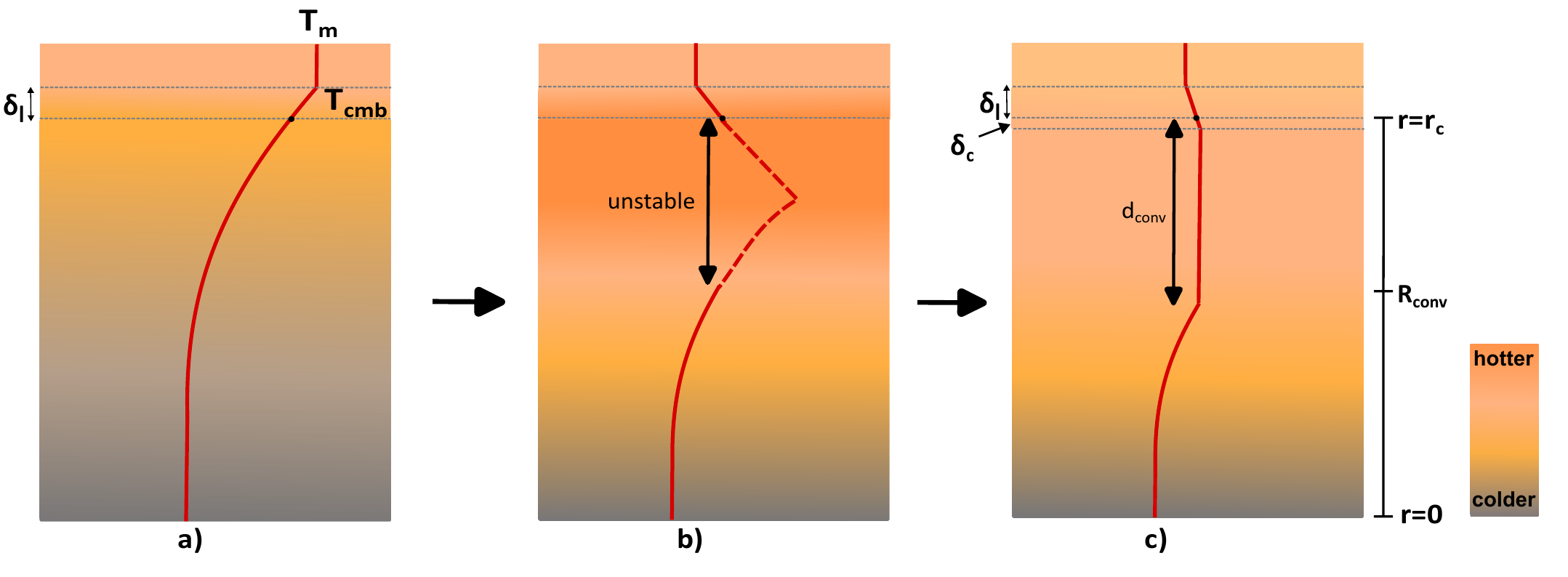}
    \caption{Schematic of the erosion of core thermal stratification by convective mixing of the unstable layer at the top of the core. The red lines indicate temperature profiles and temperature increases to the right. The dashed red line indicates the portion of the core that is unstable to convection. The top of the figure is the base of the convecting mantle above the CMB boundary layer in the mantle, $\delta_l$, and CMB boundary layer in the core, $\delta_c$. Time proceeds from left to right (a--c). Dashed lines indicate the mantle and core CMB boundary layers. The core CMB boundary layer (lower dashed layer) is only present when the core is convecting. Core thermal stratification is eroded either by mantle cooling (shifts the mantle temperature profile to the left relative to that of the core) or heating by \fe in the core (shifts the core temperature profile to the right relative to the mantle). All symbols are defined in Section \ref{model-core}.}
    \label{fig:erosion}
\end{figure}
Once the CMB temperature is below the core temperature, either due to mantle cooling or core heating by \fep, the top of the core begins to cool (\fcmb$>0$) and thermal stratification begins to erode (Figure \ref{fig:erosion}). Positive \fcmb leads to the onset of convection in the upper portion of the core that is hotter than the CMB. As such, we assume this material convectively mixes down to the level of neutral buoyancy (the depth where core temperature is equal to the CMB temperature), $d_{\rm conv}$. This forms an isothermal convecting layer, with temperature $T_\mm{c}$, above a conductive region. This convective mixing continues until core thermal stratification is removed and the entire core is isothermal and convecting.

The temperature of the convecting portion of the core evolves according to
\begin{equation}
    \rho_\mm{c} c_{p,c}V_{\rm conv}\frac{\partial T_\mm{c}}{\partial t} = -F^{\mm{conv}}_{\mm{CMB,c}}A_{\mm{CMB}} +\rho_\mm{c}V_{\rm conv}H_{\rm Fe} + 4\pi R_{\rm conv}^2F_{R_{\rm conv}},
\label{eq:core-conv}\end{equation}
where $R_{\rm conv} = r_\mm{c}-d_{\rm conv}$, $V_{\rm conv}$ is the volume of the convecting portion of the core, and $A_{\mm{CMB}}=4\pi r_\mm{c}^2$ is the CMB surface area \citep{dodds_thermal_2021}. While thermal stratification is being eroded, there is a heat flux into the convecting region from the conductive core beneath, $F_{R_{\rm conv}}=-k_\mm{c}\frac{\partial T}{\partial r}\big|_{r=R_{\rm conv}}$, and the final term in Equation \ref{eq:core-conv} is non-zero. When the core is convecting, a boundary layer, $\delta_\mm{c}$, forms at the top of the core. Although core rotation has a known effect on the convective boundary layer scaling law \citep{cheng_laboratory-numerical_2015}, the combined effect of rotation, magnetic field generation and spherical geometry on heat transfer is unclear. Therefore, for simplicity, we assume the thickness of this boundary layer can be described by the non-rotating scaling \citep{schubert_mantle_2001,dodds_thermal_2021} \begin{equation}
    \delta_\mm{c} = \left(\frac{\kappa_\mm{c} \eta_\mm{c}Ra_{c,iso}}{\rho_\mm{c}\alpha_\mm{c}g_\mm{c}(T_\mm{c}-T_{\mm{CMB}})}\right)^{\frac{1}{3}},
\label{eq:deltac}\end{equation}
where $\eta_\mm{c}$ and $\kappa_\mm{c}$ are the viscosity and thermal diffusivity of the core, respectively, and $Ra_{c,iso}=1000$ \citep{schubert_mantle_2001} is the critical Rayleigh number for isoviscous convection. 
The corresponding CMB heat flux is given by \begin{equation}
    F_{\mm{CMB,c}}^{\mm{conv}}=-k_\mm{c}\frac{T_{\mm{CMB}}-T_\mm{c}}{\delta_\mm{c}}.
\label{eq:Fcmb-cconv}\end{equation}
 
\subsubsection{Core solidification}\label{model-core-solid}
\begin{figure}
    \centering
    \includegraphics[width=1\textwidth]{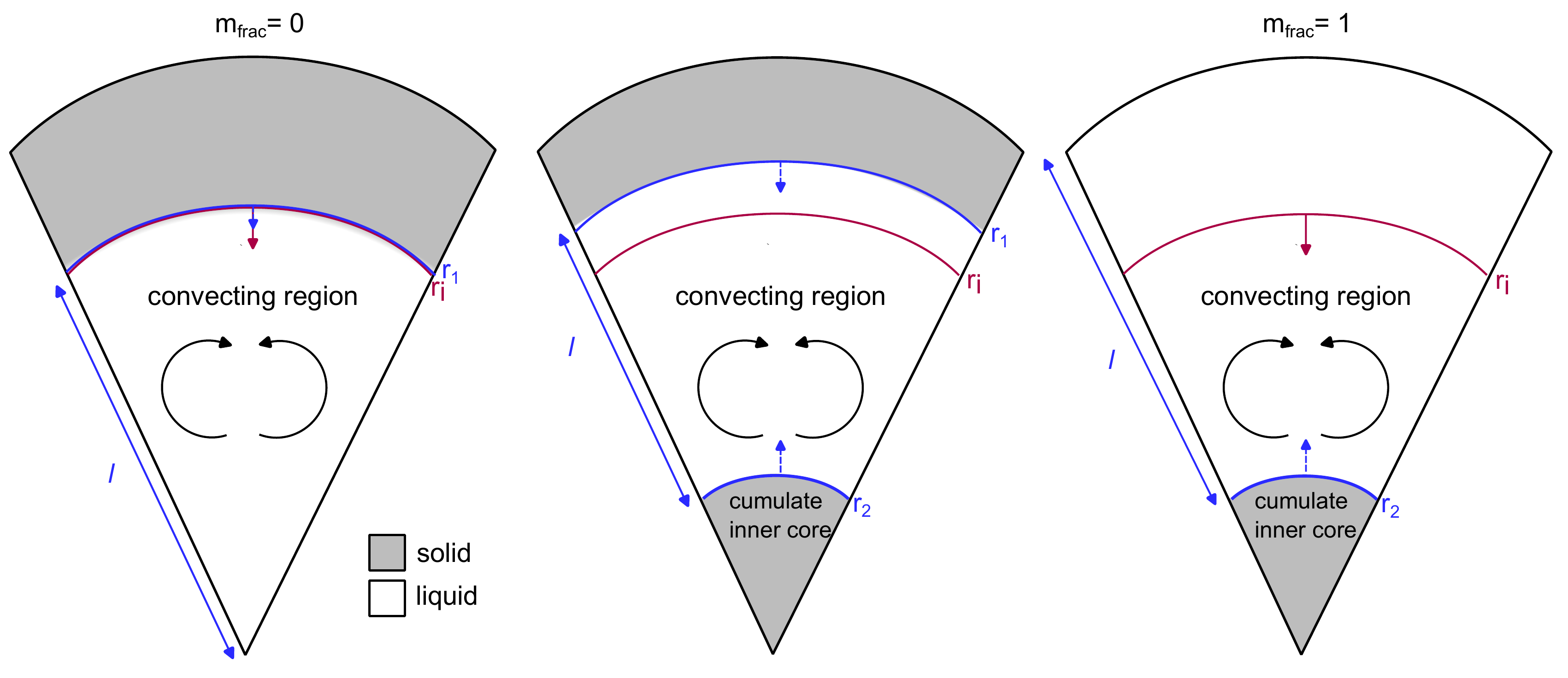}
    \caption{Schematic showing the geometry for core solidification. The amount of solidifying material is parametrised by $r_i$ (magenta line) and the lengthscale for convection is give by $l=r_1-r_2$. To account for the uncertainty in the mechanism of planetesimal core solidification we consider two endmember geometries (the middle scenario in the figure also depicts an intermediate geometry). In one endmember, \mfracp=0, the stable, solid portion of the core grows from the CMB inwards. In the other endmember, \mfracp=1, all solidified iron falls without remelting and forms a cumulate inner core. For \mfracp=0, the convective lengthscale decreases the slowest as the core solidifies, and for \mfracp=1, the convective lengthscale decreases the fastest. Due to the uncertainty in the mechanism for planetesimal core solidification, we do not prescribe how convection is driven in each scenario and just focus on the change in lengthscale. In both endmembers, the solid fraction in the solidified region is assumed to be 100\%. For more details and discussion see Sections \ref{model-core-solid} and \ref{csm-l}. }
    \label{fig:core-solid}
\end{figure}
Planetesimal CMBs are at lower pressures compared to Earth, so are likely to solidify inwards \citep{williams_bottom-up_2009,dodds_direction_2024}. The mechanism of this inward solidification and its implications for dynamo generation are still being investigated, but could include formation of iron snow \citep{ruckriemen_fe_2015,davies_iron_2018} formation of dendrites \citep{scheinberg_core_2016} or viscous delamination of solid iron  from the CMB \citep{neufeld_top-down_2019}. Previous models for mantled planetesimals that incorporate convection either ended before core solidification \citep{elkins-tanton_chondrites_2011,SterenborgCrowley2013,dodds_thermal_2021} or used eutectic core solidification to estimate the time period over which a compositional dynamo could be possible \citep{bryson_constraints_2019}. To avoid the specificity of a particular core solidification model, we have chosen to use a parametrised model that captures two key changes in an inward solidifying core that will affect dynamo generation: (i) the decrease of convective lengthscale; (ii) the increase in sulfur content of the bulk liquid, assuming perfect partitioning of sulfur into the liquid \citep{goldstein_iron_2009}. This enables us to calculate a core solidification timescale and magnetic field strengths during sub-eutectic core solidification. 

Core solidification is parametrised by a time-varying, inward moving boundary $r_i$, $\frac{\mm{d}r_i}{\mm{d}t}<0$. The total material solidified is given by $M_{solid}=\frac{4}{3}\pi\rho_\mm{c}(r_\mm{c}^3-r_i^3)$. 
This solidified material can decrease the convective lengthscale of the core either due to formation of a solidified layer at the top of the core and/or due to the formation of a solid cumulate inner core formed by solidified material falling from $r_i$ without remelting to the centre of the core \citep{scheinberg_core_2016,neufeld_top-down_2019}. To capture both of these possibilities, we vary the proportion of total solidified mass in the innermost core for a given time and $r_i$, $M_{ic}$, using a constant input parameter, \mfracp$=\frac{M_{ic}}{M_{solid}}$ (see Figure \ref{fig:core-solid}). Using $f=\frac{r_i}{r_\mm{c}}$ and conservation of mass, the resulting radii of the base of the solidified outer shell, $r_1$, and the solid innermost core, $r_2$, are given by
\begin{equation}
    r_1 = r_\mm{c}(m_{\text{frac}}(1-f^3)+f^3)^{\frac{1}{3}},
\end{equation}
and \begin{equation}
    r_2 = r_\mm{c}m_{\text{frac}}^{\frac{1}{3}}(1-f^3)^{\frac{1}{3}}.
\end{equation}
We show results for the two endmembers to explore the plausible range of geometric lengthscales: solidified outer layer only, \mfracp$=0$; and solid inner core only, \mfracp$=1$. Current dynamo theories suggest the convective lengthscale is some fixed fraction of this geometric lengthscale \citep{davidson_scaling_2013,aubert_spherical_2017}, but in this model we assume these lengthscales are the same (see Section \ref{csm-l}). Our model assumes the solidified portions of the core are completely solid and not a mushy layer, because self consistently calculating a solid fraction is overly complex given the uncertainties in core material properties (see Section \ref{csm-snow}).

The onset of solidification occurs when the temperature at the top of the core drops beneath the liquidus.
Because compositional convection in the core will efficiently homogenise the temperature in the liquid core, the thermal conductivity of the solid portion is high, and the adiabatic gradient is minimal, the entire core is assumed to be isothermal at temperature $T_\mm{c}$ during solidification.
During solidification, the heat flux across the CMB is balanced by secular cooling, radiogenic heating, and release of latent heat. Energy released due to changes in gravitational potential during solidification has a negligible effect on the time of core solidification (see Section S3.1) and contributions due to planetary contraction are also negligible. 
The resulting thermal evolution of the core is expressed as 
\begin{equation}
    F_{\mm{CMB}}A_{\mm{CMB}} = M_\mm{c}H_{\mm{Fe}} - M_\mm{c}c_{p,c}\frac{\mm{d}T_\mm{c}}{\mm{d}t} - 4\pi r_i^2\rho_\mm{c} L_\mm{c}\frac{\mm{d}r_i}{\mm{d}t}, 
    \end{equation}
where $L_\mm{c}$ is the latent heat from solidification of iron and $M_\mm{c}$ is the total mass of the core \citep{nimmo_energetics_2007}.  
For an isothermal solidifying core, the rate of change of inner core boundary is given by 
\begin{equation}
    \frac{\mm{d}r_i}{\mm{d}t} = -\frac{1}{\rho_\mm{c} g_\mm{c} \frac{\mm{d}T_{\mm{c,l}}}{dP}}\frac{\mm{d}T_\mm{c}}{\mm{d}t},
\label{eq:dTc-solid}\end{equation} which is adapted to account for inward solidification and no core adiabat \citep{gubbins_can_2003,nimmo_energetics_2009,davies_iron_2018}. The sign of the pressure gradient of the liquidus, $\frac{\mm{d}T_{\mm{c,l}}}{\mm{d}P}$, determines whether $\frac{\mm{d}r_i}{\mm{d}t}$ is positive (outward solidification) or negative (inward solidification).
As the core solidifies, we assume as an upper limit that sulfur partitions completely into the liquid portion of the core \citep{goldstein_iron_2009,nichols_time-resolved_2021}, is homogeneously distributed, and this partitioning has a negligible effect on the total volume of the core. This enriches the liquid inner core relative to the initial sulfur content, $X_{S,0}$ according to \citep{nichols_time-resolved_2021} \begin{equation}
    X_S = \frac{r_\mm{c}^3}{r_i^3}X_{S,0} = \frac{1}{f^3}X_{S,0}.
\end{equation}
Once the core sulfur content reaches the eutectic (33\,wt\% S), the remaining liquid inner core will solidify as FeS. During eutectic solidification, there is no temperature change ($T_\mm{c} = 1260$\,K) and the fraction of core solidified is determined by the removal of latent heat, as given by
\begin{equation}
    \frac{\mm{d}r_i}{\mm{d}t}= -\frac{F_{\mm{CMB}}A_{\mm{CMB}}-M_\mm{c}H}{4\pi r_i^2L_\mm{c}\rho_\mm{c}}.
\end{equation}
Once the whole core has solidified ($r_i < 0.001$) the model run ends. 

This model uses the Fe-FeS liquidus from \citet{buono_fe-rich_2011}, which depends on pressure and bulk sulfur content,
\begin{equation}
T_{\mm{c,l}}(X_{S,mol}, P) = A(P)X_{S,mol}^4 + B(P)X_{S,mol}^3 + C(P)X_{S,mol}^2+D(P)X_{S,mol}+E(P),
\label{eq:liquidus}\end{equation} where the pressure, $P$, is in GPa and $X_{S,mol} = \frac{X_s}{100-X_s}\frac{M_{r,Fe}}{M_{r,S}}$ is the mole fraction of FeS in the core, $M_{r,i}$ are the molar masses of sulfur (32.07\,amu) and iron (55.84\,amu), respectively, and $X_S$ is the core sulfur content in wt \%. The pressure dependent coefficients are given by \citep{buono_fe-rich_2011}
\begin{equation}\begin{split}
     A & = -2.4724P^4 + 28.025P^3 + 9.1404P^2 + 581.71P + 3394.8,
    \\ B & = 1.7978P^4 -6.7881P^3 -197.69P^2 -271.69P -8219.5,
    \\ C & = -0.1702P^4 -9.3959P^3 + 163.53P^2 -319.35P + 5698.6,
    \\ D  & = -0.2308P^4 + 7.1P^3 -64.118P^2 + 105.98P -1621.9,
    \\ E & = 0.2302P^4 -5.3688P^3 + 38.124P^2 -46.681P + 1813.8.
\end{split}
\end{equation} This expression for the Fe-FeS liquidus is valid for sub-eutectic sulfur concentrations and pressures $<10$\,GPa. The pressure at a radius $r$ in the core (assuming an incompressible core and mantle) is given by 
\begin{equation}
    P_\mm{c} = \frac{2\pi G}{3}\left[r_\mm{c}^2(\rho_\mm{c}^2-\rho_\mm{m}^2)+R^2\rho_\mm{m}^2+2\rho_\mm{m}(\rho_\mm{c}-\rho_\mm{m})r_\mm{c}^2\left(1-\frac{r_\mm{c}}{R}\right)\right].
\end{equation}
For a $\mathcal{O}(100\,\rm km)$ radius body, the difference in pressure between the centre and the CMB is $\mathcal{O}(10^{-3}\,\rm GPa)$; therefore, the central pressure is always used in the expression for the liquidus. The sulfur content used to calculate the liquidus and pressure gradient of the liquidus is updated as the core solidifies.

\subsection{Magnetic field generation}\label{mag-field}
\subsubsection{Magnetic Reynolds number}\label{Rem}
The magnetic Reynolds number, \Remp, determines whether flow in a planetesimal core is vigorous enough to generate a magnetic field. This dimensionless number characterises the balance between magnetic induction and diffusion in the fluid and is given by \begin{equation}
    Re_\mm{m} = \frac{ul}{\lambda},
\label{eq:Rem}\end{equation} where $u$ and $l$ are the characteristic velocity and lengthscale of the flow and $\lambda=1.3\, \rm m^2 s^{-1}$ \citep{weiss_paleomagnetic_2010} is the magnetic diffusivity. If \Rem is greater than a critical value, a magnetic field can be generated. The minimum analytical critical value is 10 and numerical dynamo simulations suggest a critical value between 40--100 \citep{christensen_scaling_2006}. To account for this uncertainty, we calculate three sets of dynamo onset and cessation times with critical \Rem values of 10, 40, and 100. $\lambda$ depends on pressure, temperature and composition, but the appropriate value for a given set of conditions is uncertain \citep{pozzo_transport_2013,konopkova_direct_2016,tassin_geomagnetic_2021}. In addition, measurements are limited to Earth-like conditions, i.e., much higher pressures and temperatures than in planetesimals. Therefore, we take $\lambda$ to be a constant between simulations and uncertainties in this parameter are considered by adopting multiple critical \Rem values. $l$ is assumed to be the length over which convection can  occur, $l=r_i-R_{\rm conv}$, which accounts for the possible existence of a stratified layer. In this model, we use the numerical scaling from \citet{aubert_modelling_2009} for the characteristic velocity, $u$, 
\begin{equation}
   u = c_up^{0.42}\Omega l. 
\label{eq:ucomp}\end{equation} $c_u=1.31$ is a constant \citep{aubert_modelling_2009}, $\Omega$ is the rotation rate of the planetesimal, and $p$ is the convective power per unit volume. To enable comparison with previous work, we adopt a nominal rotation period of 10 hours ($\Omega=1.7\times10^{-4}\rm s^{-1}$), which is a typical value for rotation periods of asteroids at the present day \citep{bryson_constraints_2019,dodds_thermal_2021}. We use this characteristic velocity scaling law because it is not singular in the absence of an inner core so can be applied to inward solidification where a solid inner core is not present. Additionally, this law has been shown to agree with subsequent analytical derivations of dynamo scaling laws \citep{davidson_scaling_2013}. This also aligns with the work of \citet{christensen_iron_2015} who used this scaling law when determining the dimensional magnetic field strength of Ganymede. 

\subsubsection{Magnetic field strength}
We have determined that the Magnetic-Archimedean-Coriolis (MAC) force balance dominates in planetesimal cores (Table S1). For a core in MAC force balance, the magnetic dipole field at the CMB can be estimated as \begin{equation}
    B^{\rm dip}_{\mm{CMB}} = c_Bf_{ohm}^{\frac{1}{2}}p^{0.31}(\rho_\mm{c}\mu_0)^{\frac{1}{2}}\Omega l.
\label{eq:B}\end{equation} Here $f_{ohm}=1$ \citep{christensen_dynamo_2009} is the fraction of energy dissipated ohmically in the core, $\mu_0=4\pi\times10^{-7} \rm NA^{-2}$ is the vacuum permeability and $c_B=0.23$ is the median value of the scaling constant for CMB dipole field strength across the simulations assessed in \citet{davies_dynamo_2022}. The field strength decays from the CMB to the surface according to $B_{surf} = \left(\frac{r_\mm{c}}{R}\right)^3B^{\rm dip}_{\mm{CMB}}$. The exponents of the scaling law agree within error of the field strength scaling in \citet{aubert_modelling_2009}, so this is consistent with the chosen characteristic velocity scaling. Additionally, the MAC force balance has been shown to best fit a large range of simulations of the geodynamo \citep{davies_dynamo_2022} and scaling analysis suggests planetesimal dynamos are in a similar regime (see Table S1).

\subsubsection{Convective power and buoyancy flux}
The scaling laws for both $u$ and $B$ require a convective power per unit volume, $p$. This can be calculated using $p=\gamma Ra_Q$, where $\gamma = 3/5$ \citep[Equation 18 in][for no inner shell and $f_i=0$]{aubert_modelling_2009}. $Ra_Q$ is the flux based Rayleigh number given by \begin{equation}
    Ra_Q = \frac{g(r_i)Q_b}{4\pi\rho_\mm{c}\Omega^3l^4},
\label{eq:Raq}\end{equation} for a gravitational acceleration at the top of the convecting region, $g(r_i)$, and buoyancy flux, $Q_b$.
The buoyancy flux combines the two possible contributions: thermal and compositional. Adapting from \citet{buffett_thermal_1996} (see Appendix \ref{app:buoy}) for an inwardly solidifying core with $\frac{\mm{d}r_i}{\mm{d}t}<0$, the buoyancy flux at the liquid inner core boundary $r_i$ is given by
\begin{equation}
    Q_b = 4\pi r_i^2 \left[\frac{\alpha_\mm{c} k_\mm{c}}{c_{p,c}}\left(-\frac{\mm{d}T}{\mm{d}r}\bigg|_{r_i}-\frac{\alpha_\mm{c}g(r_i)}{c_{p,c}}T(r_i)\right)-\left(\Delta \rho+\frac{\alpha_\mm{c}\rho_\mm{c}L_\mm{c}}{c_{p,c}}\right) \frac{\mm{d}r_i}{\mm{d}t}\right],
\label{eq:Qb}\end{equation}
where $\alpha_\mm{c}$ is the core thermal expansivity.
The first term in this equation represents the thermal buoyancy contribution from the superadiabatic heat flux at the upper boundary of the convecting region (i.e., the heat flux available to drive convection), assuming all the heat is advected by the flow.  Although we have neglected the adiabatic gradient in the temperature structure of the convecting core, we retain the adiabatic heat flux in the expression for the buoyancy flux for consistency with previous literature \citep[e.g.][]{lister_expressions_2003,ruckriemen_fe_2015,dodds_thermal_2021}. In our simulations, \Rem becomes subcritical before the CMB heat flux becomes subadiabatic and the adiabatic gradient is an order of magnitude smaller than the CMB heat flux when \Rem is supercritical. Therefore, including the adiabatic gradient will have minimal effect on the timings of dynamo generation.   

The second term in Equation \ref{eq:Qb} is the compositional density difference and the latent heat release from the solidifying core. $\Delta \rho = \rho_{\rm Fe, s} - \rho_l(X_s)$ is the density difference between the solidified iron $\rho_{\rm Fe, s}=7800\,\rm kg m^{-3}$ \citep{bryson_long-lived_2015} and the liquid inner core, $\rho_l$, which is calculated using the relationship from \citet{morard_liquid_2018}
\begin{equation}
    \rho_l = (-3108X_{S,at}^2-5176X_{S,at}+6950)(1+\alpha_\mm{c}(1900-1600)),
\label{eq:rhoc}\end{equation}
where $X_{S,at}= \left(1+ \frac{1-0.01X_s}{0.01X_s}\frac{M_{r,S}}{M_{r,Fe}}\right)^{-1}$ is atom \% S and the core is assumed to only contain Fe and FeS. In Equation \ref{eq:rhoc}, the term in the first parentheses is the density from \citet{morard_liquid_2018} and the term in the second parentheses corrects for the fact the expression from \citet{morard_liquid_2018} was derived for ambient pressure at a temperature of 1900\,K, whereas peak core temperatures are $\sim 1600$\,K in our model.  The change in core density due to core cooling below 1600\,K to the eutectic (a further 3\% difference in density) was neglected in order to simplify the model. Although we explore geometries for two possible endmembers for core solidification (Figure \ref{fig:core-solid}), we use the same density difference for both, because it is the correct order of magnitude and $\Delta \rho$ is raised to a fractional power in both Equations \ref{eq:ucomp} and \ref{eq:B}.  

Prior to core solidification, Equations \ref{eq:Raq} and \ref{eq:Qb} are evaluated at the CMB ($r_i=r_\mm{c}$) and the second term in Equation \ref{eq:Qb} is zero. As the core solidifies inwards, these equations are evaluated at the liquid inner core boundary, $r_i$, which is the boundary of the convecting region. Once the eutectic composition is reached, the second term in Equation \ref{eq:Qb} is no longer calculated, because a compositional density difference is no longer generated by solidification. Combining thermal and compositional buoyancy flux contributions enables both drivers of convection to be considered simultaneously at all timesteps.

\subsection{Numerical Implementation}\label{implement}
Our model is 1D and spherically symmetric. The one spatial dimension is radius, $r$, which runs from $r=0$ at the centre of the planetesimal to $r=R$ at the surface. The model is run on a grid with fixed node spacing of $\Delta r=500$\,m following \citet{bryson_constraints_2019} with the first node at the centre of the planetesimal and the last at the surface. The temperature change of convecting regions was calculated using the forward Euler method. The temperature in the conductive portion was calculated using matrix methods with a forward in time, centered in space (FTCS) stencil (see S6). The core conductive timescale, $\tau_{\rm c, cond}=\frac{(\Delta r)^2}{\kappa_\mm{c}}$, is a factor of ten shorter than the mantle conductive timescale. Therefore, the timestep was fixed to $dt = 0.075\tau_{c,\rm cond}$ in order to resolve conduction in the core and mantle and ensure stability of the FTCS scheme, which is stable for $dt \leq \frac{(\Delta r)^2}{2\kappa}$.To test the gridsize and timestep, four simulations were run for a 300\,km body: one using the model gridsize and timestep; one for half the gridsize; one for two thirds the timestep; and one for one third the timestep. The results of the model were independent of the choice of timestep and gridsize (Section S6.1).

\subsection{Parameter values}\label{model-params}
The list of parameters used in the model are given in Table \ref{tab:parameters}. The viscosity parameters are poorly constrained: the values here are the median value in a wide range \citep[explored in more detail in][]{sanderson_early_2024}. The range of possible initial core sulfur contents, $X_{S,0}$, primordial \feratiop, planetesimal radii, and accretion times are shown in Table \ref{tab:constants}. The maximum \Xs is assumed to be the eutectic composition and the minimum \Xs is based on the highest liquidus temperature for which all the Fe-FeS will be molten before differentiation for \rcmf= 0.3 and $R=300$\,km. The eutectic sulfur content varies with pressure and was estimated as $33$\,wt\% by finding the intersection of the liquidus with the eutectic temperature of 1260\,K \citep{buono_fe-rich_2011} for a 300\,km radius planetesimal. Due to its presumed low concentration, previous studies have neglected radiogenic heating from \fep. However, recent studies suggest the primordial \feratio could be as high as $6\times10^{-7}$ \citep{cook_iron_2021}. To incorporate this possible range, we adopt values from 0 to 6$\times10^{-7}$. The minimum planetesimal radius is based on the lower limit for compositional dynamo generation \citep{nimmo_energetics_2009}, while the upper limit is approximately the radius of Ceres \citep[470\,km][]{russell_dawn_2016}, the largest asteroid surviving today. Planetesimal accretion times affect the strength of radiogenic heating and must be within 2.5\,Ma after CAI formation for there to be sufficient \al for full planetesimal differentiation \citep{neumann_differentiation_2012}. Accretion earlier than 1.1\,Ma after CAI formation reconciles best with our choice to neglect silicate melt migration \citep[][see Section \ref{acc-diff}]{monnereau_differentiation_2023}. 

\begin{table}
    \centering
    \begin{tabular}{|c|c|c|c|}
    \hline
      Symbol   & Meaning & Value & Reference \\ \hline
    \multicolumn{4}{|c|}{\textbf{Whole body parameters}}\\\hline
      $T_s$ & Surface temperature & 200\,K & *\citet{bryson_constraints_2019} \\
      $\Omega$ & Rotational frequency of planetesimal & $1.75\times10^{-4} \rm s^{-1}$ & *\citet{bryson_constraints_2019} \\\hline
      \multicolumn{4}{|c|}{\textbf{Undifferentiated material parameters}}\\\hline
       $\rho_{\mm{ch}}$  & Density &  $ (\rho_\mm{m}(R^3-r_\mm{c}^3)+\rho_\mm{c}r_\mm{c}^3)/R^3$ $\rm kg m^{-3}$ &  \\
       $c_{\mm{p,ch}}$ & Specific heat capacity & $c_{p,m}$& \makecell{\citet{elkins-tanton_chondrites_2011} \\\citet{bryson_constraints_2019}}\\
       $\alpha_{\mm{ch}}$ & Thermal expansivity & $\alpha_\mm{m}$ & \makecell{\citet{elkins-tanton_chondrites_2011} \\\citet{bryson_constraints_2019}}\\
       $k_{\mm{ch}}$ & Thermal conductivity & $k_\mm{m}=\kappa_\mm{m}\rho_\mm{m}c_{\mm{p,m}}$ & \makecell{\citet{elkins-tanton_chondrites_2011} \\\citet{bryson_constraints_2019}}\\\hline
       \multicolumn{4}{|c|}{\textbf{Mantle parameters}}\\\hline
       $\rho_{\mm{m}}$  &  Density & 3000\,$\rm kg m^{-3}$ & \citet{elkins-tanton_chondrites_2011} \\
       $c_{p,m}$ & Specific heat capacity & 800\,$\rm J kg^{-1} K^{-1}$ & \citet{ghosh_temperature_1999}\\
       $L_\mm{m}$ & Latent heat of fusion & $400\times10^3\, \rm J kg^{-1}$ & *\citet{ghosh_thermal_1998} \\
       $T_{\rm{m,l}}$ & Liquidus & 1800\,K & \citet{dodds_thermal_2021} \\
       $T_{\rm{m,s}}$ & Solidus & 1400\,K & \citet{dodds_thermal_2021} \\
       $\alpha_{\mm{m}}$  & Thermal expansivity & $4\times10^{-5}\,\rm K^{-1}$ & *\citet{nimmo_influence_2000} \\
       $\kappa_{\mm{m}}$  & Thermal diffusivity & $9\times10^{-7}\,\rm m^2 s^{-1}$ & \citet{dodds_thermal_2021} \\
       $Ra_\mm{c}$ & \makecell{Critical Rayleigh number \\ for isoviscous convection} & 1000 & \citet{schubert_mantle_2001}\\\hline
       \multicolumn{4}{|c|}{\textbf{Viscosity parameters}}\\\hline
       \rcmf & Critical melt fraction & 0.3 & \citet{scott_effect_2006} \\
       $\beta$ & Arrhenius slope & 0.0225\,$\rm K^{-1}$ & \makecell{E=366\,kJ\,mol$^{-1}$, $T_{\mm{ref}}=1600$\,K \\ \citet{hirth_rheology_2003}} \\
       $\alpha_\mm{n}$ & Melt weakening exponent & 30 &  \citet{hirth_rheology_2003} \\
       \etar & Reference viscosity & $10^{19}$\,Pa\,s &  \citet{lichtenberg_magma_2019} \\
       $\eta_l$ & Liquid viscosity & 10\,Pa\,s & \citet{giordano_viscosity_2008}\\\hline
       \multicolumn{4}{|c|}{\textbf{Core parameters}}\\\hline
       $\rho_{\mm{Fe},s}$ & Solid iron density & 7800\,$\rm kg m^{-3}$ & \citet{bryson_long-lived_2015} \\
       $c_{p,c}$  & Specific heat capacity & 850\,$\rm J kg^{-1} K^{-1}$ & *\citet{bartels_high_1991} \\
       $L_\mm{c}$ & Latent heat of fusion of core & $270\times10^{3}\,\rm J kg^{-1}$ & *\citet{ghosh_thermal_1998} \\
       $T_{\mm{c,s}}$ & Fe-FeS solidus & 1260\,K & \citet{buono_fe-rich_2011} \\
       $X_{s,eut}$ & Eutectic sulfur content & 33\,wt \% & \citet{buono_fe-rich_2011}, $R=300$\,km \\
       $\alpha_\mm{c}$ & Thermal expansivity & $9.2\times10^{-5}\,\rm K^{-1}$ & *\citet{nimmo_energetics_2009} \\
       $k_\mm{c}$ & Thermal conductivity & 30\,$\rm W m^{-1} K^{-1}$ & *\citet{opeil_thermal_2010}\\
       $\eta_\mm{c}$ & Viscosity & 0.01\,Pa\,s & *\citet{de_wijs_viscosity_1998} \\
       $\lambda$ & Magnetic diffusivity & 1.3\,$\rm m^2s^{-1}$ & *\citet{weiss_paleomagnetic_2010} \\
       $f_{\rm ohm}$ & \makecell{Fraction of energy flux \\ dissipated ohmically} & 1 & \citet{christensen_dynamo_2009} \\
       $c_u$ &  Convective velocity constant & 1.31 & \citet{aubert_modelling_2009} \\
       $c_b$ & \makecell{Magnetic field \\ strength constant} & 0.23 & \citet{davies_dynamo_2022} \\\hline
       \multicolumn{4}{|c|}{\textbf{Radiogenic heating parameters}}\\\hline
       $H_{\rm Al,0}$ & \makecell{Heating power of $^{26}Al$ at \\ t=0\,Ma after CAI formation.} & 0.355\,Wkg$^{-1}$  & *\citet{castillo-rogez_26al_2009}\\
       $f_{^{26}Al}$ & $^{26}Al/^{27}Al$ & $5 \times10^{-5}$ & *\citet{macpherson_distribution_1995}\\
       $X_{Al}$ &wt \% Al in accreting material & 1.4 & *\citet{doyle_early_2015}\\
       $t_{1/2,Al}$ & $^{26}$Al half life & 0.717\,Ma & *\citet[][]{neumann_differentiation_2012} \\
       $H_{\rm Fe,0}$ & \makecell{Heating power of $^{60}Fe$ at \\ t=0\,Ma after CAI formation.} & 0.0366\,Wkg$^{-1}$  & \citet{ruedas_radioactive_2017}\\
       $X_{\mm{Fe}}$ & wt \% Fe in accreting material & 22.4\,wt \% & \citet{lodders_relative_2021} \\
       $t_{1/2,Fe}$ & $^{60}$Fe half life & 2.62\,Ma & \citet{ruedas_radioactive_2017} \\\hline
\end{tabular}
        \caption{All fixed parameters used in the model and their references. Blank reference indicates the value is chosen in this work. The choice of viscosity parameters is justified in Section \ref{model-eta}. References with a * denote the that they were used in \citet{dodds_thermal_2021}, but the original reference has been given here. }
    \label{tab:parameters}
\end{table}
\begin{table}
    \centering
    \begin{tabular}{|c|c|c|c|}
    \hline
      Symbol   & Meaning & Value & Reference (if relevant) \\ \hline
      \multicolumn{4}{|c|}{\textbf{Computational parameters}}\\\hline
      $\Delta r$ & Cell spacing & 500\,m & \citet{bryson_constraints_2019} \\
    dt & Timestep & $0.075\frac{(\Delta r)^2}{\kappa_\mm{c}}\,s$ &  \\\hline
    \multicolumn{4}{|c|}{\textbf{Variable parameters}}\\\hline
      $R$ & Planetesimal radius & 100--500\,km & \\
      $X_{s,0}$ & Initial core sulfur content & 26.7--33\,wt\% & \\
      $f_{^{60}Fe}$ & $^{60}Fe/^{56}Fe$  & $0-6\times10^{-7}$& \makecell{\citet{dodds_thermal_2021} \\\citet{cook_iron_2021}}\\
      $t_{acc}$ & Accretion time & $<2.5$ Ma after CAI formation & \citet{neumann_differentiation_2012} \\\hline
    \end{tabular}
    \caption{Computational parameters and variable parameters for model runs. The choice of variable parameters is explained in Section \ref{model-params}. Accretion times are discussed further in Section \ref{acc-diff}.}
    \label{tab:constants}
\end{table}
\section{Example Run}\label{exrun}
\begin{figure}
    \centering
    \includegraphics[width=1\textwidth]{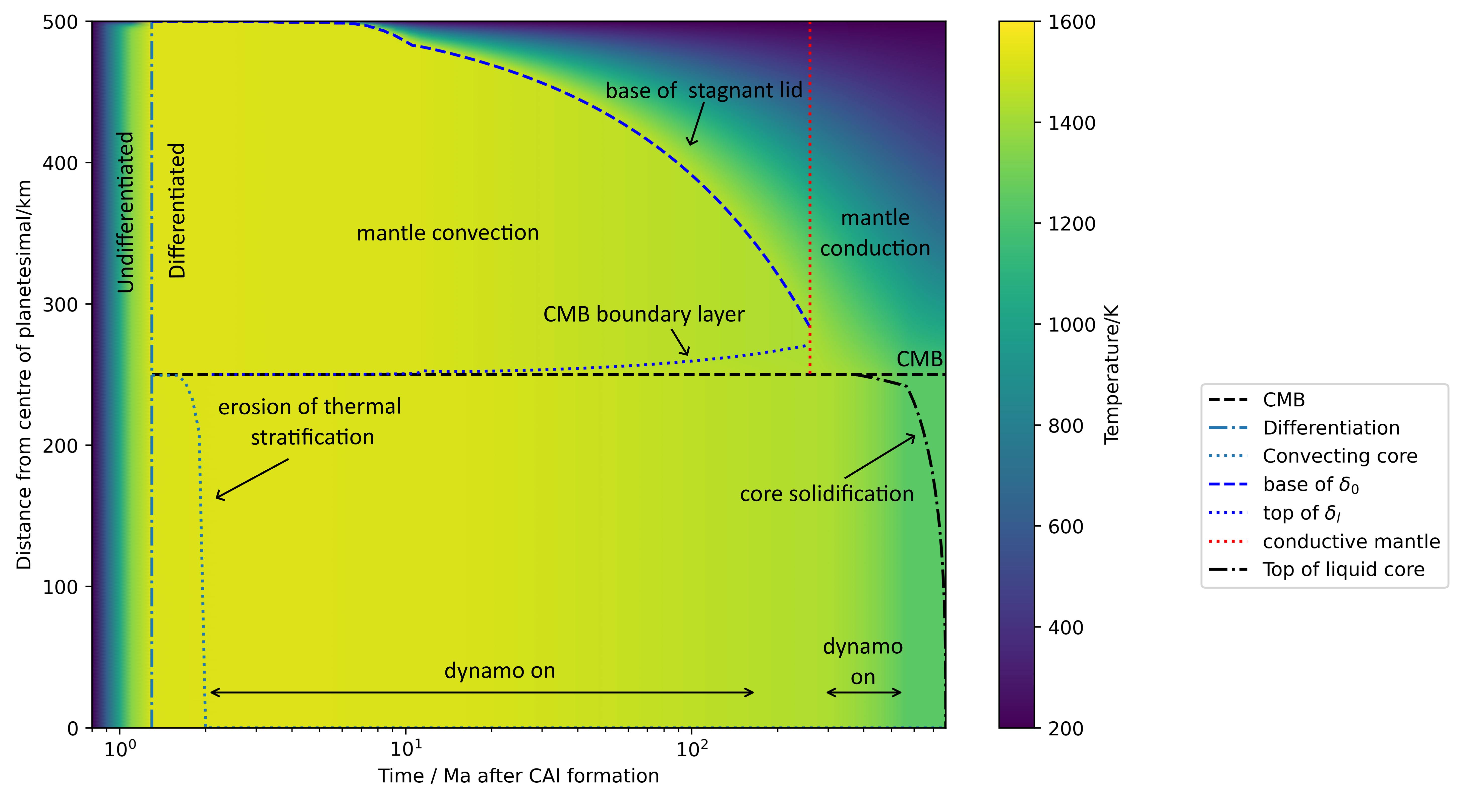}
    \caption{Annotated thermal profile for a 500\,km planetesimal accreted at 0.8\,Ma after CAI formation with \feratio=$10^{-8}$ and \Xs=29.85\,wt\%. Prior to differentiation, the body heats up due to radiogenic heating by \alp. After differentiation, core thermal stratification is eroded rapidly due to heating by \fep. The stagnant lid and CMB boundary layer thicken until the two boundary layers meet and convection ceases, which terminates the dynamo. The temperature gradient at the CMB steepens following the cessation of convection and the dynamo restarts prior to the onset of core solidification. The black horizontal arrows indicate the period of dynamo activity, their vertical position is arbitrary.}
    \label{fig:therm-prof}
\end{figure}
\begin{figure}
    \centering
    \includegraphics[width=1\textwidth]{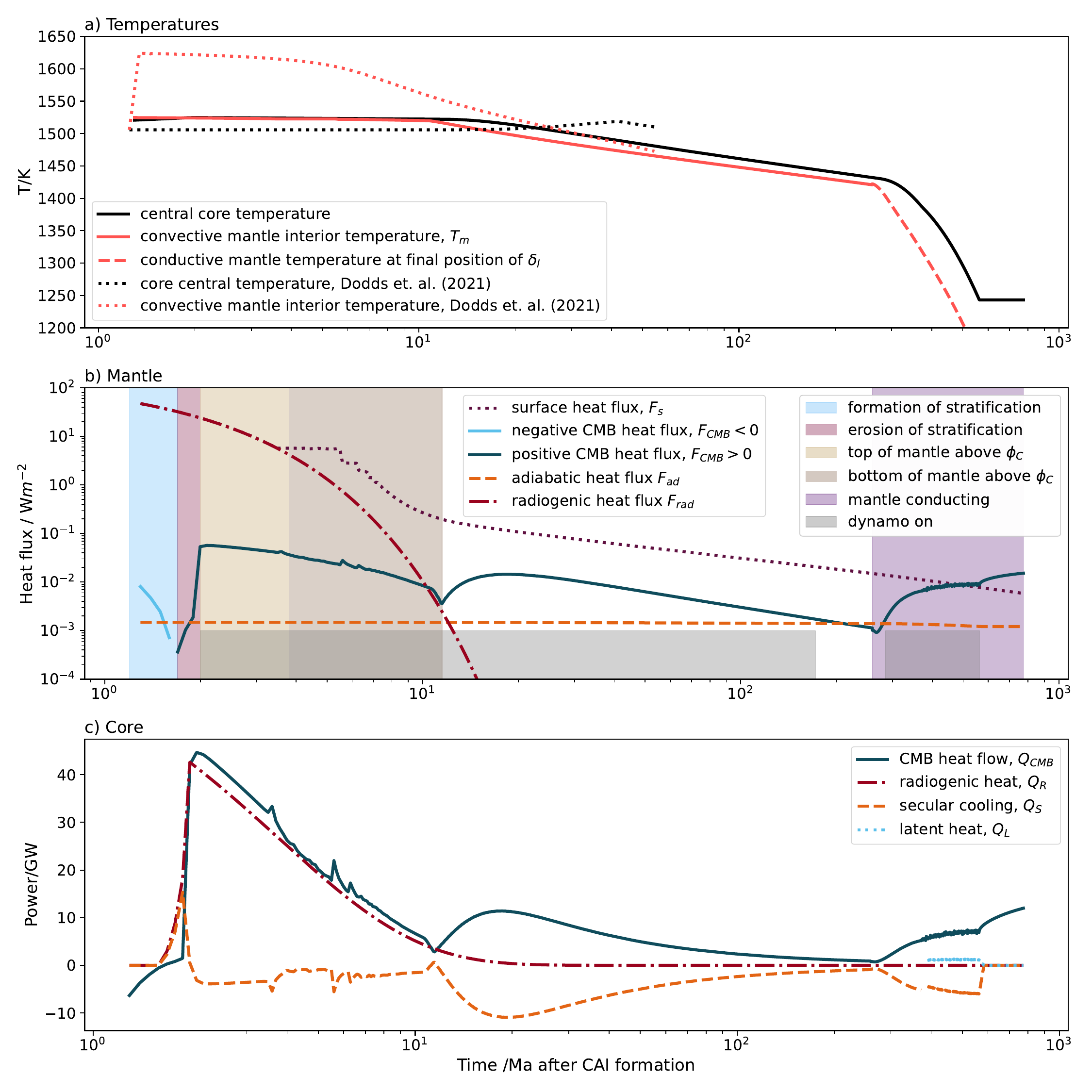}
    \caption{Temperatures (a) and mantle heat fluxes (b) and core power sources (c) for the same parameters as Figure \ref{fig:therm-prof}: 500\,km planetesimal accreted at 0.8\,Ma after CAI formation with \feratio=$10^{-8}$ and \Xs=29.85\,wt\%. In the upper panel, after the cessation of convection, the mantle temperature evolution at the position of the final mantle height of the CMB boundary layer is shown by the red, dashed line. For comparison, the central core temperature and convecting mantle temperature from \citet{dodds_thermal_2021} are shown by the dotted lines. The \citet{dodds_thermal_2021} temperature evolution is shorter, because their model ended after the cessation of mantle convection ($\sim50$\,Ma in their model).  In the middle panel, at the beginning of the thermal evolution, \fcmb is negative, because the mantle heats up faster than the core (blue box). The absolute value of \fcmb has been plotted for these times (light blue line). Before 3.8\,Ma after CAI formation, the surface heat flux, $F_S$, matches the radiogenic heat flux, $F_{rad}$, because the top of the mantle is above the critical melt fraction and there is a negative feedback loop between mantle temperature and stagnant lid thickness (yellow-brown box). The downward spike in \fcmb at 13\,Ma after CAI formation is due to the rapid thickening of the CMB boundary layer when the base of the mantle drops below the critical melt fraction. The horizontal plateaus and vertical drops in $F_S$ and spikes in \fcmb prior to 13\,Ma is a model artefact due to the discretisation of the stagnant lid, and we expect smooth curves in reality. The CMB heat flux, $F_{\mm{CMB}}$, decreases with time and the dynamo turns off just before the cessation of convection due to the thickening of the CMB boundary layer. In the lower panel, the latent heat contribution only appears once the core begins solidifying. The latent heat release and secular cooling during core solidification are shown as a rolling average over 200 timesteps (20\,Ma) to remove oscillations due to the discrete nature of the model.}
    \label{fig:flux}
\end{figure}
\begin{figure}
    \centering
    \includegraphics[width=1\textwidth]{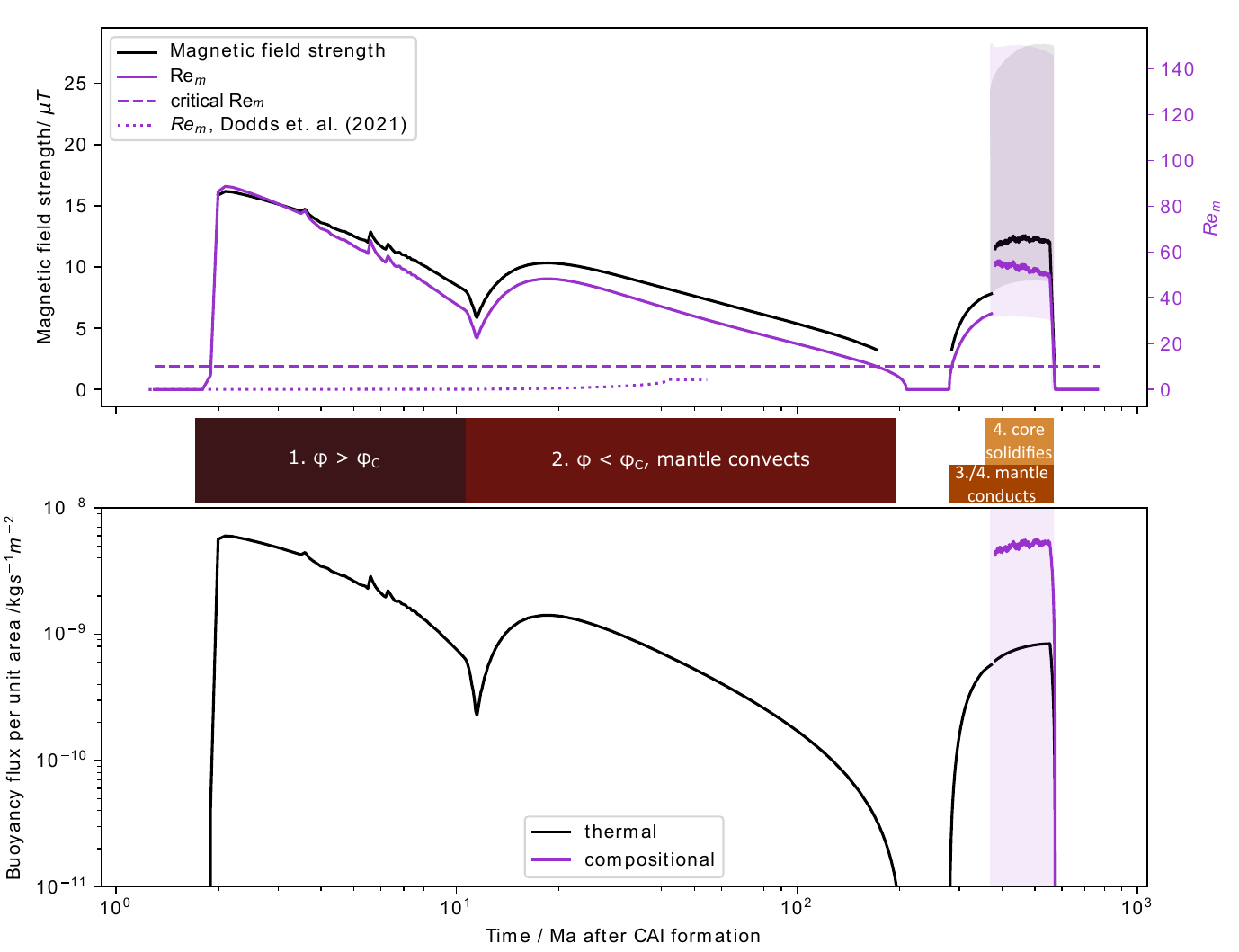}
    \caption{Predicted dipole magnetic field strength at the surface of the planetesimal and magnetic Reynolds number, \Remp, (upper panel) and buoyancy flux per unit area (lower panel) as a function of time for a 500\,km planetesimal accreted at 0.8\,Ma after CAI formation with \feratio=$10^{-8}$, \mfracp=1 and \Xs=29.85\,wt\%. In the upper panel, the purple, horizontal dashed line is a critical \Rem of 10 and the dotted, purple line is the \Rem from \citet{dodds_thermal_2021}. The longevity of core thermal stratification in \citet{dodds_thermal_2021} prevented their \Rem from becoming supercritical. The spikes in the traces prior to 13\,Ma are model artefacts due to the discretisation of the stagnant lid. Due to the discrete nature of latent heat release in the model, the magnetic field strength and \Rem oscillate during core solidification. This oscillating output is shown by the faded grey and purple traces, and the rolling average over 200 output steps (20\,Ma) is shown in bold. The small gap between the purely thermally driven dynamo and the average values for the thermo-compositional dynamo is due to the lag in the rolling average. The thermal (first term in square brackets in Equation \ref{eq:Qb}) and compositional (second term in square brackets in Equation \ref{eq:Qb}) buoyancy fluxes are shown in the lower panel. Again, the model compositional buoyancy flux (faded purple) and rolling average buoyancy flux (bright purple) are shown. Time is plotted logarithmically to make the early trends in magnetic field generation with time visible. The downward spike in magnetic field strength in the first epoch of dynamo generation occurs when the mantle cools sufficiently for its melt fraction, $\phi$, to drop below the critical melt fraction, \rcmfp. The dynamo stops when mantle convection ceases and restarts prior to the onset of core solidification once the conductive temperature gradient at the CMB has steepened sufficiently. Core solidification very slightly increases magnetic field strength and \Remp.}
    \label{fig:B}
\end{figure}
\begin{figure}
    \centering
    \includegraphics[width=1\textwidth]{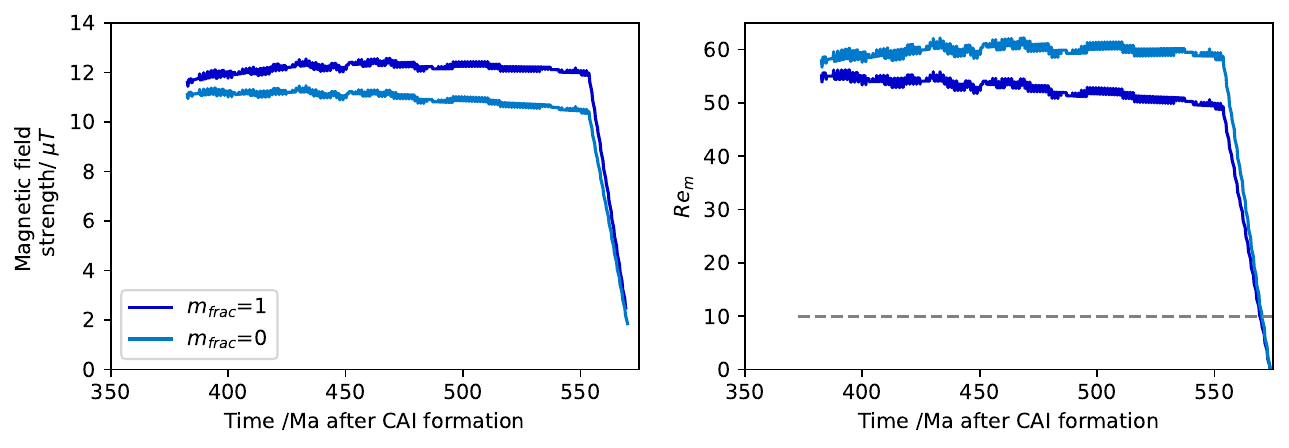}
    \caption{Magnetic field strength (left panel) and magnetic Reynolds number, \Remp, (right panel) for the two core solidification endmembers: no solidified material falls to the centre and no passive inner core forms, \mfracp=0 (dark blue line); and all solidified material falls to the centre to form a passively growing inner core, \mfracp=1 (light blue line). The lines indicate a rolling average over 20\,Ma. Magnetic field strength is slightly increased and \Rem is slightly decreased in the presence of a passively growing inner core, but the difference between the two endmembers is minimal.}
    \label{fig:Bmfrac}
\end{figure}
Results of an example run for a 500\,km body that accreted 0.8\,Ma after CAI formation with \feratio=$10^{-8}$ and \Xs=$29.85$\,wt\% are shown in Figures \ref{fig:therm-prof}--\ref{fig:Bmfrac}. The planetesimal radius and accretion time were chosen to enable comparison with the instantaneous accretion model (Case 1) from \citet{dodds_thermal_2021}. The \Xs value is the median value for the range of \Xs that can be used in our model. Following a similar argument, we chose to use \feratio=$10^{-8}$ to produce a representative simulation. This \fe value is higher than that used in previous models \citep[\feratio= 0;][]{bryson_constraints_2019,dodds_thermal_2021}, but is on the lower end of recently measured primordial \feratio values, which range from $10^{-8}$ to $6\times10^{-7}$ \citep{cook_iron_2021}. All fixed parameters are given in Table \ref{tab:parameters}. 

The thermal profile for this 500\,km radius planetesimal is shown in Figure \ref{fig:therm-prof}. The planetesimal reaches its critical melt fraction, \rcmfp, and differentiates at 1520\,K, 1.2\,Ma after CAI formation\footnote{The time resolution of this description is limited by the frequency at which model output is saved. Output is saved every 0.01\,Ma prior to differentiation and every 0.1\,Ma after differentiation. All times given are representative of the approximate time for these processes. Therefore times are given to a precision of 0.1\,Ma from 1--10\,Ma, 1\,Ma for 10--100\,Ma and 10\,Ma for >100\,Ma.} (0.4\,Ma after accretion). Once the planetesimal differentiates, the combined stagnant lid and CMB boundary layer thickness is less than the mantle thickness and the whole mantle convects. The mantle reaches a peak mantle temperature of 1525\,K within 0.1\,Ma of differentiation. At these early times, radiogenic heating from \al is so strong that the dominant controls on mantle temperature are surface heat flux, $F_S$, and radiogenic heat flux, $F_{rad}$ (see Figure \ref{fig:flux}, middle panel). When the mantle melt fraction is at or above \rcmfp, there is a negative feedback between these two heat fluxes, because the planetesimal loses heat very efficiently. An increase in mantle temperature, due to high $F_{rad}$, causes the stagnant lid to thin, which in turn increases $F_S$, causing the mantle temperature to decrease. This results in a mantle peak temperature close to $T_{\phi_\mm{C}}=1520$\,K and very thin stagnant lid ($\sim$100-700\,m) and CMB boundary layers ($\sim$100\,m) at early times. This feedback plays a role until 3.8\,Ma after CAI formation (Figure \ref{fig:flux}). From 1.2--1.7\,Ma after CAI formation the mantle is hotter than the core due to the faster decay of \al in the mantle compared to \fe in the core. From 1.7--2.0\,Ma after CAI formation, core thermal stratification is eroded as radiogenic heating from \fe increases the core temperature (Figure \ref{fig:flux}, top panel). Once core thermal stratification has been eroded, the core begins to convect. \Rem is immediately supercritical, so magnetic field generation begins 2.0\,Ma after CAI formation. At 4.3\,Ma after CAI formation, the temperature at the top of the mantle drops below $T_{\phi_\mm{C}}$ and the stagnant lid begins to thicken. At 13\,Ma after CAI formation, the temperature at the base of the mantle drops below $T_{\phi_\mm{C}}$ and the CMB boundary layer begins to thicken. As the stagnant lid and CMB boundary layer thicken, \fcmb decreases and the first epoch of magnetic field generation ends $\sim$170\,Ma after CAI formation. At $\sim$260\,Ma after CAI formation, the combined CMB boundary layer and stagnant lid thickness reaches the thickness of the mantle, so mantle convection stops (Figure \ref{fig:therm-prof}). Immediately after the cessation of mantle convection, there is a shallower temperature gradient across the portion of the mantle that was the CMB boundary compared to the portion of the mantle that was the stagnant lid (Figure S11). As a single conductive gradient is established throughout the whole mantle, the conductive temperature gradient at the CMB increases and the temperature gradient near the surface decreases. As a result, \fcmb increases and the dynamo restarts at $\sim$280\,Ma after CAI formation. Core solidification starts at $\sim$370\,Ma, which provides an additional source of buoyancy for dynamo generation. As solidification proceeds the liquid core sulfur content increases and the core reaches the FeS eutectic at $\sim$550\,Ma. This reduces the buoyancy to drive the dynamo, causing \Rem to become subcritical and the dynamo shuts off. The core solidifies completely by $\sim$770\,Ma.

\begin{table}[ht]
    \centering
    \begin{tabular}{|c|c|c|c|c|}
    \hline
       &    Our model & Previous model & Reference & Reason \\\hline
      Peak mantle temperature /K & 1525 at 1.2\,Ma & 1620 at 1.2\,Ma &    a & New viscosity model, lower \rcmf \\
      Mantle hotter than the core until /Ma  & 1.7 & 9 &  a & Inclusion of \fe \\
      Core thermal stratification eroded /Ma  & 2.0  & 40 & a & Inclusion of \fe \\
      End of mantle convection/Ma  & 260 & 56 &  a &  Mantle convection model \\
      First dynamo onset /Ma  & \makecell{2.0 \\ 1.4} & \makecell{- \\ 5.4}  & \makecell{ a \\ b} &  Inclusion of \fe \\
      First dynamo cessation /Ma & \makecell{170 \\ 150} & \makecell{- \\ 17} & \makecell{a \\ b} &  Mantle convection model \\\hline
    \end{tabular}
    \caption{Differences between key aspects of planetesimal thermal evolution between our model and previous models, and a reason for each difference. Our model run is for a 500\,km radius planetesimal accreted at 0.8\,Ma after CAI formation with \feratio=$10^{-8}$. The Case 1 run from \citet{dodds_thermal_2021} (a) is used for comparison. Dynamo generation is compared with both Case 1 from \citet{dodds_thermal_2021} and the single accretion event model from \citet{bryson_constraints_2019} (b), because thermal stratification prevented dynamo generation in \citet{dodds_thermal_2021}. A model run for a 400\,km radius planetesimal accreted at 0.5\,Ma after CAI formation was used to compare with the \citet{bryson_constraints_2019} model. All times are Ma after CAI formation and are given to a precision of 0.1\,Ma from 1--10\,Ma, 1\,Ma for 10--100\,Ma and 10\,Ma for >100\,Ma. \rcmf is the critical melt fraction. }
    \label{tab:timings}
\end{table}

\subsection{Inclusion of $^{60}$Fe}
Previous models neglected the role of radiogenic heating by \fe in the core because it only led to a $\sim$10\,K difference in peak core temperature over the lifetime of the planetesimal \citep{henke_thermal_2013}. Although this temperature difference is small compared to the overall core temperature, it is significant compared to the temperature difference across the CMB ($\sim1-5$\,K). Radiogenic heating by \fe is therefore important at early times in the thermal evolution. Indeed, radiogenic heating from \fe can increase the core temperature above the temperature at the base of the mantle, causing the core's thermal stratification to be removed far more rapidly (see Table \ref{tab:timings}) than in previous models \citep{dodds_thermal_2021}. As a result, the inclusion of \fe can lead to an earlier onset of dynamo generation \citep[by $\sim$4\,Ma compared to][]{bryson_constraints_2019} or an early epoch of dynamo generation that was not observed for an instantaneously accreting body in the model presented by \citet{dodds_thermal_2021}. Because it will cause more heat to be produced, the effect of \fe on dynamo timing and strength will be even more significant for the upper estimates of primordial \feratiop. 

\subsection{Mantle convection criterion}
Mantle convection and the first epoch of dynamo generation lasts an order of magnitude longer in our model than previous models (Table \ref{tab:timings}) due to our chosen criteria for the cessation of mantle convection. 
In previous models, solid-state convection immediately ceased when the Rayleigh number dropped below the critical Rayleigh number, $Ra_\mm{c}$, for the first time \citep{SterenborgCrowley2013,bryson_constraints_2019,dodds_thermal_2021}. This previous cessation criterion resulted in mantle convection ceasing early, at a time when the CMB boundary layer was too thin and the temperature profile near the base of the mantle was isothermal (the mantle adiabatic gradient is negligible). This isothermal profile resulted in a pause in cooling at the CMB until conductive cooling from the top of the mantle reached the CMB. We have mitigated this artefact in our model by assuming the cessation of convection occurs when the combined stagnant lid and CMB boundary layer thickness equals the mantle thickness. 
This better represents the gradual cessation of convection, because the convecting domain gradually shrinks in size until the entire domain is conductive (Figure \ref{fig:therm-prof}). This criterion has been used to calculate cessation of convection by many planetary thermal evolution studies \citep[e.g.][]{grott_thermo-chemical_2011,morschhauser_crustal_2011} and predicts similar thermal evolutions to equivalent 2D and 3D numerical simulations \citep{tosi_thermochemical_2013}. We assume a conductive profile in the CMB boundary layer, such that upon the cessation of mantle convection there is already a conductive temperature gradient at the base of the mantle. This ensures there is mantle and core cooling via conduction the moment convection ceases. The gradual thickening of the CMB boundary layer is reflected in the slow decrease in \fcmb and dynamo field strength in Figures \ref{fig:flux} and \ref{fig:B} and lack of spike in \fcmb at cessation of convection compared to \citet{bryson_constraints_2019}. The conductive gradient in the CMB boundary layer enables a second epoch of dynamo generation to begin before the onset of core solidification, because the conductive temperature gradient at the CMB is large enough to produce supercritical \Remp.

 \subsection{Magnetic field generation with time}
Planetesimal magnetic field generation can be split into four regimes (Figure \ref{fig:B}). In Regime 1, radiogenic heating from \fe is strong and the mantle is above \rcmfp, such that the CMB boundary layer is thin and \fcmb is high. The peak magnetic field strength (16\,$\mu$T for our example run) for the whole thermal evolution is reached at the onset of dynamo generation, when the temperature gradient across the CMB is largest due to heating by \fep. Magnetic field strength decreases as the CMB boundary layer thickens (see Figures \ref{fig:therm-prof} and \ref{fig:B}). Regime 1 ends $\sim$13\,Ma after CAI formation when the temperature at the base of the mantle drops below $T_{\phi_\mm{C}}$. There is a downward spike in magnetic field strength at this time, because the CMB boundary layer thickness increases rapidly (see Figure S2) and the same temperature difference is accommodated across a larger distance, which decreases \fcmbp. At the beginning of Regime 2, as the mantle continues to cool, the core-mantle temperature difference initially increases more rapidly than the CMB boundary layer thickness, which temporarily increases \fcmb and the magnetic field strength. \fcmb and magnetic field strength then decline as the CMB boundary layer thickening outweighs the change in core-mantle temperature difference. Regime 3 occurs after the mantle has stopped convecting. The conductive temperature gradient at the CMB steepens (Figure S11), which increases \fcmb enough to restart the dynamo. Magnetic field strength increases with time up to 8\,$\mu$T, because the temperature gradient in the deep mantle increases as the mantle cools. Once core solidification begins this provides an additional source of buoyancy to drive the dynamo (Regime 4) and the field strength is approximately constant at 13$\mu$T until the core reaches the eutectic composition. Once the core reaches this composition there is insufficient buoyancy flux to drive the dynamo. 

The field strengths and transition times between these regimes will change with chosen planetesimal parameters. For lower core sulfur contents, core solidification will start earlier and the compositionally driven regime (Regime 4) may overlap with preceding regimes (Figure \ref{fig:xs-mag}). Smaller planetesimals will cool more quickly, so will have a compressed magnetic field generation history. The transition between Regimes 1 and 2 is partially due to the jump in viscosity. These two regimes may become one with a smooth decrease in field strength with time if viscosity parametrisations are improved in the future. 

\subsubsection{Compositional vs thermal dynamos}
\begin{figure}
    \centering
    \includegraphics[width=1\textwidth]{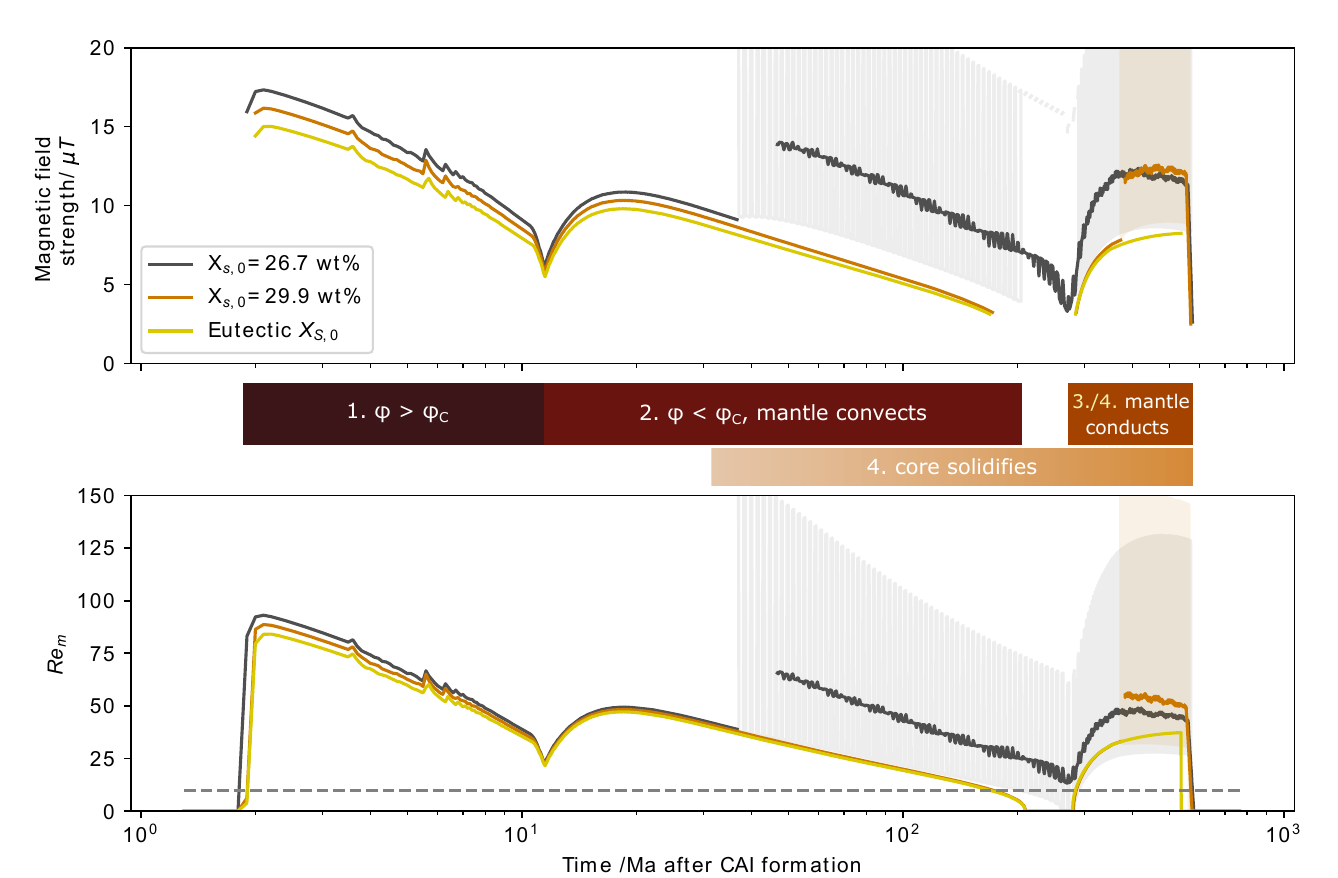}
    \caption{Magnetic field strength and \Rem with time for 500\,km planetesimals with \feratio=$10^{-8}$ and initial core sulfur contents, \Xsp, from 26.7--33\,wt\%. Due to the discrete nature of the model, the magnetic field strength and \Rem oscillate during core solidification. This oscillating output is shown by the faded traces and the rolling average over 200 output steps (20\,Ma) is shown in bold. The small gap between the purely thermally driven dynamo and the average values for the thermo-compositional dynamo is due to the lag in the rolling average. The dynamo generation regimes are the same as in Figure \ref{fig:B}. As \Xs decreases, the core liquidus temperature increases and the onset of core solidification moves to earlier times. Compositional buoyancy leads to up to a 70\% increase in magnetic field strength. Core solidification is not required to produce a second epoch of dynamo generation. }
    \label{fig:xs-mag}
\end{figure}

In our model, the primary difference between compositional and thermal dynamos is the release of gravitational potential energy in the liquid region of the core, which increases the buoyancy flux to drive the dynamo.
As a result, thermo-compositional dynamos generate stronger magnetic fields than thermal dynamos at the same point in planetesimal thermal evolution (orange/grey vs yellow lines in Figure \ref{fig:xs-mag}). For a 500\,km body at the thermo-compositional field strength peak, the magnetic field is 1.7 times stronger than the thermal dynamo alone at the same period of dynamo generation and the compositional buoyancy flux is $\sim$6 times higher than the thermal buoyancy flux (Figure \ref{fig:B}). However, the absolute difference in magnetic field strength between mechanisms ($\sim$7\,$\mu$T) would require small measurement uncertainties to be detected in the paleomagnetic record and would be degenerate with the time the magnetisation was recorded. For instance, in Regime 3, both field strengths decrease with time so a later generated thermo-compositional field could have the same strength as a thermally generated one earlier in the same regime. The timing of the different magnetic field regimes in also strongly affected by planetesimal size \citep{sanderson_early_2024}.

Although the onset of core solidification may not have a resolvable affect on field strength, it can have a strong effect on the timing of dynamo generation. For \Xs=26.7\,wt\%, the increase in buoyancy flux due to core solidification is enough to prevent cessation of the dynamo when mantle convection ceases and there is no gap in dynamo generation. 

Our model now predicts the first epoch (epoch denotes a continuous period of dynamo generation compared to regimes, which demarcate the different mantle/core states during dynamo generation) of thermal dynamo generation can last up to 170\,Ma after CAI formation for 500\,km radius planetesimals (Figure \ref{fig:B}). The dynamo restarts in 500\,km radius planetesimals with \Xs= 29.85\,wt\% and 33\,wt\% at $\sim 280$\,Ma after CAI formation prior to the onset of core solidification (Figure \ref{fig:xs-mag}). As such, core solidification is not required for a second epoch of dynamo generation. Core solidification does not prolong the life of the dynamo, but strengthens the resulting magnetic field and narrows the gap between generation epochs. The restart of the dynamo prior to core solidifcation predicted by our model suggests that core solidification may not be required to explain younger magnetic remanences \citep[e.g.][>65\,Ma after CAI formation based on current evidence]{nichols_time-resolved_2021,maurel_long-lived_2021}. This could resolve the problem of finding an inward solidification mechanism which can drive compositional convection rather than stably stratifying the core \citep{dodds_direction_2024}, because the later meteorite paleomagnetic remanences could have been generated by a purely thermal dynamo.

\subsubsection{Magnetic field strengths}
The magnetic field generation scaling laws we use enable us to predict both \Rem and dipolar, surface magnetic field strengths for the entire planetesimal thermal evolution. These magnetic field strengths can be compared to the paleomagnetic record to determine a range of possible parameters for a given meteorite parent body \citep[see][]{sanderson_early_2024}.
We have determined that a MAC scaling law for magnetic field strength is appropriate for planetesimal cores (Table S1). Previous planetesimal thermal evolution models were uncertain which core force balance and scaling law was appropriate for planetesimals, so did not predict magnetic field strengths \citep{bryson_constraints_2019,dodds_thermal_2021} or used multiple scaling laws \citep{SterenborgCrowley2013}. Additionally, we have developed a parametrised model for inward core solidification, which enables us to calculate a combined thermal and compositional buoyancy flux (Equation \ref{eq:Qb}) and use the same convective velocity and magnetic field scaling laws before and during core solidification. This means a full internally-generated magnetic field history for a given planetesimal can be predicted without having to assume a single buoyancy source at a given time. This is important for trying to understand time-resolved magnetic field generation records for a specific body, such as the Main Group pallasites \citep{tarduno_evidence_2012,bryson_long-lived_2015,nichols_time-resolved_2021} and the IIE irons \citep{maurel_long-lived_2021}, for which magnetic field strengths were previously predicted only during outward solidification with a conductive mantle. 

\clearpage 

\section{Accretion and differentiation assumptions}\label{acc-diff}

\subsection{Accretion timescale and partial differentiation}
We do not consider the effects of gradual accretion, because there are a range of possible accretion growth laws \citep{neumann_differentiation_2012} and the most appropriate law for planetesimals may depend on the initial conditions of dynamical simulations \citep{weidenschilling_accretion_2019}. 
Prolonged accretion over $\mathcal{O}$(0.1\,Ma) could erode core thermal stratification by the later addition of cool material to the top of the core, which can bring forward the onset of the dynamo relative to the instantaneous accretion case \citep{dodds_thermal_2021}. However, our model includes heating by \fe in the core, which can rapidly remove thermal stratification and will counter the effect of prolonged accretion on core thermal structure. 

Accretion over even longer timescales ($\mathcal{O}$(1\,Ma)) following an exponential accretion law  can lead to partial differentiation, smaller planetesimal cores, and weaker dynamos \citep{dodds_thermal_2021}. Partial differentiation could be included in the model in the future by adding regolith with lower conductivity to the top of the planetesimal and introducing a variable ratio of core to planetesimal radii, $\frac{r_\mm{c}}{R}$. This could lead to a weaker dynamo with shorter duration for a given planetesimal radius compared to the results in this paper due to reduced core and mantle size. These extensions will be important for applying the model to partially differentiated bodies that preserve a thick chondritic layer and have been proposed to have generated a magnetic field, such as the CV and H chondrites \citep{elkins-tanton_chondrites_2011,bryson_paleomagnetic_2019}.

\subsection{Compaction, sintering and regolith}\label{acc-diff-sint}
Our model does not include variations in planetesimal thermal conductivity between accretion and differentiation due to compaction and sintering. The planetesimal sinters at 700\,K  \citep{yomogida_multiple_1984} and rapidly reaches the thermal properties assumed in this model. For a 500\,km planetesimal the inner 496\,km is below 700\,K only for the first 0.12\,Ma after accretion and the last node sinters 0.5\,Ma after accretion, so a lower thermal conductivity prior to sintering  has a negligible effect on heat transport over the 10--500\,Ma timescales for dynamo generation. We have neglected any unsintered regolith because this layer is so thin ($<500$\,m). This may result in a very slight over-estimate of the surface heat flux and planetesimal cooling rate. 

\subsection{Water}
Following other models of differentiated planetesimals \citep{neumann_modeling_2018,sturtz_birth_2022,monnereau_differentiation_2023}, our model does not include the thermal effects of water prior to differentiation. We adopted this approach because radiogenic heating is strong enough that the thermal contributions of melting ice, hydrating, and dehydrating silicates has a minimal effect on the differentiation time. Additionally, the early accreted planetesimals considered here rapidly reach temperatures above the silicate dehydration temperature \citep[1223\,K,][]{lichtenberg_water_2019}, and after differentiation planetesimals retain very little water \citep{peterson_h_2023,newcombe_degassing_2023}.

\subsection{Differentiation mechanism}\label{acc-diff-diff}
Our model for differentiation must enable planetesimals to form a core in which a dynamo can be generated. In order for the planetesimal to differentiate, the material which will form the core, assumed to be Fe-FeS, must be able to move to the centre of the body. There are at least two possible differentiation mechanisms: percolation, where Fe-FeS melt, which has a lower solidus than silicates, moves along grain boundaries in a solid, silicate matrix; and rain-out, where more dense molten Fe-FeS falls through less dense silicate melt to the planetesimal centre.  
Percolation requires an interconnected network of melt. Metal melts have large dihedral angles \citep[$110\degree$][]{neri_reevaluation_2020} and need high volume fractions to form this interconnected network \citep[up to 20-25\%,][]{neri_reevaluation_2020}. These high volume fractions would require high planetesimal metal contents, which would result in unrealistically large core radii. Only 12.5\,vol\% metal is required to form a core that has a radius that is 50\% the radius of the planetesimal \citep{dodds_thermal_2021} and cores predicted to form from chondritic compositions have core radii that are 30--50\% of the planetesimal radius \citep{bercovici_effects_2022}. However, the need for an inter-connected network of metal for percolation could be removed by shear deformation due to solid-state convection \citep{rushmer_physical_2000}, crack formation \citep{keil_explosive_1993} or silicate melting \citep{wilson_thermal_2008}. 

Estimates of the percolation velocity, assuming Darcy flow, are linearly dependent on the permeability of the planetesimal matrix, for which model values vary by six orders of magnitude from $10^{-8}\rm m^2$ \citep{fu_fate_2014} to $10^{-14}\rm m^2$ \citep{neumann_differentiation_2012}. As a result, the time taken to differentiate via this mechanism is very uncertain ($\mathcal{O}(10$--$10^6$) years, see Section S5.1). Differentiation via rain out occurs rapidly in comparison to a model timestep, $\mathcal{O}(10^2)$ years for 1\,mm radius droplets, once $\phi=$\rcmf (i.e., once the silicate portion of the body has a liquid viscosity). Therefore, in this model we adopt differentiation via rain out when $\phi$ = \rcmfp, because the mechanics are simpler and any Fe-FeS that has not already reached the centre via percolation will move there rapidly once $\phi$ = \rcmfp.

\subsubsection{Minimum core sulfur content}\label{acc-diff-xs}
\begin{figure}
    \centering
    \includegraphics[width=1\textwidth]{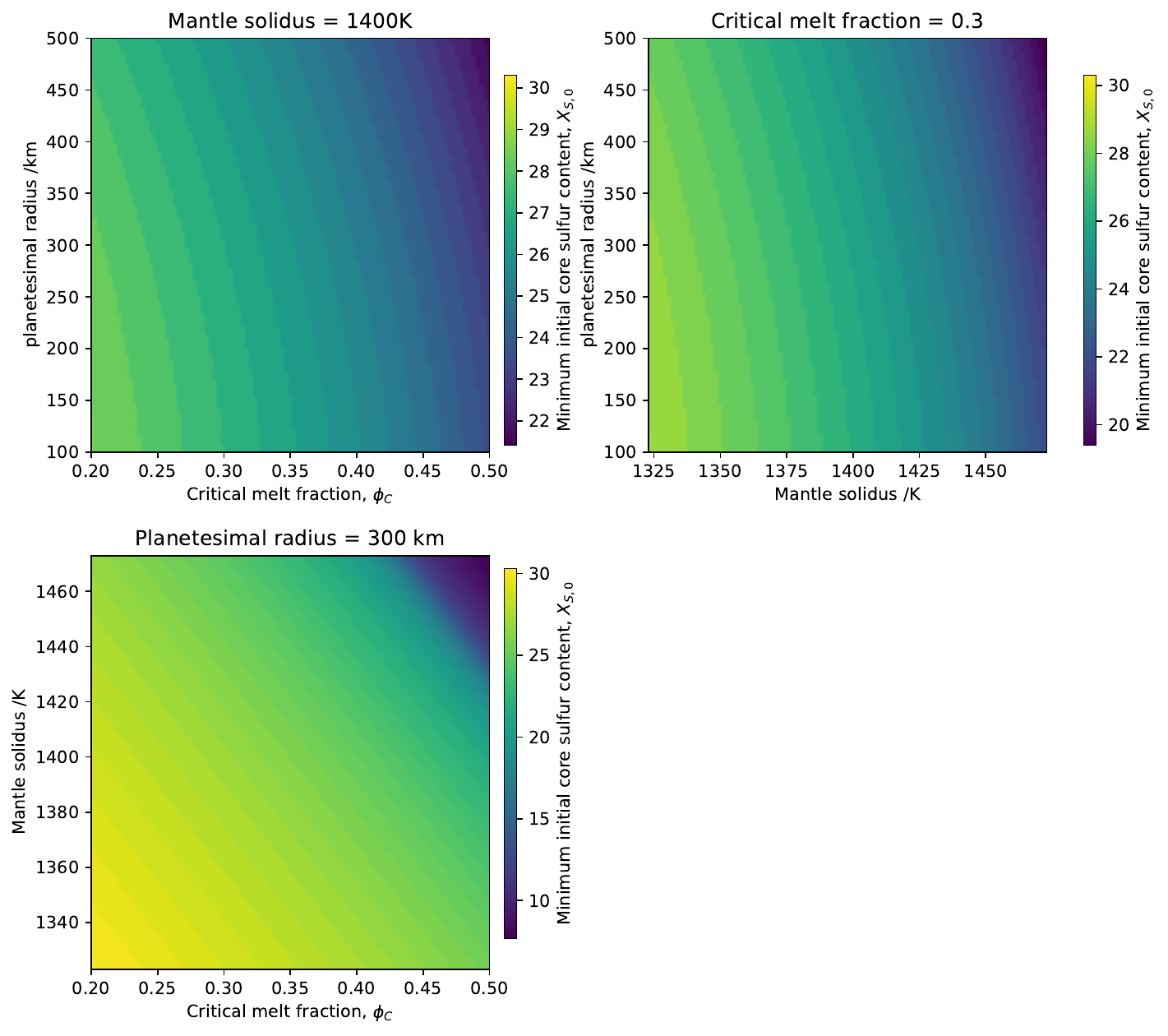}
    \caption{Minimum initial core sulfur content, \Xsp, in our model as a function of: planetesimal radius, which controls core pressure (top left, top right); critical melt fraction, which controls the temperature at differentiation (top left, bottom left); and mantle solidus, which controls the temperature at differentiation (top right, bottom left). The lowest radii, lowest critical melt fractions and lowest mantle solidus temperatures have the highest minimum \Xsp. The minimum and maximum values on the colour scale vary between subplots, in order to highlight the effect of each parameter. Variations in mantle solidus temperature combined with variations in critical melt fraction produce the largest spread in \Xsp. }
    \label{fig:min-Xs}
\end{figure}
Our model assumes differentiation occurs at the critical melt fraction and that the core of the planetesimal is completely molten at the point of differentiation, which limits the range of initial core sulfur contents we can model. For the core material to be completely molten at differentiation requires the differentiation temperature to be greater than or equal to the Fe-FeS liquidus. The liquidus temperature is a function of core sulfur content and core pressure (Equation \ref{eq:liquidus}), while the differentiation temperature is assumed to be equal to the temperature of \rcmfp, which depends on the mantle solidus (Section \ref{acc-diff-solid}) and the choice of \rcmfp. Experiments \citep[0.2,][]{scott_effect_2006} and previous planetesimal thermal evolution models \citep[0.5, e.g.][]{bryson_constraints_2019} provide a possible range of \rcmfp.

Previous studies of planetesimal dynamo generation were not limited in their value of \Xsp, because they used the linear liquidus approximation \citep{neufeld_top-down_2019,nichols_time-resolved_2021}, which has lower liquidus temperatures for \Xs in the range 0--10\,wt\% compared to the \citet{buono_fe-rich_2011} liquidus (Figure S5). Prior studies of planetesimal differentiation via percolation predicted differentiation of sulfur-poor bodies, because they only considered heat transfer via conduction not convection, which allows for higher maximum planetesimal temperatures \citep{neumann_differentiation_2012}.

For the mantle solidus used in this model (1400\,K), the minimum \Xs varies from 22.2--28.3\,wt\% (Figure \ref{fig:min-Xs}), which is higher than the sulfur contents in iron meteorites \citep[0--17\,wt\%][]{kruijer_protracted_2014}. If the mantle solidus in the model is raised to 1473\,K \citep{fu_fate_2014}, the minimum \Xs varies from 6.9--25.9\,wt\% S (Figure \ref{fig:min-Xs}). However, it is unclear if this is the most appropriate solidus value and the lower values in this range require extremal values of planetesimal size (500\,km) and critical melt fraction (0.5). Planetesimals could have initially been sulfur rich and lost their sulfur by later processes, such as immiscible fluid separation in the core \citep[e.g.][]{bercovici_effects_2022} or volatile loss \citep[e.g.][]{hirschmann_early_2021}. Bodies differentiating from CI, CM, LL, CO, CV, and CK chondritic starting compositions are predicted to have $X_S \geq 22\%$ \citep{bercovici_effects_2022}, which is in line with the minimum \Xs in the model. Further research is needed to constrain the mantle solidus and critical melt fraction and to explore these additional processes occuring during or immediately after differentiation.

As a first estimate, sulfur poor bodies would have an earlier onset and longer duration of compositional convection than those modelled here, because the core liquidus temperature is higher and more core solidification is required before the bulk liquid core sulfur content reaches the eutectic. In addition, decreasing \Xs lowers the density difference between the solidified iron and the bulk liquid core, which could decrease buoyancy flux and magnetic field strength at the beginning of core solidification. 

\section{Mantle assumptions}\label{mantle}
\begin{figure}
    \centering
    \includegraphics[width=1\textwidth]{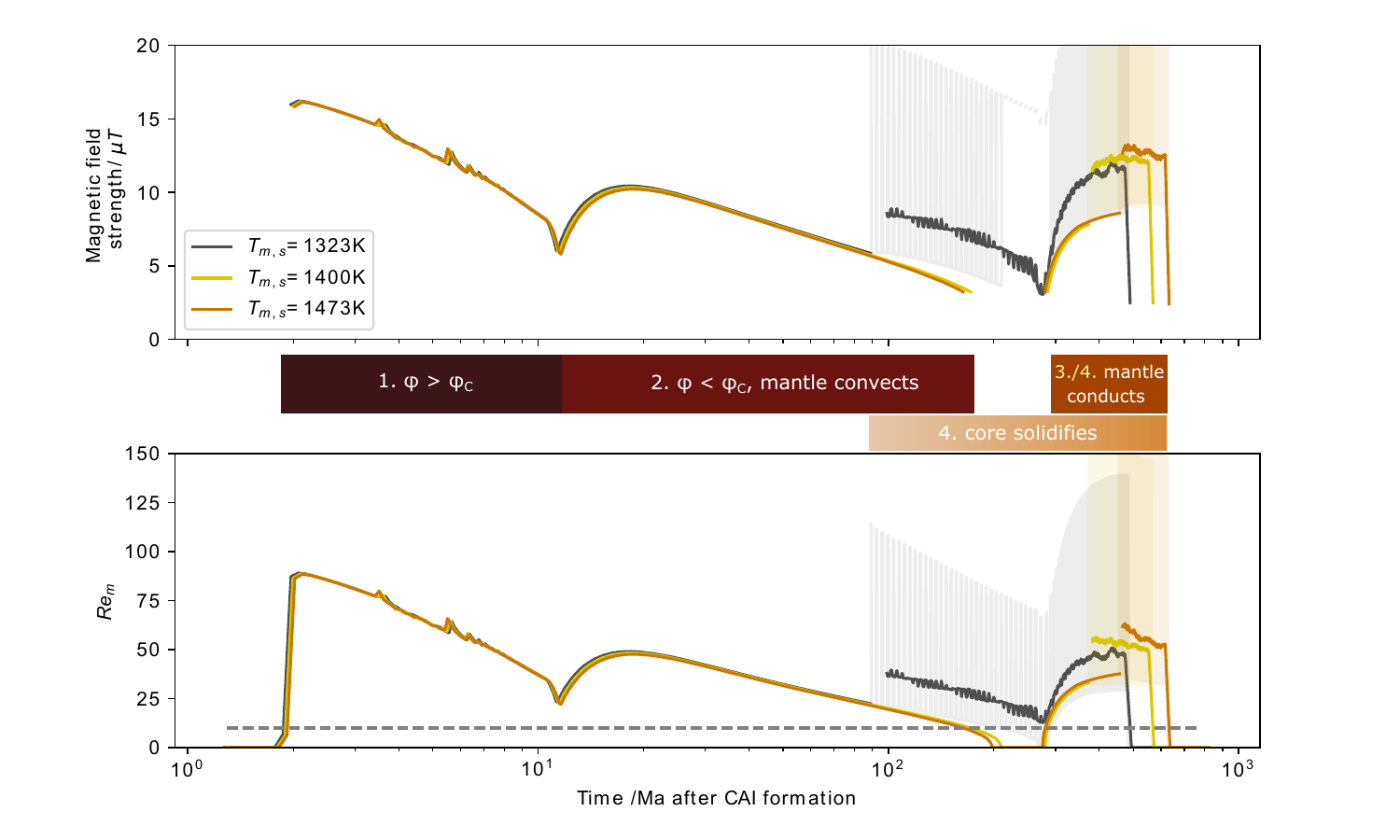}
    \caption{Magnetic field strength and \Rem with time for 500\,km planetesimals with \feratio=$10^{-8}$, \Xsp=29.85\,wt\% and three different mantle solidus temperatures. In all scenarios, the mantle liquidus temperature is 400\,K above the solidus temperature. Due to the discrete nature of the model, the magnetic field strength and \Rem oscillate during core solidification. This oscillating output is shown by the faded traces and the rolling average over 200 output steps (20\,Ma) is shown in bold. The spikes in \fcmb prior to 13\,Ma is due to the discretisation of the stagnant lid. The small gap between the purely thermally driven dynamo and the average values for the thermo-compositional dynamo is due to the lag in the rolling average. The dynamo generation regimes are the same as in Figure \ref{fig:B}. Mantle solidus temperatures do not affect thermal dynamo generation, but lower solidus temperatures bring forward the onset and completion of core solidification, because less cooling is required to reach the core liquidus. This earlier onset of core solidification can provide additional buoyancy and prevent a gap in dynamo generation at the cessation of convection.}
    \label{fig:Bsolidus}
\end{figure}
\subsection{Silicate solidus and liquidus}\label{acc-diff-solid}
The silicate solidus, $T_{\rm{m,s}}$, and liquidus, $T_{\rm{m,l}}$, depend on pressure and composition \citep{hirschmann_mantle_2000}. Meteorite melting experiments suggest the silicate solidus lies between 1323-1423\,K and the liquidus is $\sim$400\,K hotter than the solidus \citep{agee_pressure-temperature_1995,mccoy_partial_1999}. Previous planetesimal models have used values for the mantle solidus between 1400--1473\,K \citep{hevey_model_2006,elkins-tanton_chondrites_2011,fu_fate_2014,kaminski_early_2020,bryson_constraints_2019,dodds_thermal_2021} based on differing degrees of rounding of these experimental results in $\degree$C. For consistency with the studies of \citet{bryson_constraints_2019} and \citet{dodds_thermal_2021}, we have chosen to use a mantle solidus and liquidus of 1400\,K and 1800\,K, respectively. 

Increasing the silicate solidus and liquidus temperature increases the temperature of the critical melt fraction and peak planetesimal temperature. Increasing the silicate solidus has little effect on thermal dynamo generation (Figure \ref{fig:Bsolidus}), but will delay compositional dynamo generation, because a higher silicate solidus temperature requires more cooling before the core reaches the Fe-FeS liquidus, for the same initial core sulfur content. Decreasing the silicate solidus can bring forward the onset of core solidification, for a given initial core sulfur content, to before the cessation of mantle convection. In this scenario, the additional buoyancy provided by core solidification could power the dynamo when mantle convection ceases and prevent a gap in dynamo generation. The choice of mantle solidus will determine the range of \Xs over which early onset of core solidification is possible.

\subsection{Silicate melt migration}\label{mantle-melt-mig}
Silicate melt migration could lead to formation of high silicate melt fraction layers rather than global planetesimal (pre-differentiation) or mantle (post-differentiation) melting due to the preferential partitioning of \al into melt \citep[e.g.][]{neumann_differentiation_2014}. Melt migration is significant if the timescale for percolation of silicate melt through the solid matrix is shorter than the timescale for radiogenic heating \citep{lichtenberg_magma_2019,monnereau_differentiation_2023}. 
However, the timescale for melt migration varies by seven orders of magnitude between studies \citep{moskovitz_differentiation_2011,neumann_differentiation_2012,monnereau_differentiation_2023} due to different assumptions about silicate grain growth and matrix permeability. Therefore, the importance of melt migration in planetesimal differentiation is heavily debated. An additional limitation of these studies of melt migration is that they do not include the possibility of solid-state convection below the critical melt fraction, which may chemically and thermally homogenise the mantle. 
 
Due to the uncertainties in melt migration timescale, we have neglected melt migration in this model. For bodies that accrete $<1.1$\,Ma after CAI formation, the radiogenic heating timescale is short enough to neglect melt migration \citep{monnereau_differentiation_2023}. For bodies accreting after 1.1\,Ma after CAI formation, but early enough to differentiate fully \citep[i.e., $<$2.5\,Ma after CAI formation,][]{neumann_differentiation_2012}, our model may still be appropriate if solid-state convection can homogenise temperature variations resulting from melt migration. 
If melt migration can create temperature inhomogeneity in later accreting bodies after differentiation, our model may overestimate the extent of silicate melting at the base of the mantle. The effects of temperature heterogeneity on \fcmb and the dynamo are unclear, because the mantle temperature will be lower, but the conductive region will be larger (whole mantle rather than thin CMB boundary layer) compared to our model.

\subsection{Compositional homogeneity}
We do not specify a mantle composition and assume the mantle is compositionally homogeneous. A compositionally distinct, cumulate layer could form at the base of the mantle when it is convecting above the critical melt fraction \citep{sturtz_birth_2022}. This could thicken the conductive boundary layer at the base of the mantle, decreasing \fcmb and reducing dynamo strength. However, formation of this cumulate is very sensitive to crystal size and also depends on mantle composition \citep{sturtz_birth_2022}. Because these parameters are uncertain, it is unclear how widespread cumulate formation will be among planetesimals and we have not included it in our model.

\section{Core solidification model}\label{csm}
In order to calculate a full planetesimal thermal history, we had to balance detail in our core solidification model with ease of implementation. Since there is such large uncertainty in the mechanism of core solidification, we did not want to adopt an unreasonable level of specificity in our core solidification model. We chose to focus on two key changes to dynamo generation that will result from core solidification: decrease in convective lengthscale; and change in density difference due to changing composition of bulk liquid core during solidification. The assumptions within our core solidification model are discussed below. 

\subsection{Convective lengthscale}\label{csm-l}
For a planetesimal core solidifying inwards, either a stably stratified, partially solid layer may remain at shallow depths in the core, \citep[e.g. iron snow][]{ruckriemen_fe_2015,davies_iron_2018} or a portion of material may delaminate and fall to the centre of the planetesimal forming a passively growing inner core \citep[e.g.][]{scheinberg_core_2016,neufeld_top-down_2019}. The \mfrac parameter allows us to adjust our model between these two endmembers: no passively growing inner core (\mfracp=0); and all material immediately falling to the centre to form an inner core (\mfracp=1). This gives us a first estimate to the effect of core solidification for the most (\mfracp=1) and least (\mfracp=0) rapid changes in geometric lengthscale. In reality, the convective lengthscale is some fraction of the geometric lengthscale \citep{davidson_scaling_2013,aubert_spherical_2017}, which could increase the magnetic field strength and decrease \Rem (Figure S12). However, the value of this fraction for planetesimal cores is uncertain, so we approximate the convective lengthscale as equal to the geometric lengthscale. This may overestimate dynamo duration and underestimate magnetic field strength. The endmember with the passively growing inner core was used in Figures \ref{fig:B} and \ref{fig:xs-mag}, because this is the minimum estimate of geometric lengthscale and will give a minimum estimate of \Remp. Core solidification with a passively growing inner core has a slightly stronger magnetic field strength and lower \Rem than core solidification where solid without an inner core (Figure \ref{fig:Bmfrac}). However, magnetic field strength and \Rem both scale weakly with convective lengthscale (see Figure S3), so these changes are minimal.

For both scenarios, the amount of material solidified is parametrised by the inward motion of a solidification front, $r_i$, and the buoyancy flux is calculated at this position. For \mfracp=0, the radius of the base of the solidified outer layer, $r_1$ is equal to the position of the solidification front ($r_1=r_i$). For \mfracp=1, there is no outer shell ($r_1=r_\mm{c}$) and all the material is at the centre. In this case, $r_i$ differs from $r_1$ by a maximum of 6\% (Figure S4) before the core reaches the eutectic. Therefore, evaluating the buoyancy flux at $r_i$ rather than $r_\mm{c}$ for \mfracp=1 introduces a minimal effect. 

\subsection{Density difference}\label{csm-snow}
During core solidification, the density difference between the solidified, pure iron phase and the Fe-FeS liquid drives convection. The density difference between the solid and liquid varies as the core solidifies and the sulfur content of the liquid core increases. The adiabatic gradient in planetesimal cores is probably too shallow for solidified iron to remelt as it falls towards the centre of the planetesimal \citep{dodds_direction_2024}. Therefore, we assume the bulk liquid core becomes more sulfur rich as the core solidifies and the driving density difference is between solidified iron and the bulk Fe-FeS liquid. Assuming perfect partitioning of sulfur into the liquid portion of the core \citep{goldstein_iron_2009} may overestimate the density difference available to drive the dynamo. We use the same density difference formalism in both solidification endmembers, because this estimate is the correct order of magnitude for the density difference even if it may differ between solidification modes. 

Our variable density difference is up to seven times larger than in previous models (see Figure S6) which assumed a constant fractional density difference  \citep[$\frac{\Delta\rho}{\rho_\mm{c}}$ of 5\% based on Earth values, eg.][]{nimmo_energetics_2007,bryson_long-lived_2015,bryson_constraints_2019,nichols_time-resolved_2021}. This previous value underestimated the density difference due to the higher pressure and lower core sulfur content in the Earth's core compared to planetesimal cores. Our variable density difference increases the density difference available to drive a dynamo.

Our model does not consider the density difference resulting from super-eutectic core solidification as a mechanism for driving the dynamo. This is for two reasons. First, there is minimal evidence for super-eutectic planetesimal cores in the meteorite record. For instance, there are no iron meteorites with super-eutectic compositions \citep{hilton_chemical_2022} and only parent bodies formed from CM chondrite and CI chondrite compositions potentially could have formed super-eutectic cores \citep{bercovici_effects_2022}. Second, the density and liquidus of super-eutectic Fe-FeS is very poorly constrained \citep{ruckriemen_top-down_2018}. As a result, it is unclear whether crystallisation of super-eutectic Fe-FeS can drive a compositional dynamo.

\subsection{Comparison with iron snow}
The \mfracp=0 endmember is an approximation to iron snow solidification, and uses a simplified description of the energetics and solid fraction in the snow zone. The full energetics of the iron snow system is complex \citep{ruckriemen_fe_2015,davies_iron_2018} and the dynamics of iron snow solidification, including nucleation and bursts of solidification \citep{huguet_laboratory_2023}, are still poorly understood and difficult to incorporate into a numerical model. Our simplified model aims to capture the key aspects of this complicated solidification mechanism for dynamo generation. Additonally, our simplications enable easy transition between core solidification endmembers using a single parameter, \mfracp.

One key approximation in our model compared to the iron snow system is the treatment of the solid fraction in the snow layer. In a full iron snow model, stable stratification is formed at the top of the core by a slurry layer, with a very low, almost constant solid fraction \citep{davies_iron_2018}, whereas in our model this stable stratification is formed by a layer of pure, solid iron. This solid iron layer still captures the decrease in the lengthscale of convection with the downward growth of the snow layer. 

Another key approximation in our model is in the energetics; specifically, the treatment of gravitational potential energy in the solid and the release of latent heat for driving the dynamo. Since our snow layer is a pure solid, we neglect release of gravitational potential energy in the snow zone and only consider release of gravitational potential energy in the liquid portion of the core, when calculating the buoyancy flux (Equation \ref{eq:Qb}). Neglecting the gravitational potential energy release in the snow region is reasonable, because gravitational potential energy release in the liquid dominates \citep{davies_iron_2018}. Energetically, this is the same as the growth of a solid inner core, but adapted for inward core solidification \citep{gubbins_can_2003,nimmo_energetics_2007}. Our model includes the release of latent heat due to core solidification, but not absorption of latent heat by remelting of falling snow. In an iron snow system, these two latent heat contributions almost balance \citep{davies_iron_2018}, so could be neglected altogether. However, the adiabatic gradient in planetesimal cores may not be steep enough for iron snow to remelt \citep{dodds_direction_2024}. Therefore, we have omitted latent heat absorption during remelting but included latent heat release during solidification. This enables a single parameter, \mfracp, to capture variation between core solidification endmembers. The contribution of latent heat release during solidification is an order of magnitude smaller than the other contributions to core energetics during solidification (Figure \ref{fig:flux}), so including it has a minimal effect on core solidification.

\section{Outlook}\label{outlook}
Our model includes several physical and chemical features that have been neglected or simplified in previous models. For instance, our planetesimal thermal evolution model is the first to model sub-eutectic core solidification of a mantled planetesimal that also includes mantle convection. Calculating a unified buoyancy flux also enables us to use the same scaling laws for \Rem and magnetic field strength throughout the thermal evolution, and allows thermal and compositional convection to occur simultaneously. Additionally, radiogenic heating from \fe has been added to the core thermal structure. We also chose to adjust the mantle convection parametrisation to model the gradual cessation of convection and boundary layer thicknesses applicable to systems with surface cooling and internal heating. We have implemented more physically realistic values in the mantle viscosity function and added extra pieces to the viscosity function to capture the full range of viscosity behaviour as a function of temperature. The stagnant lid scaling laws have also been implemented, such that they respond to changes in mantle viscosity. This will enable future studies to choose viscosity parameters that are most applicable for a planetesimal of a given size and composition. The other parameters in the model are fully adjustable, which enables future studies of the effects of a given parameter as well as application to specific meteorite parent bodies to enable interpretation of the meteorite paleomagnetic record. In a separate study, we systematically vary mantle viscosity, primordial \feratiop, initial core sulfur content and planetesimal size to gain a deeper understanding of these parameters on dynamo generation and the implications of our model for the meteorite paleomagnetic record \citep{sanderson_early_2024}.

There are two key refinements that could be made to the model. First, the mechanism of planetesimal differentiation and initial core sulfur contents must be reconciled. Whether planetesimal cores can either form with initially low sulfur contents or lose their sulfur later, and the consequences of these factors for the onset of core solidification and dynamo generation, require further study. Second, possible mechanisms of inward core solidification in mantled planetesimals, and the implications of this on dynamo generation, should be explored in more detail. Additional improvements to the model could be made by including partial differentiation and crust formation.

\section{Conclusions}\label{conc}
Planetesimals formed during the first few Ma after Solar System formation and many were accreted into the terrestrial planets. Thermal evolution and dynamo generation models can provide insight into the interior structures and thermal histories of these planetesimals. However, previous models have focused on either early or late epochs of dynamo generation, limiting their ability to predict a complete magnetic field history.  
Our model focuses on the description of mantle convection and the parametrisation for sub-eutectic core solidification to provide a unified, more versatile model for planetesimal thermal evolution and dynamo generation.
Specific enhancements include: 
\begin{itemize} 
\item Radiogenic heating from \fe in the core. 
\item A mantle viscosity model that is self-consistent with the mantle convection parametrisation and can be adjusted to investigate the effect of mantle viscosity. 
\item Stagnant lid and CMB boundary layer parametrisations consistent with boundary heat fluxes and internal heating. 
\item Cessation of convection when the combined stagnant lid and CMB boundary layer thickness is greater than the mantle thickness. 
\item A parametrised model for sub-eutectic core solidification with a unified thermal and compositional buoyancy flux to drive magnetic field generation. 
\item Calculation of magnetic field strength during epochs of dynamo generation.
\end{itemize}

The key results of these changes implemented in our model are demonstrated by an example of the magnetic history generated for a 500\,km radius planetesimal. Compared to previous models we find that:
\begin{itemize}
\item Core thermal stratification is eroded more rapidly.
\item Mantle convection and the first epoch of dynamo generation lasts longer. 
\item Core solidification marginally increases dipole field strength, but exerts a much stronger control on dynamo duration by providing an additional buoyancy source after the cessation of mantle convection. 
\item The second epoch of dynamo generation is not triggered by core solidification. 
\end{itemize}

Our model can predict a complete magnetic field generation history for a planetesimal with magnetic field strengths for both compositional and thermal dynamos. It therefore has the potential to serve as a powerful tool for understanding the general controls on planetesimal dynamo generation, as well as recovering constraints on the properties of meteorite parent bodies from their paleomagnetic records. 

\printcredits 

\section*{Declaration of competing interest}
The authors declare that they have no known competing financial interests or personal relationships that could have appeared to influence the work reported in this paper.

\section*{Data Availability}
The code for this model is available on \citet{sanderson_refined_2024} Github repository, as are the parameter files required for the example runs in this paper.

\section*{Acknowledgements}
The authors would like to thank the reviewers for their insightful comments, and Kathryn Dodds, Ana-Catalina Plesa, Tina Rückriemen, Adina Pusok, Richard Katz and Jerome Neufeld for their helpful discussions in preparing this work. HS acknowledges funding on a NERC studentship NE/S007474/1 and an Exonian Graduate Scholarship from Exeter College, University of Oxford. JFJB acknowledges funding from the UKRI Research Frontier Guarantee program EP/Y014375/1. CJD acknowledges funding from NERC grant NE/V010867/1
For the purpose of open access, the authors have applied a Creative Commons Attribution (CC BY) licence to any Author Accepted Manuscript version arising.

\appendix

\section{Derivation of core buoyancy flux equation}\label{app:buoy}
From equations A8 and A14 in \citet{buffett_thermal_1996}, an approximation for the dissipation in the core is \begin{equation}
    \Phi = \frac{\alpha_\mm{c} Q_{\mm{CMB}}^*}{c_{p,c}}[\psi(r_\mm{c})-\bar{\psi}]+4\pi r_i^2\frac{\mm{d}r_i}{\mm{d}t}\left(\Delta \rho + \frac{\alpha_\mm{c} \rho_\mm{c} L_\mm{c}}{c_{p,c}}\right)[\bar{\psi}-\psi(r_i)],
\label{eq:buffett}\end{equation} 
where $Q_{\mm{CMB}}^*=Q_{\mm{CMB}}-Q_{ad}$ is the difference between the total CMB heat flux and adiabatic flux; $\psi$ is the gravitational potential evaluated at the CMB or solidification front, $r=r_i$; and $\bar{\psi}$ is the mass-averaged gravitational potential in the liquid portion of the core. During core solidification, the region above the solidification front is solid and does not contribute entropy to drive the dynamo. Therefore, we calculate the first term in Equation \ref{eq:buffett} at the top of the liquid inner core (dynamo region), $r=r_i$. Before core solidification, $r_i=r_\mm{c}$.  This simplifies the change in gravitational potential and the dissipation in the core becomes
\begin{equation}
    \Phi = (F_\mm{T} + F_\mm{c}) \Delta \psi =  \left[\frac{\alpha_\mm{c} Q_{r_i}^*}{c_{p,c}}-4\pi r_i^2\frac{\mm{d}r_i}{\mm{d}t}\left(\Delta \rho + \frac{\alpha_\mm{c} \rho_\mm{c} L_\mm{c}}{c_{p,c}}\right)\right](\psi(r_i)-\bar{\psi}),
\label{eq:bicb}\end{equation} where $Q_{r_i}^*=Q_{r_i}-Q_{ad}$ is the superadiabatic heat flux at the boundary between the solid outer and liquid inner core and $F_\mm{T}$ and $F_\mm{c}$ are the thermal and compositional buoyancy fluxes, respectively \citep{ruckriemen_fe_2015}. The combined buoyancy flux, $Q_b = F_\mm{T} + F_\mm{c}$ is inside Equation \ref{eq:bicb}.
Using $Q_{r_i}^*=4\pi r_i^2(F_{r_i}-F_{ad})=(-k_\mm{c}\frac{\mm{d}T}{\mm{d}r}\big|_{r_i}-\frac{\alpha_\mm{c} g(r_i) k_\mm{c}}{c_{p,c}}T(r_i))$ gives
\begin{equation}
    Q_b = 4\pi r_i^2\left[\frac{\alpha_\mm{c}k_\mm{c}}{c_{p,c}}\left(-\frac{\mm{d}T}{\mm{d}r}\bigg|_{r_i}-\frac{\alpha_\mm{c}g(r_i)}{c_{p,c}}T(r_i)\right)-\left(\frac{\alpha_\mm{c} \rho_\mm{c} L_\mm{c}}{c_{p,c}}+\Delta \rho \right)\frac{\mm{d}r_i}{\mm{d}t}\right].
\label{eq:qb2}\end{equation}

Equation \ref{eq:buffett} neglects compositional diffusion and assumes compositional fluxes are mixed evenly into the liquid portion of the core rather than assuming the combination of compositional fluxes and entropy are evenly mixed \citep{lister_expressions_2003}. This well-mixed approximation means the latent heat released at the upper boundary of the dynamo generation region during core solidification can contribute entropy to drive the dynamo. This introduces the same amount of error as the Boussinesq approximation \citep{lister_expressions_2003}, so for our model this approximation is acceptable.

For a constant density core and a solidified outer layer only (\mfracp=0), the flux-based Rayleigh number approximation to the convective power used by our model is equivalent to calculating the convective power directly. The convective power, $p$, is related to the Ohmic dissipation, $\Phi$, by 
\begin{equation}
    p = \frac{\Phi}{\rho_\mm{c}\Omega^3l^2V},
\label{eq:p}\end{equation}
where $V$ is the volume of the convecting region \citep{aubert_modelling_2009}. For a constant density core $\psi(r_i) = \frac{r_ig_i}{2}$ and $\bar{\psi} = \frac{3r_ig_i}{10}$ \citep[denoting the outer convecting boundary by $r_i$ rather than $r_o$ and setting the radius of the inner convecting boundary to 0 in Equation 16 in][]{aubert_modelling_2009}. Combining the expressions for $\psi(r_i)$ and $\bar{\psi}$ with Equation \ref{eq:bicb},  \begin{equation}
    \Phi = \frac{Q_br_ig_i}{5}.
\label{eq:phi}\end{equation}
Substituting Equation \ref{eq:phi}, the volume of the convecting core and $r_i=l$ into Equation \ref{eq:p} gives
\begin{equation}\begin{split}
    p & = \frac{g_ir_iQ_b}{5\rho\Omega^3r_i^2V}\\
    & = \frac{3}{5}\frac{g_iQ_b}{4\pi\rho\Omega^3l^4}=\gamma Ra_Q,
\end{split}
\end{equation}
which is equivalent to Equation \ref{eq:Qb}.
\bibliographystyle{cas-model2-names}
\bibliography{Paper1}
\end{document}


\maketitle

\tableofcontents
\listoffigures
\listoftables

\section{Mantle boundary layer parametrisations}
\begin{figure}
    \centering
    \includegraphics[width=0.5\textwidth]{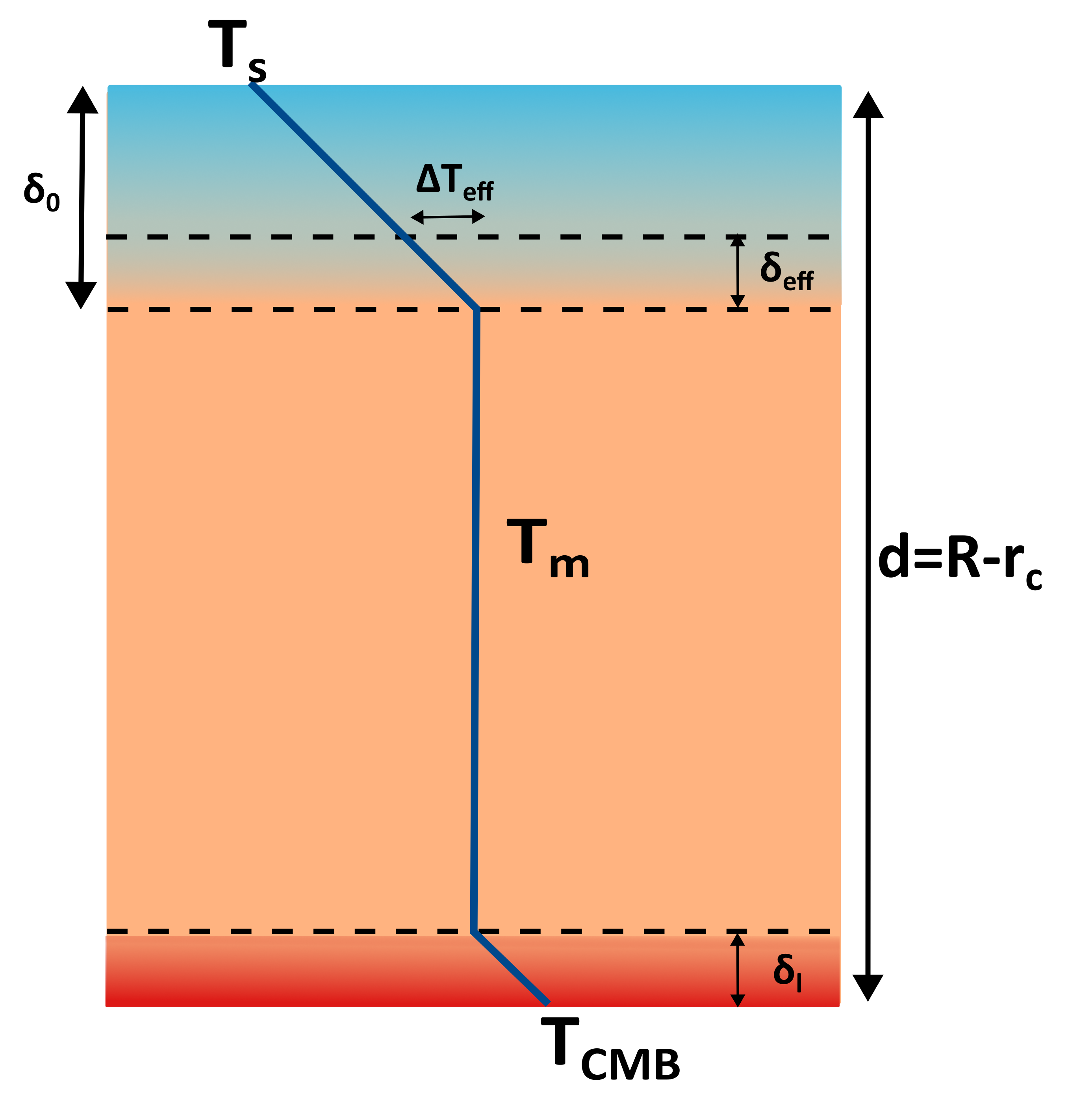}
    \caption[Stagnant lid set up]{Schematic of a stagnant lid convecting system. The blue line indicates the temperature profile in the interior and temperature increases to the right. $\delta_0$ and $\delta_l$ are the stagnant lid and CMB boundary layer thicknesses, respectively. $\delta_{eff}$ is the mobile portion of the stagnant lid.}
    \label{fig:supp-d0}
\end{figure}
The boundary layer thicknesses in our model were chosen to ensure they are self consistent with the mode of mantle heating (conduction vs convection), the chosen rheology, and adjust as the constants in the viscosity law change. In order to fulfil these requirements, the scaling laws we use have been derived from 2D and 3D simulations \citep{thiriet_scaling_2019,deschamps_scaling_2021}.
\subsection{Theoretical background}
A stagnant lid system occurs where there is a viscosity contrast $>10^4$ between the surface and the convecting interior \citep{solomatov_scaling_1995,deschamps_scaling_2021}. In this system, there is a fully convective isothermal interior with temperature $T_m$, surface temperature $T_s$, and an approximately linear temperature gradient across a stagnant lid, $\delta_0$, with temperature difference $\Delta T = T_m-T_S$ (Figure \ref{fig:supp-d0}). The lid is so thick and the viscosity contrast is so large that only the layer at the base of the lid, with thickness $\delta_{\rm eff}$, is mobile with an effective temperature difference across this layer of $\Delta T_{\rm eff}$ (assuming a linear temperature profile through the lid). The lid is assumed to become too viscous to participate in convection after a factor of $e$ increase in viscosity, which sets the ratios of lengthscales of the mobile and stagnant layers \citep{solomatov_scaling_1995}.
\begin{equation}
    \frac{\delta_{\rm eff}}{\delta_0}=\frac{1}{f_{rh}}=-\frac{\eta(T_m)}{(d\eta/dT)(T_m)\Delta T},
\label{eq:f}\end{equation}
where $\eta$ is the viscosity. $1/f_{rh}$ parametrises the proportion of $\Delta T$ which controls convection \citep[$\Delta T_{\rm eff}=\frac{\Delta T}{f_{rh}}$][]{michaut_formation_2022}. 

Analytical derivations of stagnant lid thickness are based on the stability of the boundary layer \citep{howard_convection_1966}. They calculate $\delta_{\rm eff}$ for which the mobile layer's Rayleigh number becomes supercritical and the mobile layer becomes convectively unstable. This is then related to $\delta_0$ using Equation \ref{eq:f}. The thickness of the overall stagnant lid is directly related to the change in viscosity with temperature and choice of constants in the viscosity law (Equation 1 in the main text). The stagnant lid parametrisation must be consistent with choice of viscosity law and adjust to changes in viscosity parameters. Additionally, scaling laws based on this analysis must be appropriate for the relevant heating modes and boundary conditions.

\subsection{Stagnant lid thickness}
Stagnant lid scaling laws can be obtained analytically using the approach outlined above \citep[e.g.][]{solomatov_scaling_1995} or empirically from 2D or 3D numerical models \citep[e.g.][]{deschamps_scaling_2021}. The chosen scaling law should match the boundary conditions (heat flux and tractions on the top and bottom boundaries), rheology (e.g. isoviscous, exponential viscosity), and presence/absence of internal heating. Parametrisations for systems with purely internal heating, or purely basal heating are well established \citep{ferrick_generalizing_2023}. However, planetesimals have internal heating (from radioactive decay), variable heat flux on the top (surface) and bottom (CMB) boundaries, and temperature dependent viscosity, which is more complicated to parametrise \citep{ferrick_generalizing_2023}. Previous planetesimal models have employed variations of the stagnant lid scaling from \citet{solomatov_scaling_1995} for a system with fixed top and bottom boundary temperatures, Newtonian viscosity, and no internal heating. Here, we have chosen to use the empirical stagnant lid scaling from \citet{deschamps_scaling_2021} because it has been derived for mixed heating (basal and surface heat fluxes and internal heating) and a temperature dependent viscosity, and is therefore closer to the mantle convection conditions in planetesimals. The functional form from \citet{deschamps_scaling_2021}, which is expressed as $\delta_0 \propto df_{rh}^{1.21}Ra^{-0.27}$ is similar to the analytical boundary layer form $\delta_0 \propto df_{rh}^{\frac{4}{3}}Ra^{-\frac{1}{3}}$, but with slight adjustments to the exponents and derivation of constants of proportionality from numerical simulations. 

Recently, \citet{ferrick_generalizing_2023} have developed an analytical scaling law for mixed heating systems with constants determined from numerical simulations. Their new, more complicated scaling laws agree well with numerical simulations and previous scaling laws based on boundary layer stability analysis. Since they conclude previous boundary layer stability scaling laws for convection with mixed heating are valid, we have chosen to utilise the approach of \citet{deschamps_scaling_2021} because it is simpler to implement.

\subsubsection{Location of the mobile boundary layer}
The formalism described above is one of two methods for describing stagnant lid growth. The first, adopted in our model, describes the growth of the entire lid using a scaling law and the mobile boundary layer at the base of the lid is included in the lid thickness. Models which use this approach include \citet{solomatov_scaling_1995,nimmo_influence_2000,deschamps_scaling_2021} and \citet{dodds_thermal_2021}. The second approach is to use a scaling law to determine the mobile boundary layer thickness and energy conservation at the base of the lithosphere to determine overall lid thickness \citep[e.g.][]{spohn_mantle_1991,grott_evolution_2008,michaut_formation_2022}. The second approach allows for processes in the lid, such as crust production to be resolved \citep[e.g.][]{morschhauser_crustal_2011}. However, crustal processes are below the resolution of our model so for simplicity we adopt the first formalism. In both formalisms, the heat flux out of the convecting region is determined by the temperature gradient at the boundary between the convecting interior and conductive boundary layer (full lid or mobile boundary layer). Therefore, for stagnant lids of similar thicknesses, the two formalisms shoud give similar results. Differences in overall lid thickness at a given time for the same initial conditions for the two formalisms has not been studied but could be explored in future work.

\subsection{Nusselt number and cessation criterion}
An additional justification for our convection cessation criterion comes from considering the Nusselt number, $Nu$, which is the ratio of convective to conductive heat flux out of a convecting system. These heat fluxes can be approximated using the temperature difference across the mantle and the lengthscale for the temperature change in the two scenarios to give the following expression \begin{equation}
    Nu = \frac{F_{conv}}{F_{cond}}\approx \frac{\frac{k_m\Delta T}{\delta_l+\delta_0}}{\frac{k_m\Delta T}{R-r_c}}=\frac{R-r_c}{\delta_l+\delta_0}.
\label{eq:nusselt}\end{equation} When the combined CMB boundary layer thickness and stagnant lid thickness, $\delta_l+\delta_0$, is equal to the mantle thickness, $R-r_c$ then $Nu=1$. The Nusselt number can also be written in terms of the Rayleigh number $Nu \sim \left(\frac{Ra}{Ra_c}\right)^n$ where n is a constant. Therefore, $Nu=1$ implies $Ra=Ra_c$ and the system is marginally stable. Using boundary layer thicknesses gives a different time for cessation of convection than the criterion $Ra<Ra_c$ because Equation \ref{eq:nusselt} is an approximation.

\begin{figure}
    \centering
    \includegraphics[width=1\textwidth]{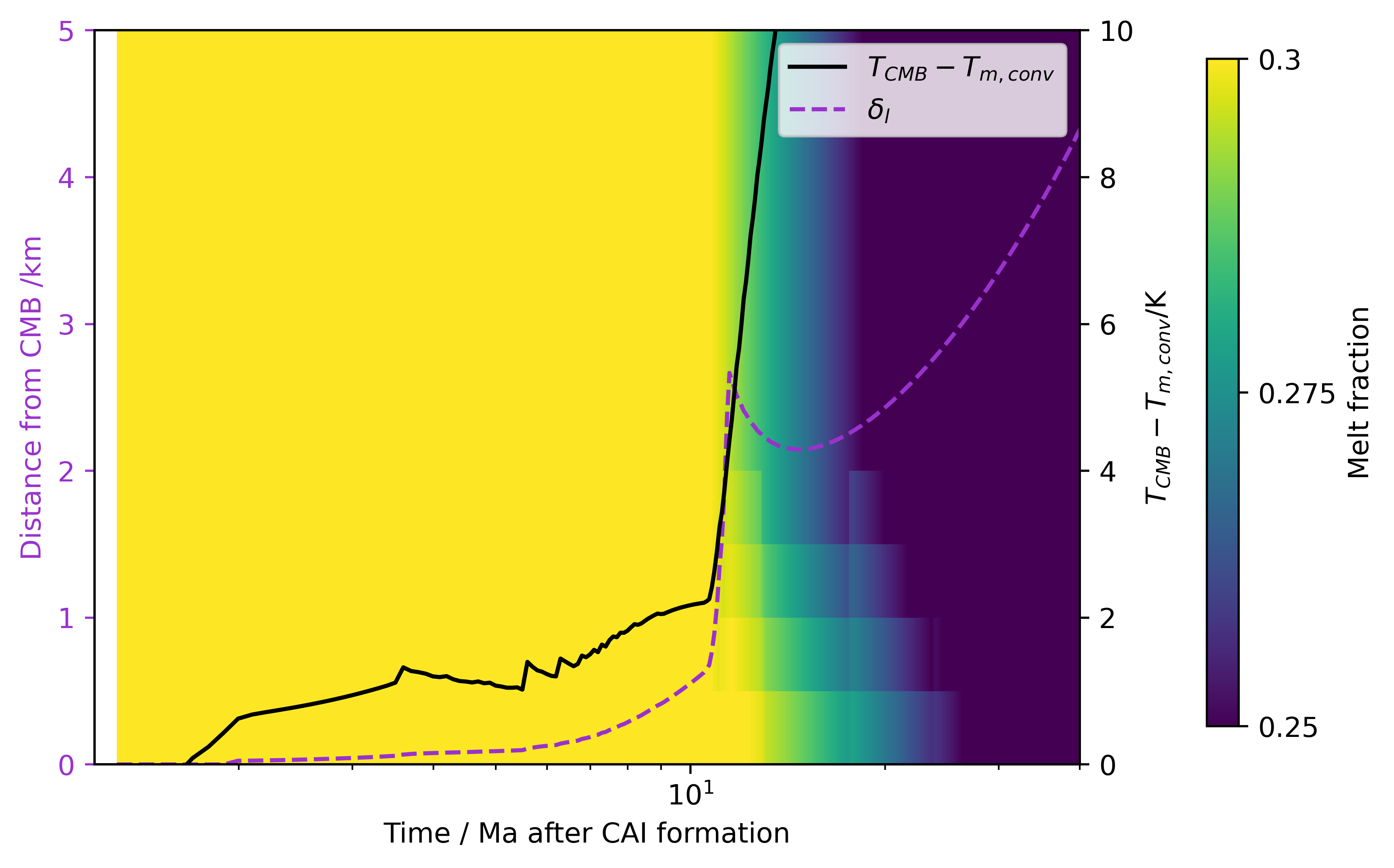}
    \caption[CMB boundary layer thickness below \rcmf]{CMB boundary layer thickness, $\delta_l$, and temperature difference between convective mantle and CMB for the first 40\,Ma of the planetesimal thermal evolution. Results are for a 500\,km planetesimal, accreted 0.8\,Ma after CAI formation with \feratio=$10^{-8}$ (same example run as Figure 6 in the main text). The background colourmap indicates the melt fraction 0--5\,km above the CMB (left hand axis). All melt fractions at or above the rheologically critical melt fraction, \rcmf=0.3, are shown in yellow. There is a jump in CMB boundary layer thickness when the melt fraction first drops below the critical melt fraction and the viscosity increases rapidly. This increases the temperature difference between the convecting mantle and CMB.}
    \label{fig:deltal}
\end{figure}
\begin{figure}
    \centering
    \includegraphics[width=1\textwidth]{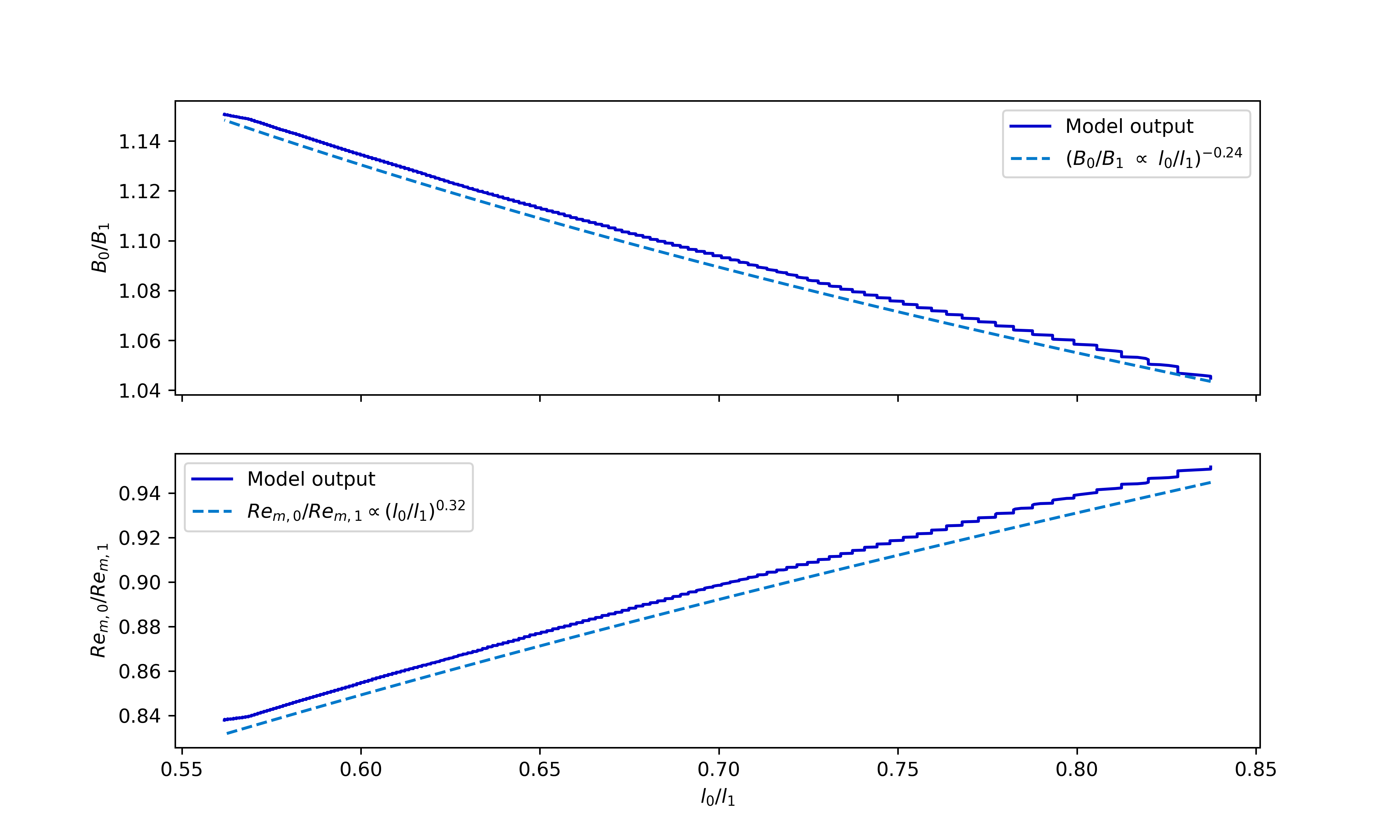}
    \caption[Ratios of magnetic field strength and \Rem for the two endmember models]{Ratios of magnetic field strength and \Rem for the two endmember models as a function of the ratio of convective lengthscales. Subscripts 0 and 1 indicate the values for the $m_{frac}=0$ and $m_{frac}=1$ endmembers, respectively. The model output (solid lines) follows the scaling predicted by combining the Equations 32, 33 and 35 for buoyancy flux, convective velocity and magnetic Reynolds number in the main text (dashed lines). Staircasing in the output is due to the discrete nature of our model. The runs shown here are for the same parameters as the example run in the main text. Magnetic field strength and \Rem depend weakly on convective lengthscale. $B\propto l^{-0.24}$ and $Re_m \propto l^{0.32}$.}
    \label{fig:Bproptol}
\end{figure}
\begin{figure}
    \centering
    \includegraphics[width=1\textwidth]{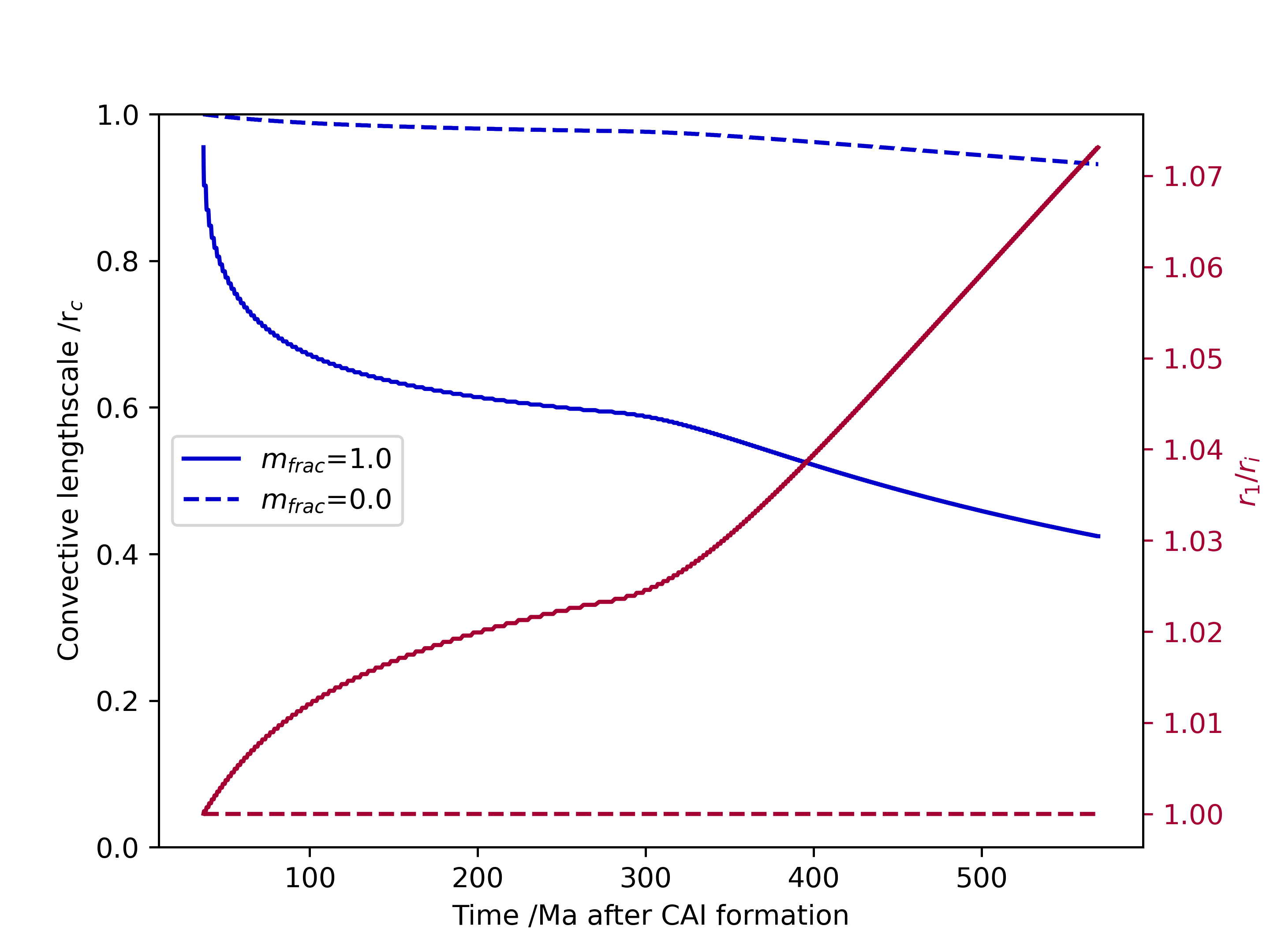}
    \caption[Convective lengthscales as a function of time]{Convective lengthscale as a function of time for the two endmember solidifcation geometries (dark blue lines). Convective lengthscale decreases more rapidly for $m_{frac}=1$. The pink lines indicate the ratio of the radius of the base of the solid outer core, $r_1$, to the solidification front, $r_i$. For $m_{frac}=1$, the maximum deviation between these positions is less than 6\%. The runs shown here are for the same parameters as the example run in the main text, but for \Xs=26.7\,wt\% in order to investigate the maximum possible change in lengthscale as lower \Xs take longer to reach the eutectic composition.}
    \label{fig:conv-l}
\end{figure}

\begin{figure}
    \centering
    \includegraphics[width=0.8\textwidth]{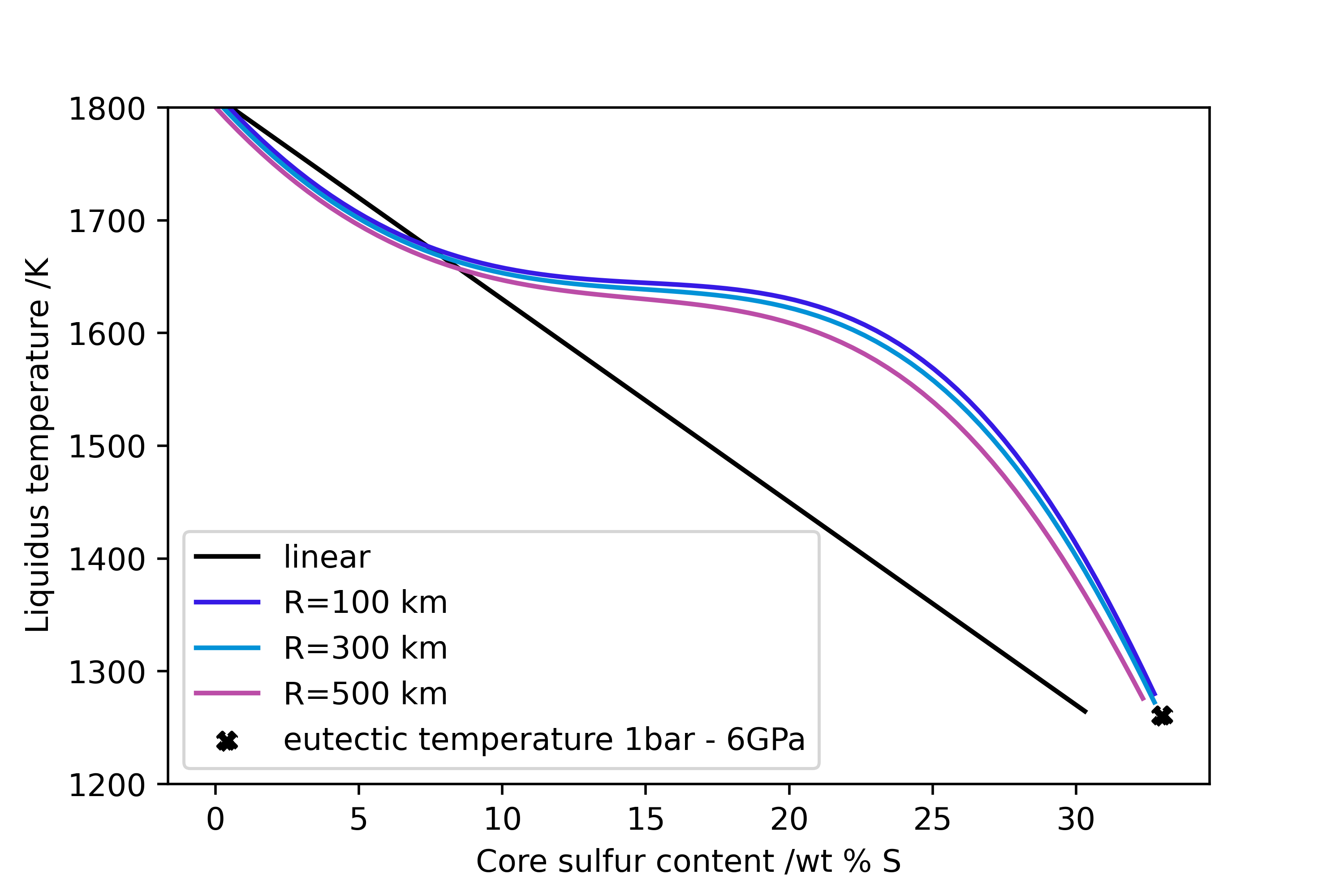}
    \caption[\citet{buono_fe-rich_2011} liquidus]{Comparison between linear liquidus approximation and liquidus from \citet{buono_fe-rich_2011}. This liquidus is a function of pressure so liquidi for the range of planetesimal radii appropriate for this model are shown. The liquidus temperatures for core sulfur contents $>10$\,wt\% are higher than temperatures for the linear liquidus approximation. The liquidus gradient is shallow from 10--25\,wt\% at just above 1600\,K. This limits the range of planetesimal initial core sulfur contents to high values $>25$\,wt\% (for a 1400\,K mantle solidus), because the maximum possible planetesimal temperatures in our model are $\sim$1600\,K for a rheologically critical melt fraction of 0.5.}
    \label{fig:liquidus}
\end{figure}

\begin{figure}
    \centering
    \includegraphics[width=0.7\textwidth]{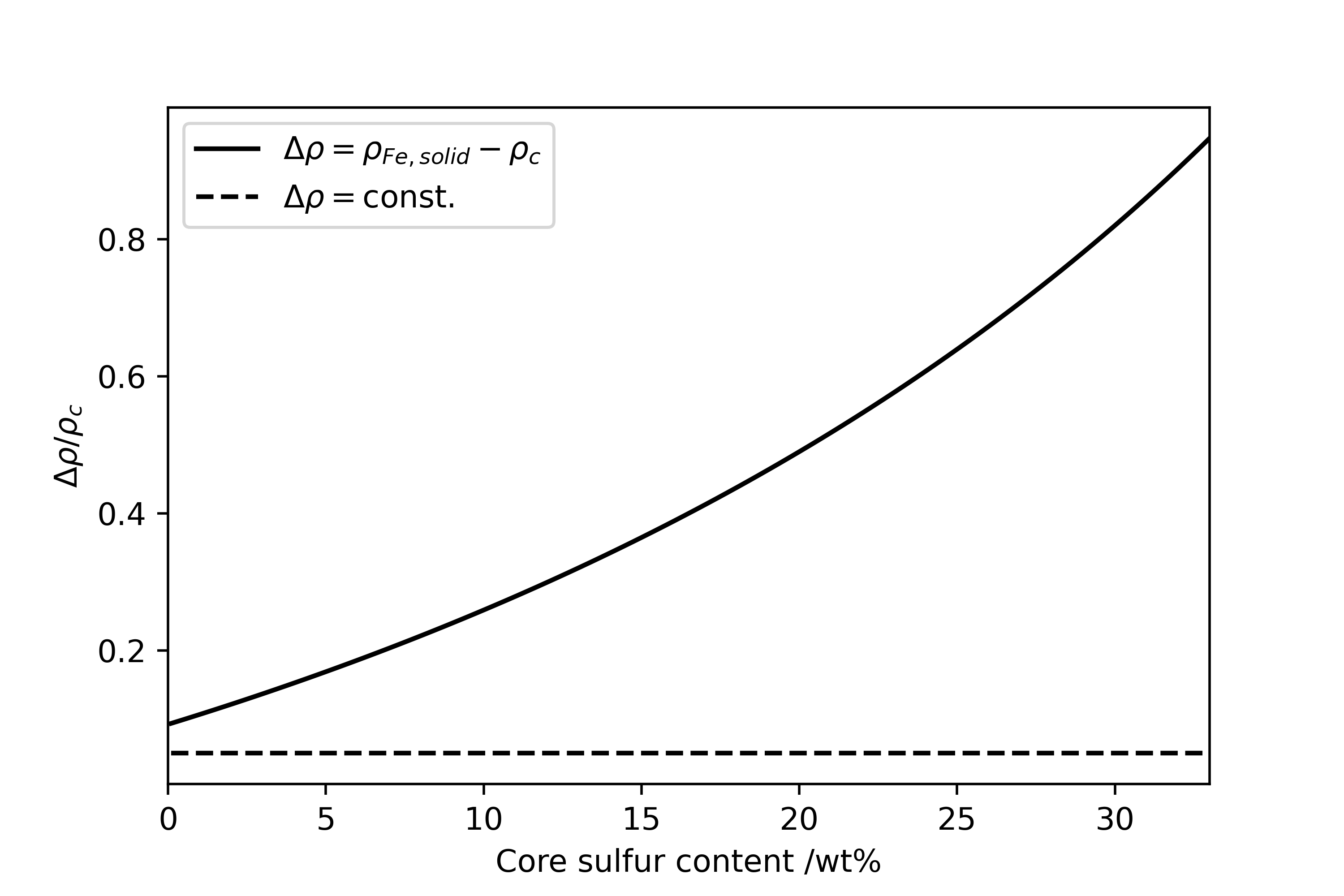}
    \caption[Density difference driving buoyancy flux in the compositional dynamo]{Density difference driving buoyancy flux in the compositional dynamo. The density difference between solid iron and a liquid inner core increases with increasing sulfur content. The density difference is higher than the previously assumed value of 0.05$\rho_c$ for all core sulfur contents.}
    \label{fig:density-difference}
\end{figure}
\begin{figure}
    \centering
    \includegraphics[width=1\textwidth]{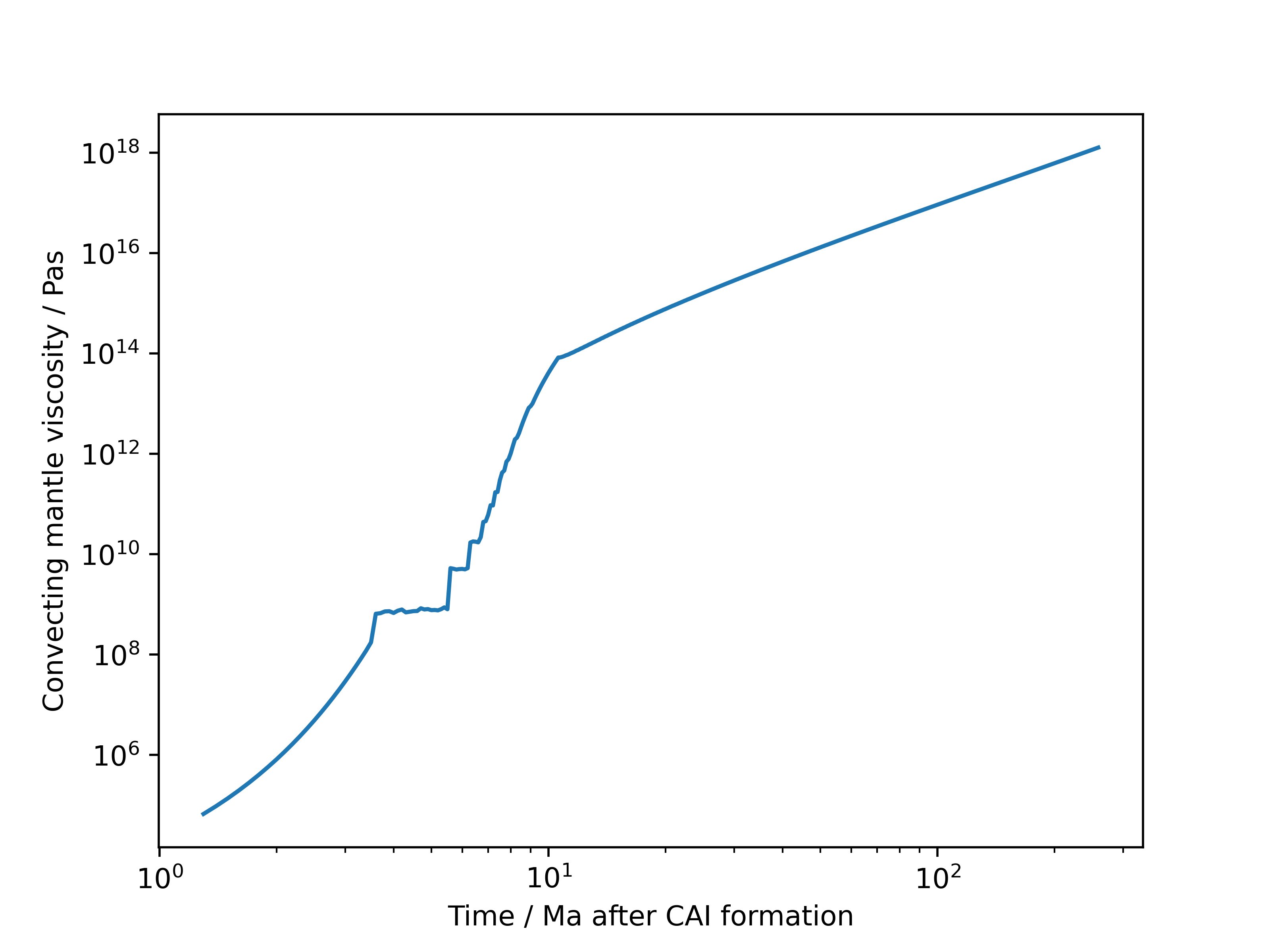}
    \caption[Mantle viscosity as a function of time for a 500\,km radius planetesimal]{Viscosity in the isothermal, convecting mantle interior as a function of time for a 500\,km planetesimal accreted at 0.8\,Ma after CAI formation with \feratio=$10^{-8}$, \Xs=29.85\,wt\%, $\eta_0=10^{19}$\,Pas, $\phi_C=0.3$, $\beta=0.0225\rm K^{-1}$, $\alpha_n=30$, and $\eta_l=10$\,Pas.}
    \label{fig:eta}
\end{figure}

\begin{figure}
    \centering
    \includegraphics[width=1\textwidth]{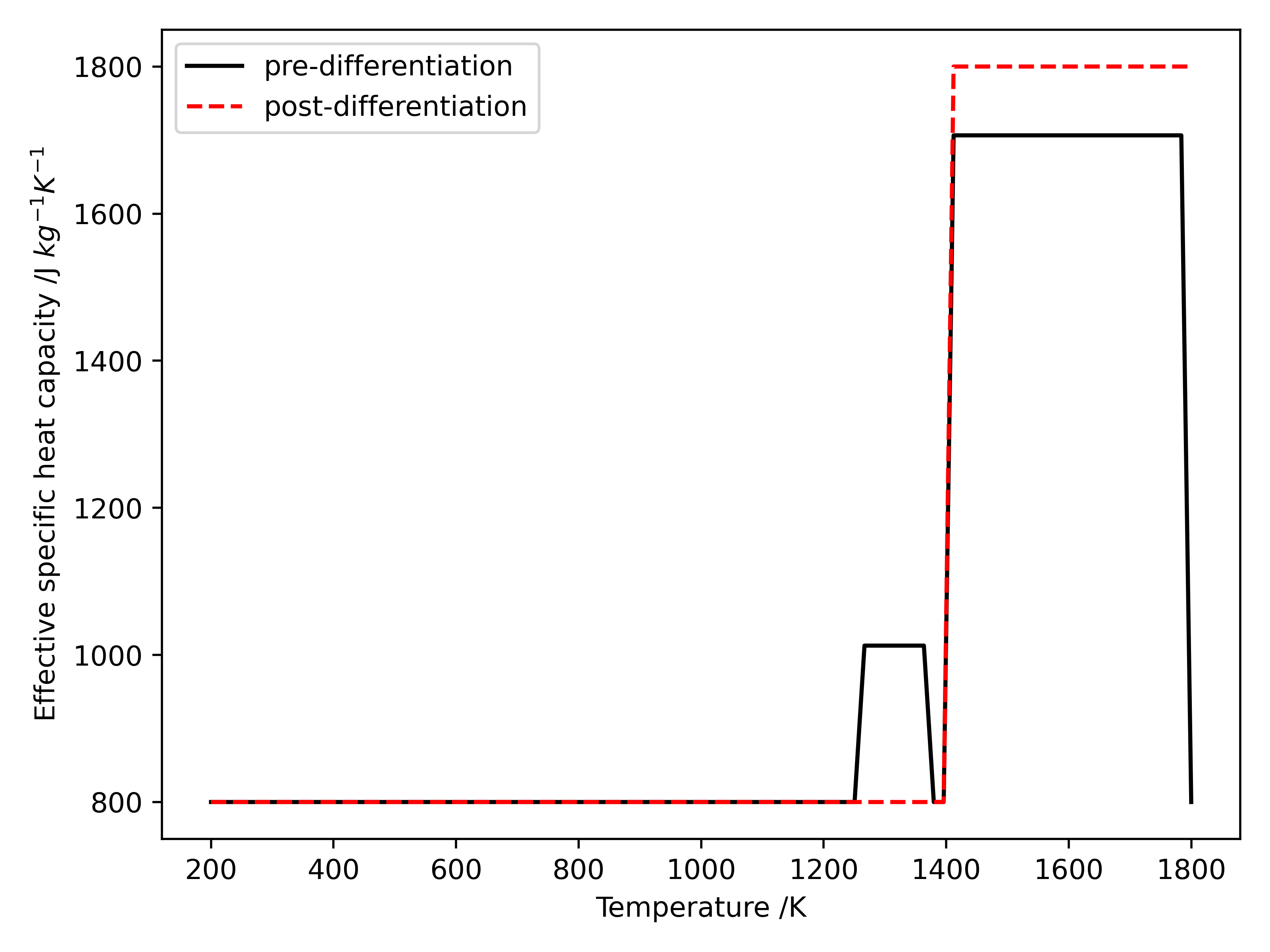}
    \caption[Modified specific heat capacity]{Chondritic (pre-differentiation) and silicate (post-differentiation) effective specific heat capacity as a function of temperature for 29.85\,wt\% sulfur in the Fe-FeS. Prior to differentiation, there are two periods of increased specific heat capacity as the Fe-FeS and silicate phases melt. After differentiation, there is one period of increased specific heat capacity when the silicate melts. The post-differentiation increase in effective specific heat capacity is higher, because the material is entirely silicate and does not contain any Fe-FeS volume fraction. }
    \label{fig:cp}
\end{figure}

\section{Increase in \al in the mantle during differentiation}
Upon differentiation, all the iron in the planetesimal is assumed to go into the core and all the aluminium is assumed to stay in the mantle. The asteroid's initial sulfur content is assumed to partition between the core and mantle to give the desired core sulfur content $X_s$, but the initial sulfur content of the body is not prescribed and the sulfur content of the mantle and undifferentiated body are not tracked. These assumptions can then be used to calculate the new iron and aluminium abundances of the core and mantle, respectively. The core is assumed to be purely iron and sulfur, so the abundance of iron in the core in wt \% is given by \begin{equation}
    X_{Fe,c} = 100 - X_s,
\end{equation} where $X_s$ is the core sulfur content in wt \%, which is specified at the beginning of the model. All aluminium is assumed to remain in the mantle, so by mass conservation the abundance of aluminium in the mantle is given by \begin{equation}
    X_{Al,m} = \frac{\rho_{ch}V}{\rho_m V_m}X_{Al,ch} = \frac{\rho_{ch}R^3}{\rho_m (R^3-r_c^3)}X_{Al,ch}
\end{equation}

\section{Core}
\subsection{GPE release during core solidification} 
During core solidification, the heat across the CMB, $Q_{CMB}$, is balanced by radiogenic heating, $Q_R$, secular cooling, $Q_S$, latent heat release, $Q_L$, and release of gravitational potential energy, $Q_G$, 
\begin{equation}
    Q_{CMB} = Q_R + Q_S + Q_L + Q_G.
\end{equation}
Substituting for each term following \citet{nimmo_energetics_2007}
\begin{equation}\begin{split}
    F_{CMB}A_{CMB} & = M_cH - M_cc_{p,}\frac{dT_c}{dt} - 4\pi r_i^2\rho_l L\frac{dr_i}{dt}+\frac{8}{3}\pi^2G\rho_c\Delta\rho r_i^4\frac{dr_i}{dt}
    \\ & = M_cH +\left[- M_cc_{p,c} + \frac{4\pi r_i^2L_c}{g_c\frac{dT_l}{dP}}+\frac{32}{15}\pi^2G\frac{\Delta\rho r_i^4}{ g_c\frac{dT_l}{dP}}\right]\frac{dT_c}{dt},
\end{split}\end{equation}
where $M_c$ is the mass of the core; $H$ is the radiogenic heating from \fep; $c_{p,c}$ is the core specific heat capacity; $r_i$ is the boundary between the solid outer and liquid inner core; $L_c$ is the core's latent heat; $g_c$ is the gravitational field strength at the CMB; $\rho_c$ is the core density; $\frac{dT_l}{dP}$ is the pressure derivative of the Fe-FeS liquidus; $\Delta \rho$ is the density difference between solidified iron and the liquid inner core; and $T_c$ is the core temperature. This can then be rearranged to give an expression for $\frac{dT_c}{dt}$
\begin{equation}
    \frac{dT_c}{dt}=\frac{Q_{CMB}-Q_R}{Q_{ST}+Q_{LT}+Q_{GT}},
\end{equation} where $Q_{ST}$ etc. denotes the relevant heat terms divided by $\frac{dT_c}{dT}$. Including $Q_G$ in the expression for core cooling during core solidification does not change the core solidification time within the resolution of the model. 

\begin{figure}
    \centering
    \includegraphics[width=1\textwidth]{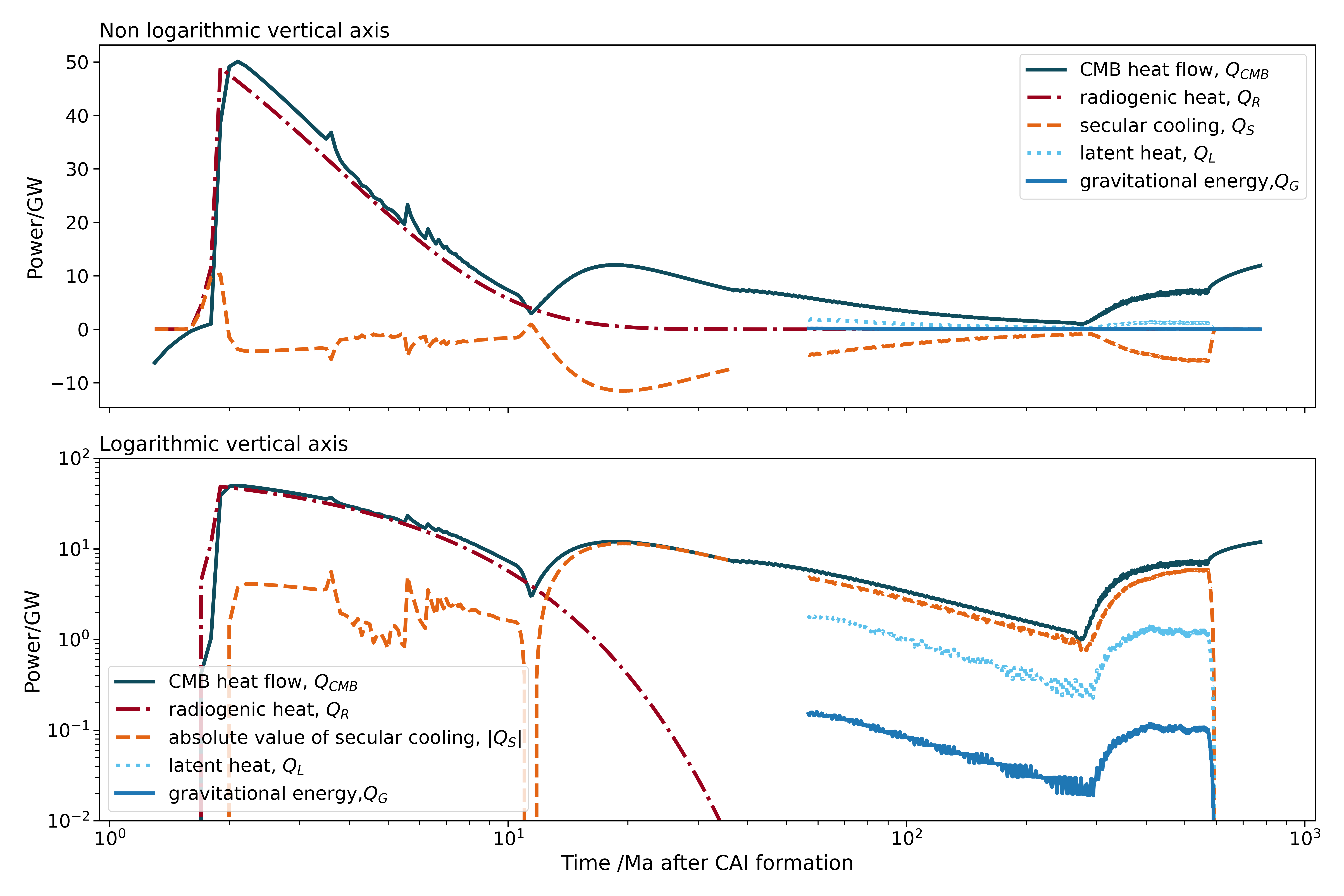}
    \caption[Gravitational energy release in the core]{Power sources in the core for a 500\,km radius planetesimal, with a 250\,km radius core, accreted at 0.8\,Ma after CAI formation with \feratio=$10^{-8}$ and \Xs=29.85\,wt\%. The lower panel shows the same sources as the upper panel, but with a logarithmic vertical axis so that the gravitational contribution, $Q_G$, can be seen. The absolute value of the secular cooling is plotted on the logarithmic axis to enable the values to be log-scaled. Due to the discrete nature of the model, $Q_G$, $Q_L$ and $Q_S$ oscillate during core solidification and are plotted as a rolling average over 200 output steps (20\,Ma). The small gap between the purely thermally driven dynamo and the average values for the thermo-compositional dynamo is due to the lag in the rolling average. The power contribution of gravitational energy release during solidification is two orders of magnitude smaller than other contributions and can be neglected in the thermal evolution of the core.}
    \label{fig:qg}
\end{figure}
\subsection{Core density}
The central pressure in a planetesimal is low enough (2\,MPa to 50\,MPa) that there is little difference in density compared to ambient pressure \citep[Fig. 10 from][]{morard_liquid_2018}, so no correction was made for the pressure difference.

\section{Magnetic field generation}
\begin{table}[h]
    \centering
    \begin{tabular}{|c|c|c|c|c|c|}
    \hline
        Dimensionless number & Symbol & Force balance & Equation & Earth value & Planetesimal value \\\hline
        Ekman number & $E$ & $F_\nu$/$F_C$ & $\frac{\nu}{\Omega l^2}$ & $10^{-15}$ & $10^{-14}$ \\ Rossby number & $Ro$ & $F_I$/$F_C$ & $\frac{U}{\Omega l}$ & $10^{-6}$ & $10^{-6}$\\ Buoyancy number & $Bu$ & $F_B$/$F_C$ & $\frac{\alpha_c \Delta T' g_c}{\Omega U}$ & $10^{-1}$ & $10^{-3}$ \\ Dynamic Elsasser number & $\Lambda_d$ & $F_L$/$F_C$ & $\frac{B^2}{\rho_c \mu_0 \Omega U l}$ & $10^{-2}$ & $10^{-2}-10^{-3}$ \\ Elsasser number & $\Lambda_t$ & $F_L$/$F_C$ & $\frac{B^2}{\rho \mu_0 \lambda \Omega}$ & $10$ & $1-10$ \\ \hline
    \end{tabular}
    \caption[Force balance in planetesimal dynamos]{Dimensionless numbers and their values for Earth \citep{schwaiger_relating_2021} and planetesimals arranged in order of size in planetesimals from smallest to largest. Planetesimal scaling assumes a rotational period of 10\,hrs, a core size of 250\,km (half the radius of a 500\,km planetesimal) and a field strength 0.1--1 times the strength of the Earth. All other quantities are assumed to be the same order of magnitude as Earth. $\nu$ is the kinematic viscosity, $U$ is the convective speed, $\Delta T'$ is a temperature perturbation and all other symbols are defined in the main text. The Elsasser number assumes the current density, $J$, is given by $J=\sigma UB$ and the dynamic Elsasser number assumes $J=\frac{\nabla\times B}{\mu_0}$ and is thought to be a better estimate of the strength of the Lorentz force \citep{aurnou_cross-over_2017,schwaiger_relating_2021}. The table indicates that inertial, $F_I$, and viscous, $F_{\nu}$, forces are negligible. The Coriolis force, $F_C$, is slightly larger than the buoyancy force, $F_B$, and the Lorentz force, $F_L$. Therefore, a MAC balance is the next most important force balance after the quasi-geostrophic (QG) force balance that dominates in planetary cores \citep{schwaiger_relating_2021}}
    \label{tab:dim-numbers}
\end{table}
\subsection{Regime 2}
\begin{figure}
    \includegraphics[width=0.5\textwidth]{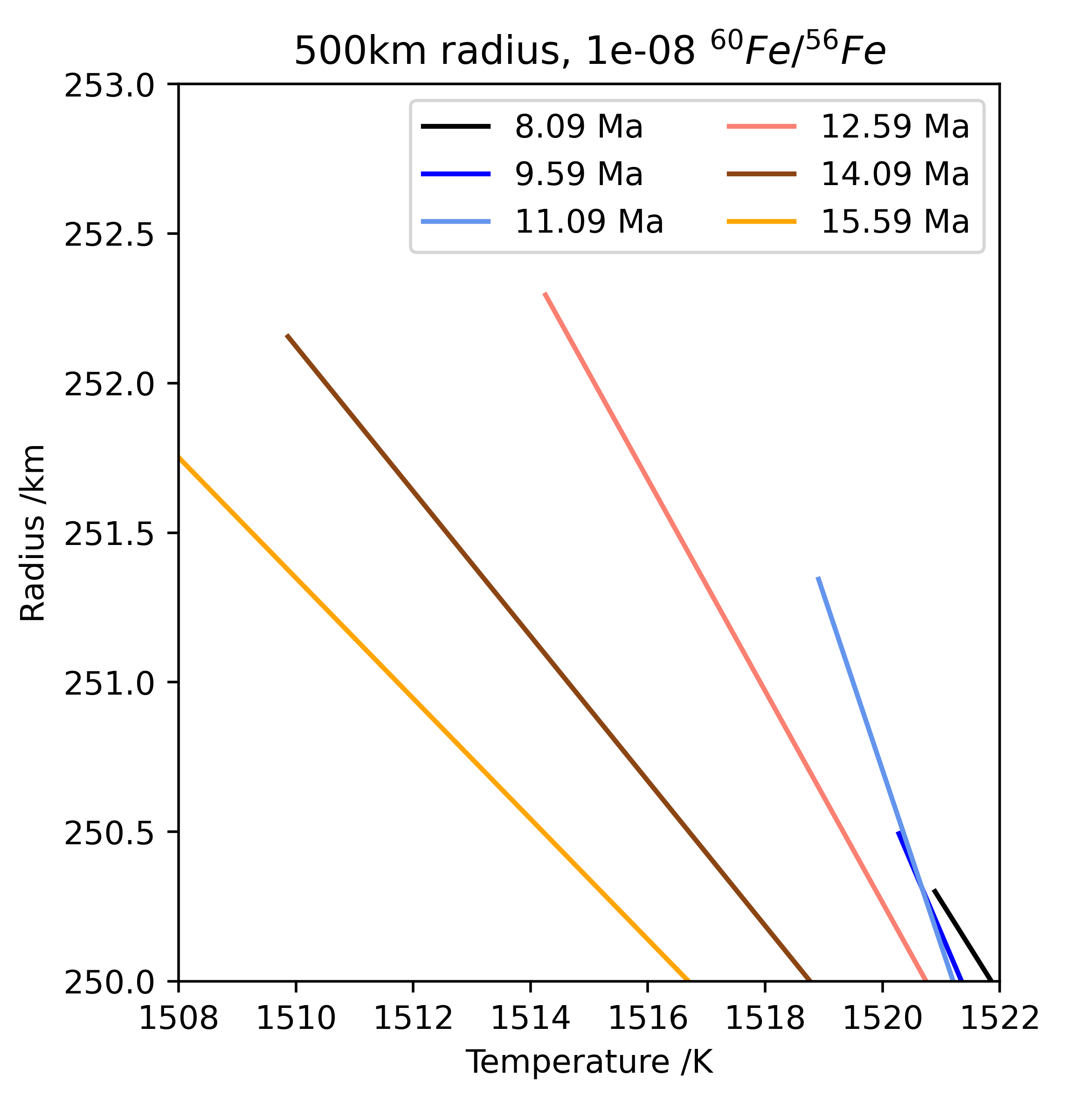}
    \caption[CMB temperature profiles when the mantle drops below the critical melt fraction (regime 2)]{Temperature profile across the CMB boundary layer, $\delta_l$ before (9.59\,Ma) and after (11.09\,Ma) the basal mantle temperature drops below the temperature of the critical melt fraction. The profiles here are for a 500\,km radius planetesimal, with a 250\,km core radius accreted at 0.8\,Ma after CAI formation with \feratio=$10^{-8}$ and \Xs=29.85\,wt\%.}
    \label{fig:regime2}
\end{figure}
When the mantle is convecting $F_{CMB}=-k_m \frac{T_m-T_{CMB}}{\delta_l}$.
When $T<T_{\phi_C}$ (i.e. $T<1520$K) there is a jump in lid thickness. This results in a shallower temperature gradient at the CMB (9.59 Ma vs 11.09 Ma, Figure \ref{fig:regime2}), because the same temperature difference is accommodated across a larger boundary layer. This decreases $F_{CMB}$.
As the mantle continues to cool, the temperature at the top of the thermal boundary layer drops. This increases the temperature difference across the boundary layer (12.59 to 14.09 Ma, Figure \ref{fig:regime2}) and increases $F_{CMB}$. 
\begin{figure}
    \centering
    \includegraphics[width=1\textwidth]{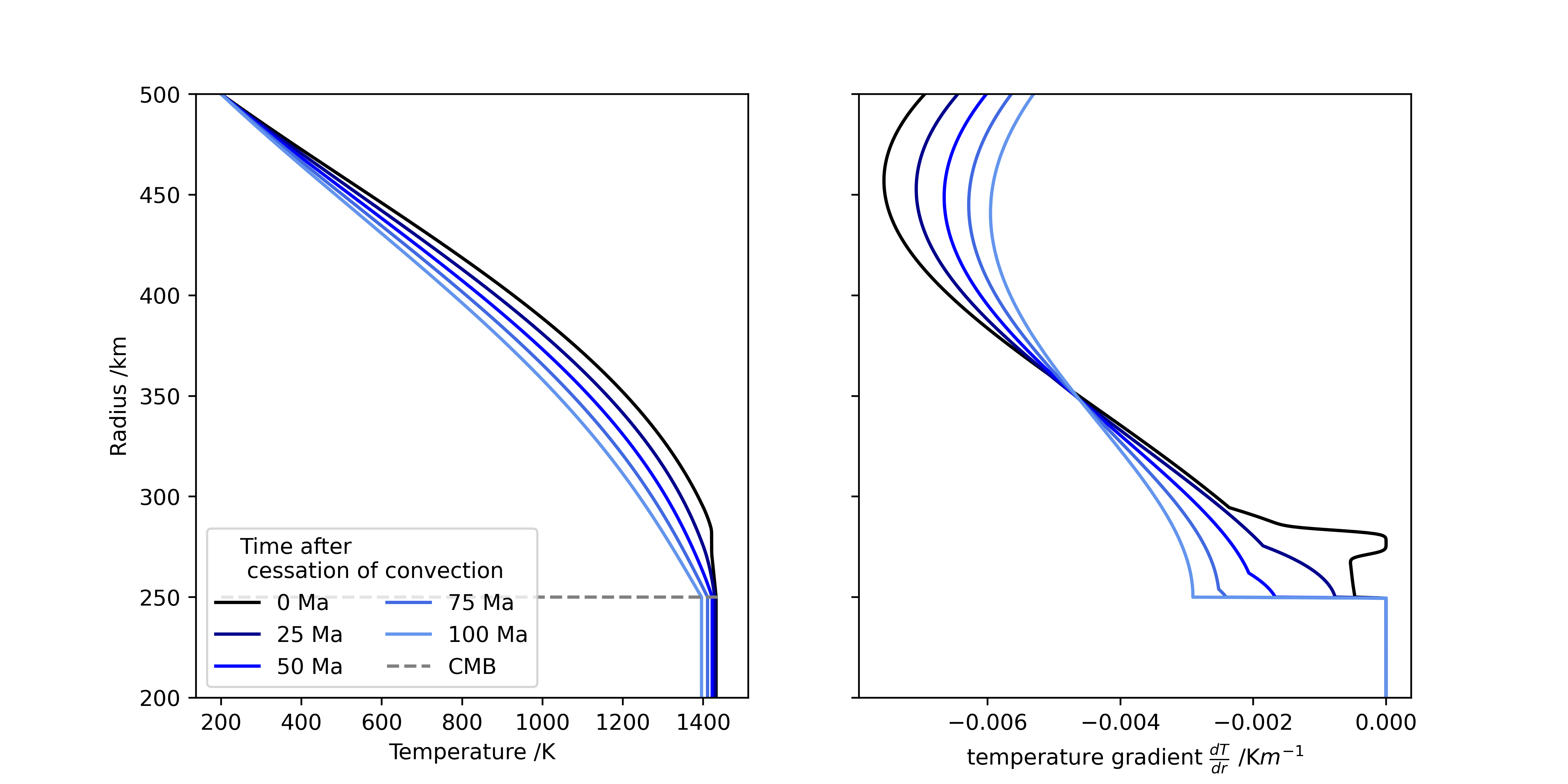}
    \caption[CMB temperature profiles after the cessation of mantle convection (regime 3)]{Planetesimal temperature profiles and temperature gradients following the cessation of convection. The profiles here are for a 500\,km radius planetesimal, with a 250\,km core radius and accreted at 0.8\,Ma after CAI formation with \feratio=$10^{-8}$ and \Xs=29.85\,wt\%. The grey, dashed line indicates the postion of the CMB. Immediately following the cessation of convection, a small portion of the mantle is isothermal and there is a shallow temperature gradient across the CMB boundary layer. As the mantle cools, a temperature gradient is established throughout the entire mantle and the temperature gradient at the CMB increases.}
    \label{fig:regime3}
\end{figure}

\subsection{Convective lengthscale}
\begin{figure}
    \centering
    \includegraphics[width=1\textwidth]{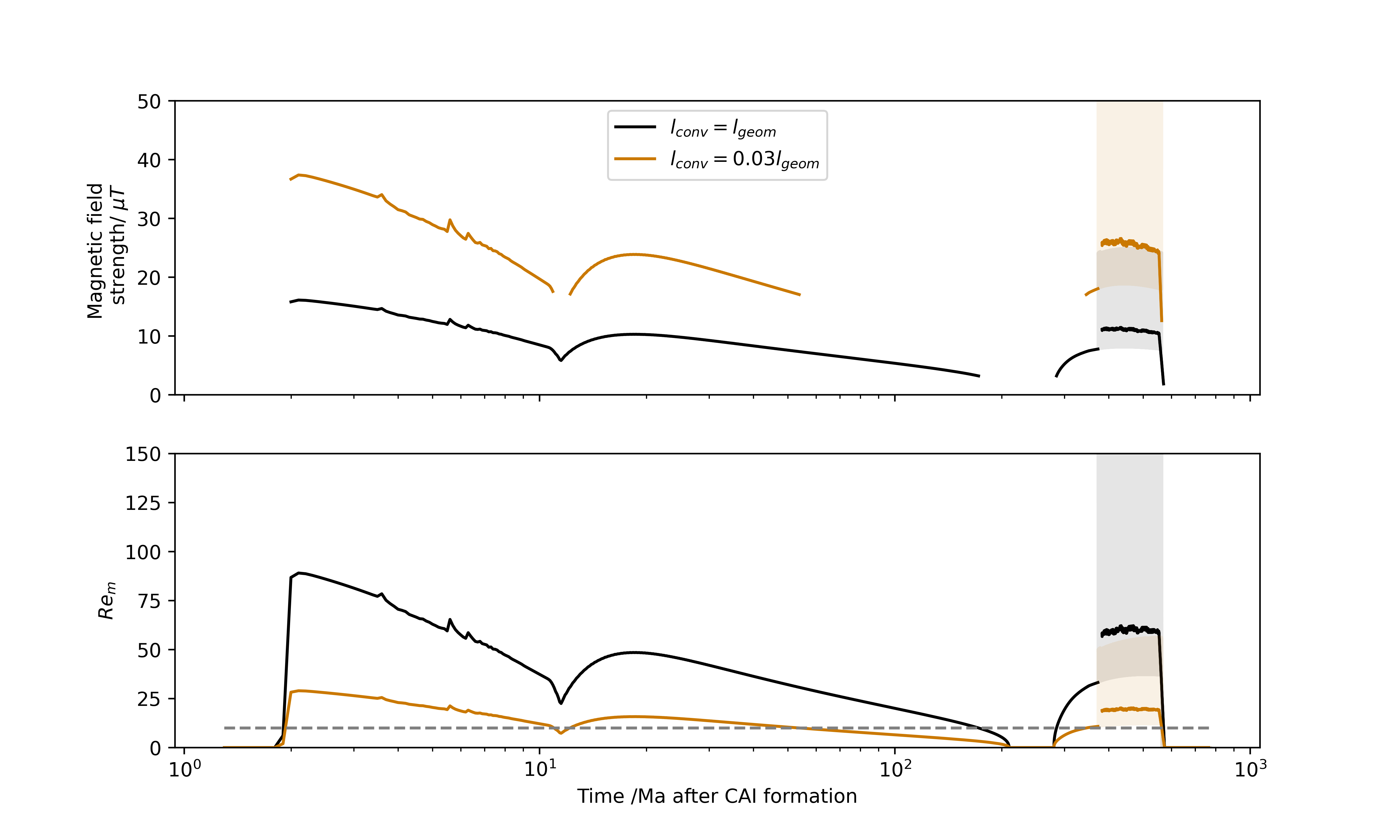}
    \caption[Effect of convective lengthscale on dynamo generation]{Magnetic field strength and \Rem with time for a 500\,km planetesimals with 250\,km radius cores, \feratio=$10^{-8}$, \Xsp=29.85\,wt\% assuming $l_{\text{conv}}=l_{\text{geom}}$ and $l_{\text{conv}}=0.03l_{\text{geom}}$. Due to the discrete nature of the model, the magnetic field strength and \Rem oscillate during core solidification. This oscillating output is shown by the faded traces and the rolling average over 200 output steps (20\,Ma) is shown in bold. The spikes in \fcmb prior to 13\,Ma are due to the discretisation of the stagnant lid. Using a smaller convective lengthscale increases the magnetic field strength, but decreases \Remp, which leads to shorter epochs of dynamo generation.}
    \label{fig:lgeom}
\end{figure}

A possible way to calculate the ratio of the convective lengthscale, $l_{\rm conv}$, to the geometric lengthscale of the convecting region in the core, $l_{\rm geom}$, is to use the Rossby number \citep{davidson_scaling_2013,aubert_spherical_2017} \begin{equation}
    \frac{l_{\text{conv}}}{l_{\text{geom}}}\sim Ro^{\frac{1}{4}}.
\end{equation} 
The Rossby number quantifies the ratio between inertial and Coriolis force in the core and is $\approx10^{-6}$ in planetesimal cores (Table \ref{tab:dim-numbers}). Therefore, $l_{\rm conv}\sim 0.03 l_{geom}$. Using this expression for $l_{\text{conv}}$, in the equations for buoyancy flux, \Rem and magnetic field strength can lead to elevated magnetic field strengths and smaller \Rem (Figure \ref{fig:lgeom}). Smaller \Rem leads to shorter epochs of dynamo generation, but the second epoch of dynamo generation still starts before the onset of core solidification.
\clearpage

\section{Timescales for differentiation}
Here we estimate the timescales for percolation and rain out via Stokes settling. Differentiation via sinking of large iron diapirs \citep{samuel_dynamics_2008} has not been included because the timescales are similar to Stokes settling \citep{neumann_differentiation_2014}. Additionally, diapirism is more pertinent to terrestrial planets, which have rheological boundaries where metal material can accumulate \citep{samuel_dynamics_2008}. Due to uncertainties in the percolation timescale and the complexity of the percolation mechanism, this work focuses on differentiation via Stokes' settling at $\phi$ = \rcmfp, which is much more rapid.

\subsection{Differentiation via percolation}
The percolation velocity from Darcy's law is given by \begin{equation}\begin{split}
    v & = \frac{K(\phi)\Delta \rho g}{\eta_c}
   \\ & = 50 \left(\frac{K(\phi)}{10^{-8}m^2}\right) \,\rm km\, yr^{-1},
\label{eq:darcy}\end{split}\end{equation} where $\Delta \rho \approx 3000\rm kg m^{-3}$ is the density difference between molten Fe-FeS and silicate material, $g\approx0.5 \,\rm ms^{-2}$ is the gravitational field at the surface of a 500\,km planetesimal, $\eta_c = 0.01\,\rm Pas$ is the viscosity of molten iron, and $K(\phi)$ is the permeability of the planetesimal matrix, for which model values vary by six orders of magnitude from
$10^{-8}\rm \, m^2$ \citep{fu_fate_2014} to $10^{-14}\rm \, m^2$ \citep{neumann_differentiation_2012}. An upper estimate for the timescale for differentiation can be obtained using  $t_{\rm diff}=\frac{R}{v}$ and the upper limit on planetesimal radius, 500\,km. This gives $t_{diff}10\left(\frac{10^{-8} m^2}{K(\phi)}\right)$ years and differentiation times of $10^{-5}$--1 Ma over our permeability range. The literature bounds on differentiation time via percolation are similarly wide from 0.001\,Ma \citep{sahijpal_numerical_2007} to 1\,Ma \citep{neumann_differentiation_2012}.

\subsection{Differentiation via rain-out}
The Stokes settling velocity is given by
\begin{equation}
    v = \frac{dr}{dt} = - \frac{2 \Delta \rho g a^2}{\eta_m}
\end{equation} where $a$ is the radius of the sinking metal droplet and $\eta_m \approx 10\,Pas$ is the viscosity of the body above \rcmfp. Integrating with respect to time gives a settling time of $10^2\left(\frac{a}{10^{-3}\rm \,m}\right)^{-2}\, \rm yr$ for the droplet to fall to 1\% of its initial radius ($r_{\text{final}}=0.01r_{\text{initial}}$). The radius of the metal droplets has been estimated as 1\,mm based on the size of metal grains in element maps of metal rich chondrites in \citet{krot_11_2014}, but coalesence during melting may be required for metal grains to reach large enough sizes to fall rapidly during rain-out \citep{neri_textural_2021}. Once \rcmf is reached and the silicate phase is rheologically weak, the molten metal will fall rapidly to the centre of the planetesimal and can be approximated as instantaneous.

\section{Numerical Implementation}
We use a forward in time, centered in space (FTCS) stencil to calculate the temperature change in the conductive portion of the planetesimal. The boundary conditions are $\frac{dT}{dr}\big|_{r=0}=0$, flux continuity at the CMB and a fixed surface temperature. For a given step the temperature change is calculated by
\begin{equation} \vec{T}^{i+1} = \vec{T}^i + \mathbf{M}\vec{T}^idt\end{equation}
where $\mathbf{M}$ is the FTCS matrix, $\vec{T}$ is a vector of temperatures with radius in the planetesimal, and $^i$ indicates the $i^{th}$ timestep. For an undifferentiated planetesimal, the whole body has the same thermal diffusivity and $\mathbf{M}$ and $\vec{T}$ are given by
\begin{align}\mathbf{M} = \kappa_{ch}\begin{bmatrix}
-1 & 1 &  & & & \\
1-\frac{dr}{r} & -2 & 1+\frac{dr}{r} &  &  & \\
 & \ddots & \ddots & \ddots &  \\
 &  &  1-\frac{dr}{r} & -2 & 1+\frac{dr}{r} & \\
 & & & & 0 \\
\end{bmatrix} &&
    \vec{T} = \begin{bmatrix}
        T_{r=0} \\
        T_{r=dr} \\
        T_{r=2dr}\\
        \vdots \\
        T_s
    \end{bmatrix}.
\end{align}
For the differentiated body, conduction in the core and mantle are calculated separately using $\mathbf{M_c}$ and $\vec{T}_{core}$ in the core
\begin{align}\mathbf{M}_{c} = \kappa_{c}\begin{bmatrix}
-1 & 1 &  & & & \\
1-\frac{dr}{r} & -2 & 1+\frac{dr}{r} &  &  & \\
 & \ddots & \ddots & \ddots &  \\
 &  &  1-\frac{dr}{r} & -2 & 1+\frac{dr}{r} & \\
 & & & & 0 \\
\end{bmatrix} && 
    \vec{T}_{core} = \begin{bmatrix}
        T_{r=0} \\
        T_{r=dr} \\
        \vdots \\
        T_{CMB}
    \end{bmatrix}
\end{align}
and $\mathbf{M_m}$ and $\vec{T}_{mantle}$ in the mantle
\begin{align} \mathbf{M}_m = \kappa_{m}\begin{bmatrix}
0 & & & & & \\
1-\frac{dr}{r} & -2 & 1+\frac{dr}{r} &  &  & \\
 & \ddots & \ddots & \ddots &  \\
 &  &  1-\frac{dr}{r} & -2 & 1+\frac{dr}{r} & \\
 & & & & 0 \\
\end{bmatrix} && 
    \vec{T}_{mantle} = \begin{bmatrix}
        T_{CMB} \\
        T_{r_c+dr} \\
        \vdots \\
        T_{s}
    \end{bmatrix}.
\end{align} The last entry in $\mathbf{M}_c$ is 0 alongside the first and last entry in $\mathbf{M}_m$ because the CMB temperature is calculated separately by balancing the core and mantle heat fluxes and the surface temperature is fixed to 200\,K. The boundary condition at the top of the core is a fixed temperature boundary condition using the CMB temperature from the previous timestep. This CMB temperature is then updated by balancing the heat fluxes across the CMB. The small timestep ($0.075\tau_{cond,core}$) ensures consistency between the flux condition and the fixed temperature boundary condition for an individual timestep.
\subsection{Timestep and gridsize testing}
\begin{figure}
    \centering
    \includegraphics[width=1\textwidth]{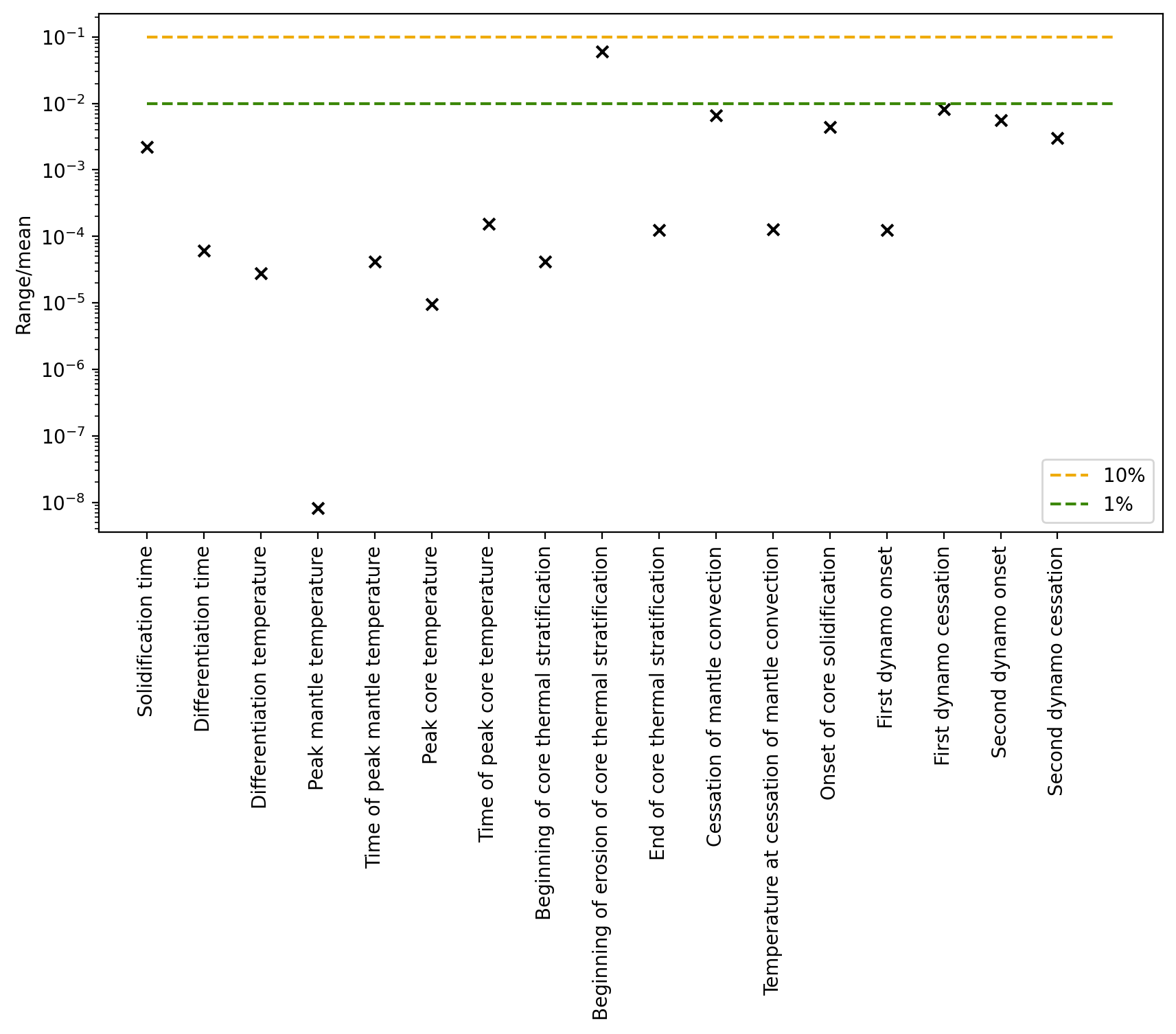}
    \caption[Timestep tests]{Variation in temperatures and times of key points in thermal evolution and dynamo generation for changes in gridsize. For each output value, the range in output divided by the mean output value is plotted. Changing the timestep and gridsize results in $<1$\% variation in model outputs, except for the time for beginning of the erosion of core thermal stratification. The percentage range in the time for beginning of the erosion of core thermal stratification is larger, because variation by 0.1\,Ma (one output step) is a large percentage of the time of this process (1.6--1.7\,Ma after CAI formation). }
    \label{fig:timestep-test}
\end{figure}
To test the gridsize and timestep, four simulations were run for a 300km body: one using the model gridsize and timestep; one for half the gridsize; one for two thirds the timestep; and one for one third the timestep. Changing the timestep and gridsize results in $<$1\% variation in model outputs, except for the time for beginning of the erosion of core thermal stratification. There is a 0.1\,Ma (one output step) difference in the time for beginning of the erosion of core thermal stratification. Core stratification begins to be eroded very early in the thermal evolution (1.6--1.7\,Ma after CAI formation). Therefore, this one output step time difference manifests as a large percentage variation, but in absolute terms is not significant. The small variation in model outputs indicates our choice of timestep and gridsize is small enough to not influence any our model results.

\bibliography{Paper1}